\documentclass[aos,preprint]{imsart}
\pdfoutput=1  

\RequirePackage{amsthm,amsmath,amsfonts,amssymb}
\RequirePackage[authoryear]{natbib}
\RequirePackage[colorlinks,citecolor=blue,urlcolor=blue]{hyperref}
\RequirePackage{graphicx}
\usepackage{comment}
\usepackage{xcolor}
\usepackage{enumerate}
\usepackage{mathtools}
\usepackage{multirow}
\usepackage{graphicx}
\usepackage{pifont}
\usepackage{float}
\DeclarePairedDelimiter{\ceil}{\lceil}{\rceil}
\definecolor{navyblue}{rgb}{0.0, 0.0, 0.5}
\definecolor{bleudefrance}{rgb}{0.19, 0.55, 0.91}
\definecolor{darkpastelgreen}{rgb}{0.01, 0.75, 0.24}
\definecolor{darkpink}{rgb}{0.91, 0.33, 0.5}
\definecolor{bluegreen}{rgb}{0.0, 0.87, 0.87}

\startlocaldefs
\theoremstyle{plain}

\newtheorem{theorem}{Theorem}[section]
\newtheorem{lemma}{Lemma}[section]
\newtheorem{proposition}{Proposition}[section]
\newtheorem{corollary}{Corollary}[section]
\theoremstyle{remark}

\newtheorem{assumption}{Assumption}[section]
\newtheorem{remark}{Remark}[section]

\newcommand{\customfootnotetext}[2]{{
		\renewcommand{\thefootnote}{#1}
		\footnotetext[0]{#2}}}

\newcommand{\cmark}{\ding{51}}%
\newcommand{\xmark}{\ding{55}}%
\def\independenT#1#2{\mathrel{\rlap{$#1#2$}\mkern2mu{#1#2}}}

\def\rR{\mathbb{R}}

\def\kK{\mathbb{K}}

\def\I{{\cal I}}

\def\G{{\cal G}}

\def\Z{{\cal Z}}

\def\boxit#1{\vbox{\hrule\hbox{\vrule\kern6pt  \vbox{\kern6pt#1\kern6pt}\kern6pt\vrule}\hrule}}

\def\sumi{\hbox{$\sum_{i=1}^n$}}

\def\diag{\hbox{diag}}
\def\wh{\widehat}

\def\diag{\hbox{diag}}
\def\log{\hbox{log}}

\def\var{\hbox{var}}
\def\cov{\hbox{cov}}

\def\trace{\hbox{trace}}

\def\be{\begin{eqnarray}}
	\def\ee{\end{eqnarray}}
\def\bq{\begin{equation}}
	\def\eq{\end{equation}}

\def\wh{\widehat}
\def\trans{^{\rm T}}

\def\b1e{{\mathbf e}}

\def\bzero{{\mathbf 0}}

\newcommand{\bgamma}{\mbox{\boldmath $\gamma$}}
\newcommand{\bxi}{\mbox{\boldmath $\xi$}}
\newcommand{\bzeta}{\mbox{\boldmath $\zeta$}}
\newcommand{\bchi}{\mbox{\boldmath $\chi$}}

\newcommand{\balpha}{\mbox{\boldmath $\alpha$}}
\newcommand{\bbeta}{\mbox{\boldmath $\beta$}}

\newcommand{\bphi}{\mbox{\boldmath $\phi$}}

\def\bfa{{\bf a}}

\def\bfg{{\bf g}}

\def\bfr{{\bf r}}

\def\bft{{\bf t}}

\def\bfU{{\bf U}}
\def\bfv{{\bf v}}
\def\bfV{{\bf V}}

\def\bfZ{{\bf Z}}

\def\trans{^{\rm T}}

\def\b1e{{\mathbf e}}

\def\W{{\mathbf W}}
\def\w{{\mathbf w}}
\def\x{{\mathbf x}}
\def\X{{\mathbf X}}
\def\s{{\mathbf s}}
\def\S{{\mathbf S}}
\def\P{{\mathbb P}}

\def\Z{{\mathbf Z}}

\def\bzero{{\mathbf 0}}

\def\tmu{\widetilde{m}}

\renewcommand{\hat}{\widehat}
\renewcommand{\tilde}{\widetilde}

\newcommand{\hD}{\hat{D}_N}
\newcommand{\hl}{\hat{\ell}_{n,k}}
\newcommand{\hlo}{\hat{\ell}^{(1)}_{n,k}}
\newcommand{\hlz}{\hat{\ell}^{(0)}_{n,k}}
\newcommand{\hlt}{\hat{\ell}^{(t)}_{n,k}}

\newcommand{\hateo}{\hat{e}^{(1)}_{n,k}}
\newcommand{\hatez}{\hat{e}^{(0)}_{n,k}}
\newcommand{\hatet}{\hat{e}^{(t)}_{n,k}}
\newcommand{\lnk}{\ell_{n,k}}
\newcommand{\enk}{e_{n,k}}
\newcommand{\hf}{\hat{f}_n}

\newcommand{\phis}{\phi^*}
\newcommand{\tphi}{\tilde{\phi}}
\newcommand{\phihatn}{\hat{\phi}_n}
\newcommand{\phihatnk}{\hat{\phi}_{n,k}}

\newcommand{\pihatN}{\hat{\pi}_N}
\newcommand{\pihatn}{\hat{\pi}_n}
\newcommand{\pihatnk}{\hat{\pi}_{n,k}}
\newcommand{\pis}{\pi^*}

\newcommand{\vt}{\theta_0}

\newcommand{\bfeta}{{\boldsymbol\eta}}

\newcommand{\hvt}{\hat{\theta}}
\newcommand{\thetahatsup}{\hvt_{\mbox{\tiny SUP}}}
\newcommand{\thetahatss}{\hvt_{\mbox{\tiny SS}}}
\newcommand{\thetahatora}{\hvt_{\mbox{\tiny ORA}}}
\newcommand{\muhatsup}{\hat{\mu}_{\mbox{\tiny SUP}}}
\newcommand{\muhatss}{\hat{\mu}_{\mbox{\tiny SS}}}
\newcommand{\muhatora}{\hat{\mu}_{\mbox{\tiny ORA}}}

\newcommand{\thetahatinit}{\hvt_{\mbox{\tiny INIT}}}
\newcommand{\lams}{\lambda_{\mbox{\tiny SUP}}}
\newcommand{\lamss}{\lambda_{\mbox{\tiny SS}}}

\newcommand{\lamate}{\lambda_{\mbox{\tiny ATE}}}
\newcommand{\zes}{\zeta_{\mbox{\tiny SUP}}}
\newcommand{\zess}{\zeta_{\mbox{\tiny SS}}}

\newcommand{\oms}{\omega_{\mbox{\tiny SUP}}}
\newcommand{\omss}{\omega_{\mbox{\tiny SS}}}

\newcommand{\sigsup}{\sigma_{\mbox{\tiny SUP}}}
\newcommand{\sigss}{\sigma_{\mbox{\tiny SS}}}

\newcommand{\sigqte}{\sigma_{\mbox{\tiny QTE}}}

\newcommand{\mhatn}{\hat{m}_n}
\newcommand{\mhatnk}{\hat{m}_{n,k}}

\newcommand{\tvt}{\tilde{\theta}}

\newcommand{\hpsi}{\hat{\psi}_{n,k}}

\newcommand{\nn}{n^{-1}}

\newcommand{\bgammahat}{\wh{\bgamma}}
\newcommand{\bxihat}{\wh{\bxi}}
\newcommand{\hg}{\wh{g}}

\newcommand{\mbG}{\mathbb{G}}
\newcommand{\mbP}{\mathbf{P}}
\newcommand{\hmbP}{\hat{\mathbf{P}}_k}
\newcommand{\E}{\mathbb{E}}

\def\ms{\mathcal{S}}
\def\mp{\mathcal{P}}
\def\md{ {\cal D}}
\def\mx{ {\cal X}}
\def\mb{ \mathcal{B}}

\def\cl{ \mathcal{L}}
\def\cu{ \mathcal{U}}
\def\ct{ \mathcal{T}}

\def\sl{\hbox{$\sum_{i=1}^n$}}
\def\slk{\hbox{$\sum_{i\in\I_{k}^-}$}}

\def\sk{\hbox{$\sum_{k=1}^\kK$}}

\def\mm{\mathcal{M}}

\def\bze{\mathbf{0}}
\def\bon{\mathbf{1}}
\def\mbtv{\mb(\vt,\v)}

\def\sg{\hbox{$\sup_{\bgamma\in\ct}$}}

\def\sgx{\hbox{$\sup_{\x\in\mx,\,\bgamma\in\ct}$}}

\def\sb{\hbox{$\sup_{\theta\in\mbtv}$}}

\def\sbx{\hbox{$\sup_{\x\in\mx,\,\theta\in\mbtv}$}}
\def\sx{\hbox{$\sup_{\x\in\mx}$}}
\def\sxx{\hbox{$\sup_{\x,\X\in\mx}$}}

\def\ss{\hbox{$\sup_{\s\in\ms}$}}
\def\ssx{\hbox{$\sup_{\s\in\ms\,\x\in\mx}$}}

\def\hL{\hat{L}_n}

\def\mn{\mathcal{N}}
\def\Enk{\E_{n,k}}
\def\bmu{\mathbf{M}}
\def\bPsi{\mbox{\boldmath $\Psi$}}
\def\bbetahat{\hat{\bbeta}}
\def\Xarrow{\overrightarrow{\X}}
\def\btheta{\boldsymbol \theta}

\def\bthetahatss{\hat{\btheta}_{\mbox{\tiny SS}}}
\def\bthetahatinit{\hat{\btheta}_{\mbox{\tiny INIT}}}
\def\bpsi{\boldsymbol{\psi}}

\def\bphi{\boldsymbol{\phi}}
\def\bphihat{\hat{\bphi}}
\def\bH{\mathbf{H}}
\def\bHhat{\hat{\bH}}
\def\bomegass{\boldsymbol{\omega}_{\mbox{\tiny SS}}}
\def\bomegasup{\boldsymbol{\omega}_{\mbox{\tiny SUP}}}

\makeatletter

\def\trans{^{\rm T}}

\def\boxit#1{\vbox{\hrule\hbox{\vrule\kern6pt  \vbox{\kern6pt#1\kern6pt}\kern6pt\vrule}\hrule}}

\def\rcom#1 {{\color{red}\bf#1} }
\def\bcom#1 {{\color{blue}\bf#1} }

\def\v{{\varepsilon}}

\def\bse{\begin{eqnarray*}}
	\def\ese{\end{eqnarray*}}
\def\be{\begin{eqnarray}}
	\def\ee{\end{eqnarray}}
\newlength{\gnat}
\baselineskip=\gnat

\def\ghat{\widehat{g}}
\newcommand\ind{\protect\mathpalette{\protect\independenT}{\perp}}
\def\independenT#1#2{\mathrel{\rlap{$#1#2$}\mkern4mu{#1#2}}}

\def\bbeta{\boldsymbol{\beta}}
\def\mbY{\mathbb{Y}}
\def\R{\mathbb{R}}
\def\mq{\mathcal{Q}_{n,N,k}}
\def\mh{\mathcal{H}_{N}}

\def\tcr{\textcolor{red}}
\def\tcm{\textcolor{magenta}}

\def\tcn{\textcolor{teal}}
\def\tcg{\textcolor{bluegreen}}
\def\tcn{\textcolor{navyblue}}

\def\cred{\color{red}}
\def\cmag{\color{magenta}}

\def\tcmAC{\textcolor{blue}} 

\def\tcr{\textcolor{black}}
\def\tcg{\textcolor{black}}
\def\tcm{\textcolor{black}}  
\def\tcmAC{\textcolor{black}}

\def\cred{\color{black}}
\def\cmag{\color{black}}

\endlocaldefs

\begin{document}
	
	\begin{frontmatter}
		\title{A General Framework for Treatment Effect Estimation in Semi-supervised and High Dimensional Settings}
		
		\runtitle{Semi-supervised Treatment Effect Estimation}
		
		\begin{aug}
			\author[A]{\fnms{Abhishek} \snm{Chakrabortty}\ead[label=e1]{abhishek@stat.tamu.edu}}
			\and
			\author[B]{\fnms{Guorong} \snm{Dai}\ead[label=e2]{guorongdai@fudan.edu.cn}}
			
			\address[A]{Department of Statistics, Texas A\&M University, \href{mailto:abhishek@stat.tamu.edu}{abhishek@stat.tamu.edu}}
		
		\address[B]{Department of Statistics and Data Science, School of Management, Fudan University, \href{mailto:guorongdai@fudan.edu.cn}{guorongdai@fudan.edu.cn}}

\customfootnotetext{\tcr{$^1$}}{\tcr{AC's research was supported in part by the National Science Foundation grant NSF DMS-2113768.}}
\customfootnotetext{$^2$}{Guorong Dai is a joint first author. Guorong Dai is the corresponding author. 
	\vspace{0.05in}
}
\end{aug}

\vspace{-0.05in} 
\begin{abstract}
In this article, we \tcr{aim to provide a general and complete understanding of {\it semi-supervised} (SS) causal inference for treatment effects, using two such estimands as prototype cases. Specifically,} we consider estimation of\tcr{:} (a) \tcr{the} {\it average treatment effect} and (b) \tcg{the} {\it quantile treatment effect}\tcr{,} in an \tcr{SS} 
setting, which is characterized by two available data sets: (i) a {\it labeled data set} of size $n$, providing 
observations for a response and a set of potentially high dimensional covariates, \tcr{as well as 
	a binary treatment indicator;} and
(ii) \emph{an unlabeled data set} of size $N$, \emph{much \tcr{larger}} than $n$, 
\tcr{but without the response observed.} 
Using these two data sets, we develop a \tcr{\it family} of SS estimators which are guaranteed to \tcr{be:  (1)} more robust \tcr{\emph{and} (2) more} efficient\tcr{,} than their supervised counterparts \tcr{based on the} 
the labeled data set only. \tcg{\tcr{Moreover, b}eyond the ``standard'' double robustness results \tcr{(in terms of consistency)} that can be achieved by supervised methods as well,} 
we further establish  \tcr{\it root-$n$ consistency and asymptotic normality} of our SS estimators whenever the propensity score in the model is correctly specified, \tcg{\emph{without requiring specific forms of the nuisance functions involved}.} 
Such an improvement \tcr{in} 
robustness arises from the use of \tcr{the} massive unlabeled data, so it is generally not attainable in a purely supervised setting. In addition, our estimators are shown to be semi-parametrically efficient \tcr{also} as long as all the nuisance functions are correctly specified. 
Moreover, as an illustration of the nuisance function estimation, we consider \tcg{{inverse-probability-weighting type}} kernel smoothing estimators involving possibly unknown 
covariate 
\tcg{transformation} mechanisms,
and establish in high dimensional scenarios novel results \tcr{on} their uniform convergence rates. \tcr{These results should be of independent interest.} 
\tcg{Numerical results on both simulated and real data validate the advantage of our methods over their supervised counterparts with respect to both robustness and efficiency.}
\end{abstract}

\par\bigskip

\end{frontmatter}

\section{Introduction}\label{seci}
Semi-supervised (SS) learning has received increasing attention as one of the most promising areas in statistics and machine learning in recent years.
We refer interested readers to \citet{zhu2005semi} and \citet{chapelle2010semi} for a detailed overview on this topic, including its definition, \tcr{goals}, 
applications and \tcr{the} fast growing literature. Unlike traditional supervised or unsupervised learning settings, a\tcr{n} SS setting, as the name suggests, represents a \tcr{confluence} 
of these two kinds of settings, in the sense that it involves two data sets: (i) a {\it labeled data set} $\cl$ containing observations for an outcome $\mbY$ and a set of covariates $\X$ (that are possibly high dimensional), and (ii) a \emph{much larger} {\it unlabeled data set} $\cu$ where only $\X$ is observed. Such situations arise naturally when $\X$ is easily available for a large number of individuals while the corresponding observations for $\mbY$ are much harder to collect owing to cost or time constraints.  The SS setting is common
to a broad class of practical problems in the modern era of ``big data'', including \tcr{machine learning applications like} text mining, web page classification, speech recognition, natural language processing etc.

\tcr{Among} biomedical \tcr{applications}, SS settings have turned out to be increasingly relevant \tcr{in} 
modern integrative genomics, especially \tcr{in} 
expression quantitative trait loci \tcr{(eQTL)} studies \citep{michaelson2009detection} combining genetic association studies with gene expression profiling. 
\tcr{These} have become instrumental in understanding various important questions in genomics, including gene regulatory networks \citep{gilad2008revealing, hormozdiari2016colocalization}. However, one issue with such studies is that they are often under-powered due to the limited size of the gene expression data which are expensive \citep{flutre2013statistical}. On the other hand, records on the genetic variants are cheaper and often available for a massive cohort, thus naturally leading to SS settings while necessitating robust and efficient strategies that can leverage this extra information to produce more powerful association mapping tools as well as methods for detecting the causal effects of the genetic variants. \tcr{Moreover, SS settings also have great relevance in the analysis of electronic health records data, which are popular resources for discovery research but also suffer from a major bottleneck in obtaining validated outcomes due to logistical constraints; see\tcr{, e.g.,}
\citet{chakrabortty2018efficient} and \citet{cheng2020robust} for more details.} 

\subsection{Problem setup}\label{sec:psetup}
\tcr{In this paper, we consider causal inference problems in SS settings. To characterize the \tcr{basic} setup,} suppose our sample consists of two independent data sets: the labeled \tcr{(or supervised)} data $\cl:=\{(\mbY_i,T_i,\X_i\trans)\trans:i=1,\ldots,n\}$, and the unlabeled \tcr{(or unsupervised)} data $\cu:=\{(T_i,\X\trans_i)\trans:i=n+1,\ldots,n+N\}$ \tcr{(with $N \gg n$ possibly)}, containing \tcr{$n$ and $N$}
independent copies of $\Z:=(\mbY,T,\X\trans)\trans$ and $(T,\X\trans)\trans$, respectively, where $T\in\{0,1\}$ serves as
a {\it treatment indicator}, i.e., $T=1$ or $0$ represents whether an individual is treated or not. 
The covariates \tcr{(often also called confounders)} $\X\in\mx\subset\rR^p$ \tcr{are (possibly) high dimensional, with dimension} 
$p\equiv p_n$ allowed to diverge and \tcr{possibly exceed} 
$n$ \tcr{(including $p \gg n$)}, while the {\it observed outcome} \tcr{is given by:}
\bse
\mbY~:=~TY(1) + (1-T)Y(0),
\ese
where $Y(t)$ is the {\it potential outcome} of an individual with $T=t\in\{0,1\}$ \citep{rubin1974estimating, imbens2015causal}. \tcr{Thus, $(\mbY \mid T = t) ~\equiv~ Y(t)$ (also called the consistency assumption).} In this work, we mainly focus on the setup where in addition to the covariates, the treatment indicator is observed in the unlabeled data as well. This is the case when the treatment can be considered \emph{inherent} in the individuals and $T$ is thereby recorded in both $\cl$ and $\cu$ as a baseline feature along with $\X$. An example is the genetic study in Section \ref{sec_data_analysis} where $T$ indicates the occurrence of mutations on some position of the HIV reverse transcriptase, which is known for individuals in both the labeled and unlabeled data. Though not the main focus, we \tcr{also} consider in Section \ref{sec_ate_u_dagger} the setting where $T$ is unobserved in $\cu$.

\tcr{A} major challenge \tcr{(and a key feature)} in the above 
\tcr{framework} arises from the \tcr{(possibly)} {\it disproportion\tcr{ate} \tcr{sizes}} of 
$\cl$ and $\cu$, \tcr{namely $|\cu| \gg |\cl|$}, \tcr{an issue} widely \tcr{encountered} 
in modern \tcr{(often digitally recorded) observational} datasets \tcr{of massive sizes}, such as electronic health records
\citep{cheng2020robust}. We therefore assume \tcr{(rather, allow for)}:
\be
\nu~:=~\hbox{$\lim_{n,\tcr{N}\to\infty}$}n/(n+N)~=~0,
\label{disproportion}
\ee
as in \citet{chakrabortty2018efficient} and \citet{gronsbell2018semia}. An example \tcr{of \eqref{disproportion}} is the {\it ideal SS setting} where $n<\infty$ and $N=\infty$ (i.e., the distribution of $(T,\X\trans)\trans$ is known). Essentially, the condition \eqref{disproportion} distinguishes our framework from that of traditional missing data theory, which typically requires the proportion of complete cases in the sample to be bounded away from zero 
\tcr{-- often known as the ``positivity condition'' \citep{imbens2004nonparametric, tsiatis2007semiparametric}. The natural violation of this condition in SS settings is what makes them 
unique and more challenging than traditional missing data problems. 
\tcg{On the other hand, we \tcr{do assume} 
throughout this paper 
\tcr{that $\cl$ and $\cu$ have the same underlying distribution (i.e., $\mbY$ in $\cu$ are missing completely at random)} which is the typical (and often implicit) setup in the traditional SS literature \citep{zhu2005semi, chapelle2010semi}. 
\tcr{We formalize this below.}}} 
\begin{assumption}\label{ass_equally_distributed}
\tcg{\tcr{The observations in $\cl$ and $\cu$ have the same underlying distribution, so that $\{(\mbY_i,T_i,\X_i\trans)\trans: i=1,\ldots,n\}$ and $\{(T_i,\X_i\trans)\trans: i=n+1,\ldots,n+N\}$ respectively
	are $n$ and $N$ independent realizations from 
	the distributions of $(\mbY,T,\X\trans)\trans$ and $(T,\X\trans)\trans$.}} 
\end{assumption}

\paragraph*{Causal parameters of interest} Based on the available data $\cl\cup\cu$, we aim to estimate:
\vskip0.05in
\tcr{(i)} the {\it average treatment effect} (ATE)\tcr{:}
\be
\mu_0(1) - \mu_0(0)~:=~\E\{Y(1)\}-\E\{Y(0)\}, \tcr{~~\mbox{and}}
\label{ate}
\ee
\hspace{0.15in} \tcr{(ii)} the {\it quantile treatment effect} (QTE)\tcr{:}
\be
\vt(1,\tau)-\vt(0,\tau)~\equiv~\vt(1)-\vt(0),
\label{qte}
\ee
where $\vt(t,\tau)\equiv\vt(t)$ represents the $\tau$\tcr{-}quantile of $Y(t)$ for some fixed and known $\tau\in(0,1)$, defined as the solution to the equation\tcr{:}
\be
\E[ \psi\{Y(t),\vt(t,\tau)\}]  ~:=~  \E[I\{ Y(t) < \vt(t,\tau)\} - \tau] ~=~ 0 \quad (t=0,1)\tcr{,}
\label{defqte}
\ee
with $I(\cdot)$ \tcr{being} the indicator function. It is \tcr{worth noting that} 
by setting $T\equiv 1$ and $\mu_0(0)=\vt(0)\equiv 0$, the above problems \tcr{also}
cover SS estimation of \tcr{the} response mean \citep{zhang2019semi, zhang2019high} and quantile \citep{chakrabortty2022semi} as special cases.
\tcr{The ATE and the QTE are both well-studied choices of causal estimands in supervised settings; see Section \ref{sec_literature} for an overview of these literature(s). 
While the ATE is perhaps \tcr{the more common choice,} the QTE is \tcr{often} more useful and informative\tcr{, especially} in settings where the causal effect of the treatment is heterogeneous and/or the outcome distribution\tcr{(s)}
is highly skewed so that the average causal effect may be of limited value.}
%

\vskip0.05in
\tcr{Our goal here, in general, is to investigate how, when, and to what extent, one can exploit the full data $\cl \cup \cu$ to develop SS estimators of these parameters that can ``improve'' standard supervised approaches using $\cl$ only, where the term ``improve'' could be in terms of efficiency or robustness or both. The rest of this paper is dedicated to a thorough understanding of such questions via a complete characterization of the possible SS estimators.} 

\vskip0.05in
\tcr{We also clarify that we choose the ATE and QTE as two representative causal estimands -- presenting diverse methodological and technical challenges -- to exemplify the key features of our SS approach and its benefits, without compromising much on the clarity of the main messages. Extensions to other more general functionals (\tcmAC{such as} 
those based on general estimating equations)
are indeed possible -- as we discuss later in Section \ref{sec_conclusion_discussion} \tcmAC{and Appendix  \ref{sm_Z_estimation}} -- though we skip
\tcmAC{a detailed technical analysis}
for the sake of brevity and minimal obfuscation.}

\paragraph*{\tcr{Basic assumptions}} To 
\tcr{ensure} parameters 
\tcr{$\{\mu_0(t),\vt(t)\}_{t = 0}^1$} are identifiable and estimable \tcr{from the observed data}, we make the \tcr{following standard assumptions \citep{imbens2004nonparametric}:}
\be
T \ind \{Y(0), Y(1)\} \mid \X,\quad \mbox{and} \quad \pi(\x)~:=~\E(T\mid\X=\x)~\in(c,1-c)\tcr{,} \label{mar_positivity}
\ee
for any $\x\in\mx$ and some constant $c\in(0,1)$. \tcr{The quantity $\pi(\x)$ is also known as the \emph{propensity score} for the treatment. \eqref{mar_positivity} encodes some well known conditions \citep{imbens2015causal}.} The first part of \eqref{mar_positivity} is \tcr{often} known as the \emph{no unmeasured confounding} assumption, equivalent to the {\it missing at random} assumption in the context of missing data \citep{tsiatis2007semiparametric, little2019statistical}, while the second part is the {\it positivity} (or {\it overlap}) assumption on the treatment. 

\paragraph*{Clarification}
Considering the \tcr{corresponding case} 
of $Y(0)$ is analogous, we would henceforth focus on the mean and quantile estimation of $Y(1)$ without loss of generality,
\tcr{and} 
\be
\tcr{\mbox{let~$\{Y,\mu_0,\vt\}$~~generically denote~~$\{Y(1),\mu_0(1),\vt(1)\}$.}} \label{generic_notation}
\ee

\subsection{\tcr{Related literature} }\label{sec_literature}
\tcr{The setup and contributions of our work naturally relate to three different facets of existing literature, namely: (a) ``traditional'' (non-causal) SS inference, (b) supervised causal inference, and finally, (c) SS causal inference. Below we briefly summarize the relevant works in each of these areas, followed by a detailed account of our contributions.}

\paragraph*{SS learning and inference}
For estimation in an SS setup, the primary and most critical \tcr{goal} 
is \tcr{to investigate} when and how its robustness and efficiency can be improved, compared to {supervised} methods using the labeled data $\cl$ only, by exploiting the unlabeled data $\cu$. Chapter 2 of \citet{Chakrabortty_Thesis_2016} provided an elaborate discussion on this question, claiming that the answer is generally determined by the \tcr{nature of the relationship} 
between the parameter of interest and the marginal distribution, $\P_\X$, of $\X$\tcr{,} as $\cu$ provides information regarding $\P_\X$ only. Therefore\tcr{,} many existing algorithms \tcr{for} 
SS learning \tcr{that target} 
$\E(\mbY\mid\X)$, including, for instance, generative modeling \citep{nigam2000text, nigam2001using}, graph-based methods \citep{zhu2005semi} and manifold regularization \citep{belkin2006manifold}, rely to some extent on assumptions relating $\P_\X$ to the conditional distribution of $\mbY$ given $\X$. When these assumptions are violated, \tcr{however,} they may perform even worse than the corresponding supervised methods \citep{cozman2001unlabeled, cozman2003semi}. Such undesirable degradation highlights the need for safe usage of the unlabeled data $\cu$. To achieve this goal, \citet{chakrabortty2018efficient} advocated the {\it robust} and {\it adaptive} property for SS approaches, i.e., being consistent for the target parameters while \tcr{being} at least as efficient as their supervised counterparts and more efficient whenever possible. Adopting such a perspective explicitly or implicitly, robust and adaptive 
\tcr{procedures for} SS estimation and inference have been developed under the semi-parametric framework recently for various problems\tcr{,} including mean estimation \citep{zhang2019semi,zhang2019high}, linear regression \citep{azriel2016semi, chakrabortty2018efficient}, general $Z$-estimation \citep{kawakita2013semi, Chakrabortty_Thesis_2016}, prediction accuracy evaluation \citep{gronsbell2018semia} and covariance functionals \citep{tony2020semisupervised, chan2020semi}. \tcr{However,} different from our work considering causal inference and treatment effect estimation, most of this recent progress focused on relatively ``standard'' \tcr{(non-causal)} problems defined {\it without} the potential outcome framework \tcg{(and its ensuing challenges, e.g., confounding\tcr{,} and \tcr{the}
missingness of one of the potential outcomes 
induced by the treatment assignment \tcr{$T$})}.


\paragraph*{Average treatment effect} Both \tcr{the} ATE and \tcr{the} QTE are fundamental and popular causal estimands which have been extensively studied in the context of supervised causal inference based on a wide range of approaches; see \citet{imbens2004nonparametric} and \citet{tsiatis2007semiparametric} for an overview of the ATE literature. In particular, these include inverse probability weighted (IPW) approaches \citep{rosenbaum1983central, rosenbaum1984reducing, robins1994estimation, hahn1998role, hirano2003efficient,  ertefaie2020nonparametric} involving approximation of the propensity score $\pi(\X)$, as well as \tcr{\emph{doubly robust}} (DR) methods \citep{robins1994estimation, robins1995semiparametric, rotnitzky1998semiparametric, scharfstein1999adjusting, kang2007demystifying, vermeulen2015bias} which require estimating both $\E(Y\mid\X)$ and $\pi(\X)$. As the name implies, the DR estimators are consistent whenever one of the two nuisance models is correctly specified, while attaining the semi-parametric efficiency bound for the unrestricted model,  as long as both are correctly specified. When the number of covariates is fixed, semi-parametric inference via such DR methods has a rich literature; see \citet{bang2005doubly}, \citet{tsiatis2007semiparametric}, \citet{kang2007demystifying} and \citet{graham2011efficiency} for a review. In recent times, there has also been substantial interest in the extension of these approaches to high dimensional scenarios,
leading to a flurry of work\tcr{, e.g., \citet{farrell2015robust, chernozhukov2018double, athey2018approximate, smucler2019unifying}}. \tcg{\tcr{Most of t}hese papers generally impose one of the following two conditions on the nuisance function\tcr{s'} estimation to attain $n^{1/2}$\tcr{-}consistency and asymptotic normality for valid (supervised) inference \tcr{based on their ATE estimators}:}
\begin{enumerate}[(a)]
\item \tcg{Both $\E(Y\mid\X)$ and $\pi(\X)$ are correctly specified, and the product of  their estimators' convergence rates vanishes fast enough \tcr{(typically, faster than $n^{-1/2}$)}
\citep{belloni2014inference, farrell2015robust, belloni2017program, chernozhukov2018double}.} 

\item \hspace{-0.056in}
\tcg{Either $\E(Y\mid\X)$ or $\pi(\X)$ is correctly specified by a linear/logistic regression model\tcr{, while} some \tcr{carefully tailored} 
bias corrections are applied\tcr{,}
and some rate conditions are satisfied \tcr{as well} \citep{smucler2019unifying, tan2020model, dukes2021inference}.}
\end{enumerate}
\tcg{However, we will show that, \tcr{under our SS setup,} through
	using the massive unlabeled data, \tcr{there are some striking \emph{robustification benefits} that ensure} these requirements can be substantially relaxed, 
	\tcr{and that $n^{1/2}$-rate inference on the ATE (or QTE) can be achieved in a \emph{seamless} way, without requiring any specific forms of the nuisance model(s) or any sophisticated
		bias correction techniques under  misspecification}; see Point (I) in Section \ref{sec_contributions}.}




\paragraph*{Quantile treatment effect} \tcr{The} marginal QTE, though technically a more challenging parameter due to the \tcr{inherently} {non-smooth} nature of the quantile estimating equation \eqref{defqte}, provides a \tcr{more complete} 
picture of the causal effect on \tcr{the} outcome distribution, beyond just its mean\tcr{.}
%
\tcr{There is a fairly rich literature on \tcr{(supervised)} QTE estimation as well.}
For example, \citet{firpo2007efficient} developed an IPW estimator \tcr{that attains} 
semi-parametric efficiency under some smoothness assumptions. \citet{hsu2020qte} viewed the quantile $\vt$ \tcr{from the
perspective of the conditional distribution,} as the solution to the equation $\tau=\E\{F(\vt\mid\X)\}$\tcr{,} 
where $F(\cdot\mid\x):=\P(Y<\cdot\mid\X=\x)$. Their method thus requires estimating the whole conditional distribution
of $Y$ given $\X$. To avoid such a burdensome task, \citet{kallus2019localized} recently proposed the localized debiased machine learning approach, which only involves estimation of $F(\cdot\mid\X)$ at a preliminary estimate of the quantile and can leverage a broad range of machine learning methods besides kernel smoothing used by \citet{hsu2020qte}. Moreover, \citet{zhang2012causal} compared methods based on the propensity score $\pi(\X)$ and the conditional distribution $F(\cdot\mid\X)$. They also devised a DR estimator for the QTE under parametric specification of $\pi(\X)$ and $F(\cdot\mid\X)$. Nevertheless, all \tcr{these} aforementioned work\tcr{s are still} 
restricted to the supervised domain involving only \tcr{the} labeled data $\cl$.

\paragraph*{SS inference for treatment effect\tcr{s}} Although there \tcr{has} 
been \tcr{work} on a variety of problems in SS settings, as listed in the first paragraph of Section \ref{sec_literature}, less attention, however, has been paid to causal inference and treatment effect estimation \tcr{problems}, except \tcr{for some 
(very recent)} progress \citep{zhang2019high, kallus2020role, cheng2020robust}. When there exist post-treatment surrogate variables that are potentially predictive of the outcome, \citet{cheng2020robust} combined imputing and inverse probability weighting, building on the\tcr{ir} technique of ``double-index'' propensity scores \citep{cheng2020estimating}, to devise \tcr{an IPW-type} SS estimator for \tcr{the} ATE, which is doubly robust. \tcr{Though not explicitly stated, their approach, however, only applies to low dimensional $(p \ll n)$ settings, and more importantly, their estimator being of an IPW type,  does not have a naturally ``orthogonal'' structure (in the sense of \citet{chernozhukov2018double}), and therefore, is not first order insensitive to estimation errors of the nuisance functions, unlike our proposed approach. This feature is particularly crucial in situations involving  high dimensional and/or non-parametric nuisance estimators.} \citet{kallus2020role} also considered the role of surrogates in SS estimation of the ATE, but \tcr{mostly} 
in cases where the labeling fractions are bounded below. Further, with a largely theoretical focus, their main aims were characterizations of efficiency and optimality\tcr{,} rather than \tcr{implementation.}
%
\tcg{In a setting similar to \citet{kallus2020role}, with surrogates available, \citet{hou2021efficient}, a very recent work we noticed at the final stage\tcr{s} of our preparation \tcr{of} 
this paper,
\tcr{also} developed 
SS estimators \tcr{for} 
the ATE. Unlike our data structure\tcr{,} where $\cu$ provides observations for both $\X$ and $T$, \citet{hou2021efficient} 
assumed 
the treatment indicator is missing in the unlabeled data, \tcr{and} so their estimators have fairly different robustness guarantees from ours. This case, with $T$ unobserved in $\cu$, is not of our \tcr{primary} interest\tcr{.} 
But we will briefly address \tcr{it as well} in Section \ref{sec_ate_u_dagger}.}
Lastly, \citet{zhang2019high} extended their SS mean estimation method using a linear working model for $\E(Y \mid \X)$ to the case of the ATE. While all these articles mostly investigated the efficiency of their approaches, none of them clarified the potential gain of \tcr{\it robustness} from leveraging \tcr{the} unlabeled data $\cu$. In addition, \tcg{\citet{zhang2019high} and \citet{cheng2020robust}
mainly focused on some specific working models for $\E(Y\mid\X)$ and/or $\pi(\X)$}, and \citet{zhang2019high} 
\tcr{only} briefly discussed the
ATE estimation \tcr{problem --}
as an illustration of their SS mean estimation approach; see Remark \ref{remark_comparison_zhang2019} for a more
detailed comparison of our work \tcr{with} 
\citet{zhang2019high}.

\vskip0.05in
\tcr{As for} \tcr{the} QTE, its SS estimation has, to the best of our knowledge, not been studied in \tcr{any of} the existing works. \tcr{Our work here appears to be the first contribution in this regard.}

\subsection{Our contributions}\label{sec_contributions}
\tcr{This paper aims to bridge some of these major gaps in the existing literature, towards a better and unified understanding -- both methodological and theoretical -- of SS causal inference and its benefits. We summarize our main contributions below.}

\begin{enumerate}[(I)]

\item We develop under the SS setting \eqref{disproportion} a {family} of DR estimators for\tcr{:} (a) the ATE \tcr{(Section \ref{secos})} and (b) the QTE \tcr{(Section \ref{secqte})}, which take the {whole} data $\cl\cup\cu$ into consideration and enable us to employ arbitrary methods for estimating the nuisance functions as long as some high level conditions are satisfied.  \tcr{These estimators, apart from affording a {flexible} and {general} construction (\tcr{involving} 
imputation and IPW strategies, along with \tcr{the}
use of cross fitting, applied to $\cl\cup\cu$), also enjoy several desirable properties and advantages.} In addition to \tcr{being} DR in \tcr{terms of} consistency, we further prove that, whenever the propensity score $\pi(\X)$ is correctly \tcg{specified and estimated at a suitably fast rate} \tcr{-- something that is indeed {achievable} under our SS setting as clarified in Remark \ref{remark_sN},}
%
our estimators are {$n^{1/2}$-consistent and asymptotically normal} {even if the outcome model is misspecified \tcg{and none of the nuisance functions has a specific (\tcr{e.g.,} linear$/$logistic) form}}; see Theorems \ref{thate} and \ref{thqte} as well as Corollaries \ref{corate} and \ref{corqte}, along with the discussions in the \tcr{subsequent Remarks \ref{remark_ate_robustness} and \ref{remark_qte_property}.} 
{Agnostic to the construction of nuisance function estimators}, this \tcr{robustness} property \tcr{-- a {$n^{1/2}$-rate robustness property} of sorts --} is particularly desirable for inference, \tcg{while \emph{generally not achievable in purely supervised settings} \tcr{without extra targeted (and nuanced) bias corrections which \tcr{do}
require specific (linear$/$logistic) forms of the nuisance function estimators along with 
\tcr{other} conditions, as discussed in our review of (supervised) ATE estimation in Section \ref{sec_literature}.}}
%
\tcg{In contrast, our \tcr{SS approach is} 
much more flexible \tcr{and seamless}, allowing for {any} reasonable strategies (parametric, semi-parametric or non-parametric) 
\tcr{for estimating} the nuisance functions.}
Moreover, \tcr{even if this improvement in robustness is set aside}, 
our \tcr{SS} estimators are 
ensured to be \emph{more efficient} than their supervised counterparts, 
and are \tcr{also} semi-parametrically \emph{optimal} when correctly specifying both the propensity score $\pi(\X)$ and the outcome model, i.e., $\E(Y\mid\X)$ or $F(\cdot\mid\X)$ for the ATE or the QTE, respectively\tcr{; see Remarks \ref{remark_ate_efficiency} and \ref{remark_qte_efficiency}, in particular, regarding these efficiency claims, and Table \ref{table_ate_summary} for a full 
characterization of the robustness and efficiency benefits of our SS estimators.}
%

\vskip0.1in
\item Compared to the case of the ATE, the QTE estimation is substantially more {challenging} in both theory and implementation due to the non-separability of $Y$ and $\theta$ in the quantile estimating equation \eqref{defqte}. To overcome these difficulties, we establish novel results of
empirical process theory for deriving the properties of our QTE estimators; see Lemma \ref{1v2} in \tcr{Appendix} \ref{sm_lemmas}.
In addition, we adopt the strategy of {one-step update} \citep{van2000asymptotic, tsiatis2007semiparametric} in the construction of our QTE estimators to facilitate computation. This strategy also avoids the laborious task of recovering the conditional distribution function $F(\cdot\mid\X)$ for the whole parameter space of $\theta_0$. Instead, we {only} need to estimate $F(\cdot\mid\X)$ at one {single} point. Such an advantage was advocated by \citet{kallus2019localized} as well. \tcr{Our QTE (as well as ATE) estimators thus have 
{simple implementations}, in general.}

\vskip0.1in
\item \tcr{Finally, a}nother major contribution \tcr{of this work, though of a somewhat different flavor,} 
\tcr{are our results on} the {nuisance function\tcr{s'} estimation} \tcr{(Section \ref{secnf})} \tcr{-- an important component in all our SS estimators' implementation --} for which we consider a {variety} of reasonable and flexible approaches\tcr{,} including kernel smoothing \tcr{(with possible use of dimension reduction)}, parametric regression and random forest. \tcr{In particular,} as a \tcr{detailed} illustration, we verify the high-level conditions required by our methods for \emph{IPW type kernel smoothing estimators with \tcr{so-called} ``generated'' covariates} \citep{mammen2012nonparametric, escanciano2014uniform, mammen_rothe_schienle_2016} \tcr{involving (unknown) transformations of covariates}. Specifically, we investigate in detail their {uniform \tcr{($L_{\infty}$)} convergence rates, extending the existing theory to cases involving high dimensionality and \tcg{IPW schemes that need to be estimated}}; see Theorems \ref{theorem_ks_ate} and \ref{thhd}. \tcg{These 
results are novel to the best of our knowledge, and can be 
applicable more generally in other problems. Thus they should be 
of independent interest.}
%
\end{enumerate}

\subsection{Organization \tcr{of the rest} of the article} We introduce our \tcr{family of}
SS estimators for (a) the ATE and (b) the QTE, as well as establish 
their asymptotic properties, in Sections \ref{secos} and \ref{secqte}, respectively. Then the choice and estimation of \tcr{the} nuisance functions \tcr{involved} in our approaches, \tcr{along with their theoretical properties}, are discussed in Section \ref{secnf}. Section \ref{sec_simulations} presents \tcr{detailed} simulation results \tcr{under various data generating settings to validate the claimed properties and improvements} of our proposed methods, followed by an empirical data example in Section \ref{sec_data_analysis}. 
Concluding remark\tcr{s} along with discussion\tcr{s} on possible extension\tcr{s} of our work \tcr{are} provided 
\tcr{in} Section \ref{sec_conclusion_discussion}. \tcm{
\tcmAC{Further details on extending our SS approaches} to more general causal estimands, as well as} \tcr{all} technical materials, \tcr{including proofs of all results}, and further numerical results, can be found in the Supplementary Material \tcr{(Appendices \ref{sm_Z_estimation}--\ref{sm_data_analysis})}.

\section{SS estimation for the ATE}\label{secos}
\tcr{Following our clarification at the end of Section \ref{sec:psetup}, it suffices to focus only on the SS estimation of $\mu_0$, as in \eqref{generic_notation}, which will be our primary goal in Sections \ref{sec_ate_sup}--\ref{sec_ate_u_dagger}, after which we formally address SS inference for the ATE in Section \ref{sec_ate_difference}.}

\vspace{-0.02in}
\subsection*{Notation\tcr{s}}
\tcr{We first introduce some notations that will be used throughout the paper.} 
\tcr{We} use the lower letter $c$ to represent a generic positive constant, including $c_1$, $c_2$, etc, which may vary from line to line. For a $d_1\times d_2$ matrix $\mbP$ whose $(i,j)$th component is $\mbP_{[ij]}$, \tcr{we} let
\bse
&&\hbox{$\|\mbP\|_0~:=~\max_{1\leq j\leq d_2}\{\sum_{i=1}^{d_1}I(\mbP_{[ij]}\neq 0)\},~~ \|\mbP\|_1~:=~\max_{1\leq j\leq d_2}(\sum_{i=1}^{d_1}|\mbP_{[ij]}|)$}, \\
&&\|\mbP\|~:=~\hbox{$\max_{1\leq j\leq d_2}\{(\sum_{i=1}^{d_1}\mbP_{[ij]}^2)^{1/2}\},~~ \tcr{\mbox{and}}~~  \|\mbP\|_\infty~:=~\max_{1\leq i\leq d_1, 1\leq j\leq d_2}|\mbP_{[ij]}|$}.
\ese
The bold numbers $\bon_d$ and $\bze_d$ refer to $d$-dimensional vectors of ones and zeros, respectively. \tcr{We d}enote $\mb(\balpha,\v):=\{\bfa:\|\bfa-\balpha\|\leq\v\}$ as a generic neighborhood of a vector $\balpha$ with some \tcr{radius} $\v>0$. We use $\balpha_{[j]}$ to 
\tcr{denote} the $j$th component of a vector $\balpha$. For two data sets $\ms_1$ and $\ms_2$, \tcr{we} define $\P_{\ms_1}(\cdot\mid\ms_2)$ as the conditional probability with respect to $\ms_1$ given $\ms_2$. For any random function $\hg(\cdot,\theta)$ and a random vector $\W$ with copies $\W_1,\ldots,\W_{n+N}$, \tcr{we} denote
\bse
\E_{\W}\{\hg(\W,\theta)\}~:=~ \hbox{$\int$} \hg(\w, \theta) d\,\P_{\W}(\w)
\ese
as the expectation of $\hg(\W,\theta)$ with respect to $\W$\tcr{,} treating $\hg(\cdot,\theta)$ as a non\tcr{-}random function, where $\P_{\W}(\cdot)$ is the distribution function of $\W $. For $M\in\{n,n+N\}$, \tcr{we} write
\bse
&&\E_M\{\hg(\W,\theta)\}~:=~ M^{-1}\hbox{$\sum_{i=1}^M$} \hg(\W_i,\theta),\\
&&\mbG_M\{\hg(\W,\theta)\}~:=~ M^{1/2}[\E_M\{\hg(\W,\theta)\}-\E_\W\{\hg(\W,\theta)\}], ~~\tcr{\mbox{and}}\\ &&\var_M\{\hg(\W,\theta)\}~:=~\E_M[\{\hg(\W,\theta)\}^2]-[\E_M\{\hg(\W,\theta)\}]^2.
\ese
Also, \tcr{we} define
\bse
&&\E_N\{\hg(\W,\theta)\}~:=~ N^{-1}\hbox{$\sum_{i=n+1}^{n+N}$} \hg(\W_i,\theta), ~~\tcr{\mbox{and}}\\
&&\mbG_N\{\hg(\W,\theta)\}~:=~ N^{1/2}[\E_N\{\hg(\W,\theta)\}-\E_\W\{\hg(\W,\theta)\}].
\ese
Lastly, we \tcr{let} $f(\cdot)$ and $F(\cdot)$ \tcr{denote} 
the density and distribution functions of $Y$, while 
$f(\cdot\mid\w)$ and $F(\cdot\mid\w)$ represent the conditional density and distribution functions of $Y$ given $\W=\w$.

\subsection{Supervised \tcg{estimator}} \label{sec_ate_sup}
As noted earlier, 
for estimating \tcr{the} ATE, we \tcr{can simply} focus on $\mu_0\equiv\E(Y)$ with $Y\equiv Y(1)$. To this end, we first observe the following representation \tcr{(and identification)} of $\mu_0$. Let $m(\X):=\E(Y\mid\X)$ \tcr{and recall $\pi(\X) \equiv \E(T\mid \X)$. 
We then have:}
\bse
\mu_0 &~=~& \E\{m(\X) \}+ \E[\{\pis(\X)\}^{-1} T\{Y-m(\X)\}] \\
& ~=~& \E\{ m^*(\X) \} + \E[\{\pi(\X)\}^{-1} T\{Y-m^*(\X)\}]\tcr{,}
\ese
for some \emph{arbitrary} functions $\pis(\cdot)$ and $m^*(\cdot)$, implying that the equivalence\tcr{:}
\be
\mu_0 &~=~& \E\{m^*(\X) \}+ \E[\{\pis(\X)\}^{-1} T\{Y -   m^*(\X) \}]
\label{ate_dr_representation}
\ee
holds given either $\pis(\X)=\pi(\X)$ or $ m^*(\X) = m(\X) $ but {\it not} necessarily both. The equation \eqref{ate_dr_representation} is thus a DR representation of $\mu_0$\tcr{, involving the nuisance functions $\pi(\cdot)$ and $m(\cdot)$}. Using the empirical version of \eqref{ate_dr_representation} based on $\cl$ precisely leads to the traditional DR estimator of the mean $\mu_0$ \citep{bang2005doubly, chernozhukov2018double}, i.e., the \emph{supervised estimator}
\be
\muhatsup~:=~\E_n\{\mhatn(\X)\}+\E_n[\{\pihatn(\X)\}^{-1}T\{Y-\mhatn(\X)\}], ~~\tcr{\mbox{where}}
\label{sup_ate}
\ee
$\{\pihatn(\cdot),\mhatn(\cdot)\}$ are \tcr{some} estimators of $\{\pi(\cdot),\mu(\cdot)\}$ from $\cl$ with possibly misspecified limits $\{\pis(\cdot),m^*(\cdot)\}$. Apart from \tcr{being} DR, the estimator $\muhatsup$ also possesses the two nice properties below as long as \tcr{the} models for $\{\pi(\cdot),\mu(\cdot)\}$ are \tcr{both} correctly specified and 
\tcr{certain rate conditions \citep{chernozhukov2018double} on the convergence of $\{\pihatn(\cdot),\mhatn(\cdot)\}$} are satisfied.
\begin{enumerate}[(i)]
\item First-order insensitivity 
\tcr{-- When both nuisance models are correctly specified, t}he influence function of $\muhatsup$ is not affected by the estimation errors of $\{\pihatn(\cdot),\mhatn(\cdot)\}$ \citep{robins1995semiparametric, chernozhukov2018double, chakrabortty2019high}. This feature is directly relevant to the {\it debiasing} term $\E_n[\{\pihatn(\X)\}^{-1}T\{Y-\mhatn(\X)\}]$ in \eqref{sup_ate} and is desirable for inference, particularly when the construction of $\{\pihatn(\cdot),\mhatn(\cdot)\}$ involves non-parametric calibrations or \tcr{if} $\X$ is high dimensional \tcr{(leading to rates slower than $n^{-1/2}$)}.


\par\smallskip
\item Semi-parametric optimality among all regular and asymptotically linear estimators for
$\mu_0$ \tcr{--} $\muhatsup$ attains the semi-parametric efficiency bound for estimating $\mu_0$ under a fully non-parametric (i.e., unrestricted up to the condition \eqref{mar_positivity}) family \tcr{of} 
distributions of $(Y,T,\X\trans)\trans$ \citep{robins1994estimation, robins1995semiparametric, graham2011efficiency}\tcr{.}
\end{enumerate}
In the sense of the above advantages, $\muhatsup$ is the ``best'' achievable estimator for $\mu_0$ under a purely supervised setting \tcr{\citep{robins1995semiparametric, chernozhukov2018double}}.

\subsection[A family of SS estimators \tcr{for mu0}]{A family of SS estimators \tcr{for $\mu_0$}}\label{sec_ate_ss}
Despite the above desirable properties, the supervised DR estimator $\muhatsup$ may, however, be suboptimal when the unlabeled data $\cu$ is available, owing to ignoring the extra observations for $(T,\X\trans)\trans$ \tcr{therein}. An intuitive interpretation is that, since $\E(Y-\mu_0\mid\X)\neq 0$ with a positive probability if we exclude the trivial case where $\E(Y\mid\X)=\mu$ almost surely, the marginal distribution $\P_\X$ of $\X$ actually plays a role in the definition of $\mu_0$ and the information of $\P_\X$ provided by $\cu$ can therefore help estimate $\mu_0$; see Chapter 2 of \citet{Chakrabortty_Thesis_2016} for further insights in a more general context.

\tcr{To} utilize $\cu$, we notice that the term $\E_n\{\mhatn(\X)\}$ in \eqref{sup_ate} can be replaced by $\E_{n+N}\{\mhatn(\X)\}$ which integrates $\cl$ and $\cu$. Moreover, estimation of the propensity score can certainly be improved by using $\cu$ as well, since $\pi(\X)$ is entirely determined by the distribution of $(T,\X\trans)\trans$. \tcr{This provides a much better chance to estimate $\pi(\cdot)$ more \emph{robustly} (possibly at a faster rate!).}

\vskip0.05in 
\tcr{Thus,} with 
\tcr{{any} estimators (with possibly misspecified limits)} $\pihatN(\cdot)$ for $\pi(\cdot)$, based on $\cu$, and $\mhatn(\cdot)$  \tcr{for $m(\cdot)$} from $\cl$\tcr{,} same as before, we propose a family of \tcr{\emph{SS estimators} of $\mu_0$:} 
\be
\muhatss~:=~\E_{n+N}\{\mhatn(\X)\}+\E_n[\{\pihatN(\X)\}^{-1}T\{Y-\mhatn(\X)\}]\tcr{,} \label{ss_ate}
\ee
indexed by $\{\pihatN(\cdot),\mhatn(\cdot)\}$. Here\tcr{,} we apply the strategy of \tcr{\emph{cross fitting}} \citep{chernozhukov2018double, newey2018cross} when estimating $\mhatn(\cdot)$. Specifically, for some fixed integer $\kK\geq 2$, we divide the index set $\I=\{1,\ldots,n\}$ into $\kK$ disjoint subsets $\I_1,\ldots,\I_\kK$ of the same size $n_\kK:=n/\kK$ without loss of generality. Let $\mhatnk(\cdot)$ be an estimator for $m^*(\cdot)$ using the set $\cl_k^-:=\{\bfZ_i:i\in\I_k^-\}$ of size $n_{\kK^-}:=n-n_\kK$, where $\I_k^-:=\I/\I_k$. Then\tcr{,} we define\tcr{:}
\be
\mhatn(\X_i)&~:=~&\kK^{-1}\sk\mhatnk(\X_i)\quad (i=n+1,\ldots,n+N), \quad \tcr{\mbox{and}} \label{ds1}\\
\mhatn(\X_i)&~:=~&\mhatnk(\X_i)\quad (i\in\I_k;\ k=1,\ldots,\kK). \label{ds2}
\ee
The motivation \tcr{for the} 
cross fitting is to bypass technical challenges from the dependence of $\mhatn(\cdot)$ and $\X_i$ in the term $\mhatn(\X_i)$ $(i=1,\ldots,n)$. Without cross fitting, the same theoretical conclusions require more {stringent} assumptions in the same spirit as the stochastic equicontinuity conditions in the classical theory of empirical process. These assumptions are generally hard to verify and less likely to hold in high dimensional scenarios. Essentially, using cross fitting makes the second-order errors in the stochastic expansion of $\muhatss$ easier to control while {not} changing the first-order properties, i.e., the influence function of $\muhatss$. See Theorem 4.2 and the following discussion in \citet{chakrabortty2018efficient}, as well as \citet{chernozhukov2018double} and \citet{newey2018cross}, for more discussion concerning cross fitting. Analogously, when estimating $\pi(\cdot)$, we use $\cu$ only so that $\pihatN(\cdot)$ and $\X_i$ are independent in $\pihatN(\X_i)$ $(i=1,\ldots,n)$. Discarding $\cl$ herein is asymptotically negligible owing to the assumption \eqref{disproportion}.

\vskip0.1in
The definition \eqref{ss_ate} equips us with a {family} of SS estimators for $\mu_0$, {indexed} by $\pihatN(\cdot)$ and $\mhatn(\cdot)$. To derive their limiting properties, we need the following \tcr{(high-level)} conditions.

\begin{assumption}\label{api4}
The function $\hD(\x):=\{\pihatN(\x)\}^{-1}-\{\pis(\x)\}^{-1}$ satisfies\tcr{:}
\be
&&(\E_\X[\{\hD(\X)\}^2])^{1/2}~=~O_p(s_N), ~~ \tcr{\mbox{and}} \label{sn2} \\
&&\{\E_\Z([\hD(\X)\{Y- m^*(\X) \}]^2)\}^{1/2}~=~O_p(b_N)\tcr{,} \label{sn4}
\ee
for some positive sequences $s_N$ and $b_N$ that \tcr{can possibly diverge,} 
where $\pis(\cdot)$ is \tcr{some} function \tcr{(target of $\pihatN(\cdot)$)} such that $\pis(\x)\in(c,1-c)$ for any $\x\in\mx$ and some constant $c\in(0,1)$.
\end{assumption}

\begin{assumption}\label{ahmu}
The estimator $\mhatnk(\cdot)$ satisfies\tcr{: for {some} function $m^*(\cdot)$,} 
\be
&&\E_\X\{|\hat{m}_{n,k}(\X)- m^*(\X) |\}~=~O_p(w_{n,1}), ~~ \tcr{\mbox{and}} \label{wn1}\\
&&(\E_\X[\{\hat{m}_{n,k}(\X)- m^*(\X) \}^2])^{1/2}~=~O_p(w_{n,2})\quad (k=1,\ldots,\kK)\tcr{,}
\label{wn2}
\ee
for some positive sequences $w_{n,1}$ and $w_{n,2}$ that are possibly divergent.
\end{assumption}

\begin{remark}\label{remark_ate_assumptions}
Assumptions \ref{api4}--\ref{ahmu} impose some rather mild \tcr{(and {high-level})} regulations on the behavior of the estimators $\{\pihatN(\cdot),\mhatn(\cdot)\}$ and their possibly \tcr{misspecified} limit\tcr{s} $\{\pis(\cdot),m^*(\cdot)\}$. The
condition \eqref{sn4} is satisfied when, for example, $\hD(\X)$ is such that $(\E_\X[\{\hD(\X)\}^4])^{1/4}=O_p(b_N)$\tcr{,} while $Y$ and $m^*(\X)$ have finite fourth moments. The restriction on $\pis(\cdot)$ in Assumption \ref{api4} is the counterpart of the second condition in \eqref{mar_positivity} under model misspecification, ensuring our estimators $\muhatss$ have influence functions with finite variances; see Theorem \ref{thate}. Moreover, it is noteworthy that all the sequences in Assumptions \ref{api4}--\ref{ahmu} are allowed to \tcr{\emph{diverge},} while specifying \tcr{{only}} the rates of finite norms \tcr{(i.e., $L_r$ moments for some finite $r$)} 
\tcr{of} $\hD(\X) $ and $\{\mhatnk(\X)-m^*(\X)\}$, \tcr{which is} weaker than requiring their convergences uniformly \tcr{over} %
$\x\in\mx$ \tcr{(i.e., $L_{\infty}$ convergence)}. These assumptions will be verified for some choices of $\{\pihatN(\cdot),\mhatn(\cdot),\pis(\cdot),m^*(\cdot)\}$ in Section \ref{secnf}.

\par\smallskip
In the theorem below, we present the stochastic expansion \tcr{(and a complete characterization of the asymptotic properties)} of our SS estimators $\muhatss$ defined in \eqref{ss_ate}.
\end{remark}



\begin{theorem}\label{thate}
Under Assumptions \ref{ass_equally_distributed} and \ref{api4}--\ref{ahmu}, the stochastic expansion of $\muhatss$ is\tcr{:}
\bse
&&\muhatss-\mu_0~=~\nn\sl\zeta_{n,N}(\Z_i)~+~O_p\{n^{-1/2}(w_{n,2}+b_N)+s_N\,w_{n,2}\}~+ \\
&& \phantom{\muhatss-\mu_0~=~}~I\{\pis(\X)\neq\pi(\X)\}O_p(w_{n,1})~+~I\{ m^*(\X) \neq m(\X) \}O_p(s_N)\tcr{,}
\ese
when $\nu\geq 0$, where \tcr{$I(\cdot)$ is the indicator function as defined earlier, and}
\bse
\zeta_{n,N}(\Z)~:=~\{\pis(\X)\}^{-1}T\{Y- m^*(\X) \}~+~\E_{n+N}\{ m^*(\X) \}~-~\mu_0\tcr{,}
\ese
\tcr{with} $\E\{\zeta_{n,N}(\Z)\}=0$ \tcr{if} 
either $\pis(\X)=\pi(\X)$ or $m^*(\X) = m(\X)$ but not necessarily both.
\end{theorem}

Theorem \ref{thate} establishes the \tcr{{asymptotic linearity}} of $\muhatss$ for the \tcr{{general}} case where $\nu\geq 0$, i.e., the labeled and unlabeled data sizes are either comparable or not. Considering, \tcr{however, the typical case is that} the number of the extra observations for $(T,\X\trans)\trans$, whose distribution completely determines the propensity score $\pi(\X)$, from the unlabeled data $\cu$ is much larger than the labeled data size $n$ in the SS setting \eqref{disproportion}, \tcr{i.e., $\nu = 0$}, it is fairly reasonable to assume that $\pi(\X)$ can be correctly specified \tcr{(i.e., $\pis(\cdot) = \pi(\cdot)$) {and}} estimated \tcr{from $\cu$} at a rate \tcr{\emph{faster}} than $n^{-1/2}$. We therefore study the asymptotic behavior of our proposed estimators $\muhatss$ under such an assumption in the next corollary, which directly follows from Theorem \ref{thate}.

\begin{corollary}\label{corate}
Suppose that the conditions in Theorem \ref{thate} hold true, that $\nu=0$, \tcr{as in \eqref{disproportion}}, and that $\pis(\X)=\pi(\X)$. Then the stochastic expansion of $\muhatss$ is\tcr{:}
\bse
&&\muhatss-\mu_0~=~\nn\sl\zess(\Z_i)~+~O_p\{n^{-1/2}(w_{n,2}+b_N)+s_N\,w_{n,2}\}~+ \\
&&\phantom{\muhatss-\mu_0~=~}~I\{ m^*(\X) \neq m(\X) \}O_p(s_N),
\ese
where
\bse
\zess(\Z)~:=~\{\pi(\X)\}^{-1}T\{Y- m^*(\X) \} ~+~ \E\{ m^*(\X) \} ~-~ \mu_0\tcr{,}
\ese
satisfying $\E\{\zess(\Z)\}=0$\tcr{, and with $m^*(\cdot)$ being arbitrary (i.e., not necessarily equal to $m(\cdot)$)}. Further, if either $s_N=o(n^{-1/2})$ or $ m^*(\X) = m(\X) $ but not necessarily both, and
\bse
n^{-1/2}(w_{n,2}+b_N)+s_N\,w_{n,2}~=~o(n^{-1/2}),
\ese
the limiting distribution of $\muhatss$ is\tcr{:} 
\be
n^{1/2}\lamss^{-1}(\muhatss-\mu_0)~\xrightarrow{d}~\mn(0,1)\quad (n,\tcr{N} \to\infty),
\label{ate_normality}
\ee
where the asymptotic variance $\lamss^2:=\E[\{\zess(\Z)\}^2]=\var[\{\pi(\X)\}^{-1}T\{Y- m^*(\X) \}]$ can be estimated by $\var_n[\{\pihatN(\X)\}^{-1}T\{Y-\mhatn(\X)\}]$.
\end{corollary}

\begin{remark}\label{remark_sN}
	Corollary \ref{corate} indicates when $\pis(\cdot)=\pi(\cdot)$ but the outcome model $m(\cdot)$ is misspecified, the key to obtaining asymptotic normality \eqref{ate_normality} of $\muhatss$ is condition $s_N=o(n^{-1/2})$ with $s_N$ as defined in \eqref{sn2}. This condition is achievable only in the SS setting \eqref{disproportion}, which allows for constructing $\pihatN(\cdot)$ using the massive unlabeled data. To see this point, consider $\pihatN(\cdot)$ calculated based on logistic regression as an example and assume $\pihatN(\cdot)$ is uniformly bounded away from zero. When the dimension of $\X$ is fixed, sequence $s_N$ generally satisfies $s_N=O(N^{-1/2})$, which is of order $o(n^{-1/2})$ since $N\gg n$. In high dimensional scenarios, the typical rate of $s_N$ is $s_N=O((q\,\log p/N)^{1/2})$ under suitable conditions with $q$ representing the number of effective parameters in working model $\pis(\cdot)$  \citep{negahban2012unified, wainwright2019high}, so condition $s_N=o(n^{-1/2})$ holds whenever $nq\,\log p/N=o(1)$. In a purely supervised setting providing only a labeled data set of size $n$, the corresponding error rate of propensity score estimators should be $O(n^{-1/2})$ or $O((q\,\log p/n)^{1/2})$ given $\X$ is low or high dimensional, which cannot converge faster than $n^{-1/2}$.
\end{remark}

\begin{remark}[Robustness \tcr{benefits} and first-order insensitivity of  $\muhatss$]\label{remark_ate_robustness}
According to the conclusions in Theorem \ref{thate}, as long as the residual terms in the expansion vanish asymptotically, our proposed estimators $\muhatss$ converge to $\mu_0$ in probability given  either $\pihatN(\cdot)$ targets the true $\pi(\cdot)$ or $\mhatnk(\cdot)$ estimates the true $m(\cdot)$\tcr{,} but \tcr{not} necessarily both. Apart from such \tcr{a} DR property\tcr{,} which can be attained using only the labeled data $\cl$ as well \citep{bang2005doubly, kang2007demystifying}, Corollary \ref{corate} further establishes the $n^{1/2}$-consistency and asymptotic normality of $\muhatss$, two critical properties for inference, \tcr{\it whenever} $\pihatN(\X)$ converges to $\pi(\X)$ at a rate faster than $n^{-1/2}$, via exploiting the information regarding the distribution of $(T,\X\trans)\trans$ from the unlabeled data $\cu$. \tcr{Notably, this holds {\it regardless} of whether $m(\cdot)$ is correctly specified or not}. To attain the same \tcr{kind of}
result without $\cu$, it is generally necessary to require that $\{\pi(\cdot),m(\cdot)\}$ are both correctly specified unless additional bias corrections are applied \tcr{(and in a nuanced targeted manner)} and \tcg{specific (linear$/$logistic) forms of $\{\pi(\cdot),m(\cdot)\}$ are assumed} \citep{vermeulen2015bias, smucler2019unifying, tan2020model, dukes2021inference}. \tcg{Such a significant relaxation of the requirements demonstrates that our SS ATE estimators 
actually enjoy \tcr{much} better robustness relative to the ``best'' achievable estimators in purely supervised setups.} \tcr{These 
benefits of SS causal inference ensure {$n^{1/2}$-rate inference on the ATE (or QTE) can be achieved in a \emph{seamless}
way}, regardless of the misspecification of the outcome model, and moreover, without requiring any specific forms for either of the nuisance model(s).}
\tcg{\tcr{It should also be noted that these benefits are} 
quite different \tcr{in flavor} from \tcr{those in} many ``standard'' \tcr{(non-causal)} SS problems, such as mean estimation \citep{zhang2019semi, zhang2019high} and linear regression \citep{azriel2016semi, chakrabortty2018efficient}, where the supervised methods possess full robustness \tcr{(\tcg{as} 
the parameter needs no nuisance function for \tcr{its} identification)} and the \tcr{main goal of SS inference is efficiency improvement.} 
\tcr{For causal inference, however, 
we have a more challenging setup,
where the supervised methods have to deal with nuisance functions -- inherently required for the parameter's identification and consistent estimation -- and are} 
no longer fully robust. 
\tcr{The} SS \tcr{setup enables one to} 
to attain extra robustness, compared to purely supervised methods, from leveraging the unlabeled data.} \tcr{Thus, for causal inference, the SS setting in fact provides a {broader scope of improvement -- in both robustness and efficiency} -- we discuss the latter aspect in Section \ref{sec_ate_efficiency_comparison} below.}
%
%
\tcr{Lastly, a}nother notable feature of $\muhatss$ is \tcr{its} \tcr{\it first-order insensitivity}, i.e., the influence function $\zeta_{n,N}(\Z)$ in Theorem \ref{thate} is not affected by estimation errors or 
\tcr{any knowledge of the mode of construction} of the nuisance estimators. This is \tcr{particularly} desirable for \tcr{($n^{1/2}$-rate)} inference 
when $\{\pihatN(\cdot),\mhatn(\cdot)\}$ involves non-parametric calibrations\tcr{, or machine learning methods, with slow/unclear first order rates,} 
or \tcr{if} $\X$ is high dimensional.
\end{remark}

\subsection{Efficiency comparison}\label{sec_ate_efficiency_comparison}
In this \tcr{s}ection, we analyze the efficiency gain of $\muhatss$ relative to its  supervised counterparts. We have \tcr{already} clarified in Remark \ref{remark_ate_robustness} 
the robustness \tcr{benefits} of $\muhatss$ 
\tcr{that are} generally not attainable by purely supervised methods.
%
%
\tcr{Therefore, setting aside this already existing improvement (which is partly due to the fact that the SS setup allows $\pi(\cdot)$ to be estimated better, via $\pihatN(\cdot)$ from $\cu$), and to ensure}
a ``fair'' comparison \tcr{(with minimum distraction)}, focusing \tcr{\it solely} on efficiency, we consider \tcg{the} {\it pseudo-supervised} estimator\tcr{(s):}
\be
\muhatsup^*~:=~  \E_{n}\{\mhatn(\X)\}+\E_n[\{\pihatN(\X)\}^{-1}T\{Y-\mhatn(\X)\}],
\label{pseudo_sup_ate}
\ee
which estimates $\pi(\cdot)$ by $\pihatN(\cdot)$\tcr{,}
but does not employ $\cu$ to approximate $\E_\X\{\mhatn(\X)\}$. \tcr{(So it is essentially a version of the purely supervised estimator $\muhatsup$ in \eqref{sup_ate} with $\pihatn(\cdot)$ therein replaced by $\pihatN(\cdot)$, due to the reasons stated above.)} \tcg{Here we emphasize that, as the name ``pseudo-supervised'' suggests, {they \tcr{\it cannot} actually be constructed in purely supervised settings and are proposed just for efficiency comparison}}. \tcr{In a sense, this gives the supervised estimator its best chance to succeed -- in terms of efficiency (setting aside any of its robustness drawbacks) -- and yet, as we will discuss in Remark \ref{remark_ate_efficiency}, they are still outperformed by our SS estimator(s).}

\vskip0.05in
We state 
\tcr{the properties of these pseudo-supervised estimator(s)} in the corollary below, which can be proved analogously to Theorem \ref{thate} and Corollary \ref{corate}, \tcr{and then compare their efficiency (i.e., the ideal supervised efficiency) to that of our SS estimator(s) in Remark \ref{remark_ate_efficiency}.} 

\begin{corollary}\label{coratesup}
Under the \tcr{same} conditions \tcr{as} in Corollary \ref{corate}, the pseudo-supervised estimator $\muhatsup^*$ in \eqref{pseudo_sup_ate} \tcr{satisfies the following expansion:} 
\be
&&\muhatsup^*-\mu_0~=~n^{-1}\sl\zes(\Z_i)~+~O_p\{n^{-1/2}(w_{n,2}+b_N)+s_N\,w_{n,2}\}~+ \nonumber\\
&&\phantom{\muhatsup^*-\vt~=~}~I\{ m^*(\X) \neq m(\X) \}O_p(s_N), ~~\tcr{\mbox{and}}\nonumber\\
&&n^{1/2}\lams^{-1}(\muhatsup^*-\mu_0)~\xrightarrow{d}~\mn(0,1)\quad (n, \tcr{N} \to\infty), ~~\tcr{\mbox{where}} \label{ate_sup_normality}
\ee
$\zes(\Z,\theta):=\{\pi(\X)\}^{-1}T\{Y- m^*(\X) \}+ m^*(\X) -\mu_0$\tcr{,} satisfying $\E\{\zes(\Z)\}=0$\tcr{,} and
\bse
&&\lams^2~:=~\E[\{\zes(\Z)\}^2]~=~\var[\{\pi(\X)\}^{-1}T\{Y- m^*(\X) \}]- \var\{ m^*(\X) \}~+\\ &&\phantom{\lams^2~:=~\E[\{\zes(\Z)\}^2]~=~}~2\,\E\{ m^*(\X) (Y-\mu_0)\}.
\ese
\end{corollary}

\begin{remark}[Efficiency improvement \tcr{of $\muhatss$ and semi-parametric optimality}]\label{remark_ate_efficiency}
	If the conditions in Corollary \ref{corate} hold and the imputation function takes the form\tcr{:}
	\be
	m^*(\X)~\equiv~\E\{Y\mid \bfg(\X)\}\tcr{,}
	\label{mstarX}
	\ee
	with some \tcr{(possibly)} unknown function $\bfg(\cdot)$, the SS variance $\lamss^2$ in \eqref{ate_normality} is less than or equal to the supervised variance $\lams^2$ in \eqref{ate_sup_normality}, i.e.,
	\be
	\qquad \lamss^2~=~\lams^2-2\,\E\{ m^*(\X) (Y-\mu_0)\}+\var\{ m^*(\X) \}~=~\lams^2-\var\{ m^*(\X) \}~\leq~ \lams^2,
	\label{variance_comparison_ATE}
	\ee
	which implies $\muhatss$ is equally or more efficient compared to the pseudo-supervised estimator $\muhatsup^*$.
	%
	An example of the function $\bfg(\x)$ is the linear transformation $\bfg(\x)\equiv\mbP_0\trans\x$, where $\mbP_0$ is some unknown $r\times p$ matrix with a fixed $r\leq p$ and can be estimated\tcr{, e.g.,} by dimension reduction techniques such as 
	\tcr{sliced} inverse regression \citep{li1991sliced, lin2019sparse}\tcr{, as well as by standard parametric (e.g., linear/logistic) regression (for the special case $r=1$).}
	
	Further, if the outcome model is correctly specified, i.e., $m^*(\X)=\E(Y\mid \X)$, we have\tcr{:}
	\be
	\lamss^2&~\equiv~&\var[\{\pi(\X)\}^{-1}T\{Y- m^*(\X) \}]\nonumber\\
	&~=~&\E[\{\pi(\X)\}^{-2}T\{Y- \E(Y\mid\X) \}^2]\label{ate_eff}\\
	&~\leq~&\E[\{\pi(\X)\}^{-2}T\{Y- g(\X) \}^2]\tcr{,} \nonumber
	\ee
	for any function $g(\cdot)$ and the equality holds only if $g(\X)=\E(Y\mid\X)$ almost surely. This fact demonstrates the asymptotic \emph{optimality} of $\muhatss$ among all regular and asymptotically linear estimators of $\mu_0$, whose influence functions take the form $\{\pi(\X)\}^{-1}T\{Y-g(\X)\}$ for some function $g(\cdot)$. Under the semi-parametric model of $(Y,\X\trans,T)\trans$, \tcg{given by the following class of allowable distributions \tcr{(the most unrestricted class allowed under our SS setup)}:}
	\be
	\tcg{\{\P_{(Y,T,\X\trans)\trans}: \hbox{ \eqref{mar_positivity} is satisfied, }\P_{(T,\X\trans)\trans} \hbox{ is known and } \P_{Y\mid(T,\X\trans)\trans}  \hbox{ is unrestricted}\},}
	\label{semiparametric_model}
	\ee
	one can show that (\ref{ate_eff}) equals the efficient asymptotic variance for estimating $\mu_0$, i.e., the estimator $\muhatss$ \emph{achieves the semi-parametric efficiency bound}; \tcr{see Remark 3.1 of \citet{chakrabortty2018efficient}\tcr{, and also the results of \citet{kallus2020role},} for similar bounds}. In Section \ref{sec_nf_ate}, we would detail the above choices of $m^*(\cdot)$ and some corresponding estimators $\mhatnk(\cdot)$. \tcr{Lastly, it is worth noting that the efficiency bound here is lower compared to the supervised case, showing the scope of efficiency gain (apart from robustness) in SS setups.}
\end{remark}

\subsection[Case where T is not observed in U]{Case where $T$ is not observed in $\cu$}\label{sec_ate_u_dagger}
So far, we have focused on \tcr{the case} 
where the unlabeled data contains observations for both the treatment indicator $T$ and the covariates $\X$. We now briefly discuss settings where $T$ is \emph{not} observed in the unlabeled data. Based on the sample $\cl\cup\cu^\dag$\tcr{,} with $\cu^\dag:=\{\X_i:i=n+1,\ldots,n+N\}$, we introduce \tcr{the \emph{SS estimators $\muhatss^\dag$}:}
\be
\muhatss^\dag~:=~  \E_{n+N}\{\mhatn(\X)\}+\E_n[\{\pihatn(\X)\}^{-1}T\{Y-\mhatn(\X)\}]
\label{hatmuss_dag}
\ee
for $\mu_0$. Here $\pihatn(\cdot)$ is constructed \tcr{-- this time solely from $\cl$ --} through a cross fitting procedure similar to \eqref{ds2}\tcr{,} so that $\pihatn(\cdot)$ and $\X_i$ are independent in $\pihatn(\X_i)$ $(i=1,\ldots,n)$. Specifically, we let $\pihatn(\X_i):=\pihatnk(\X_i)$ $(i\in\cl_k)$ with $\pihatnk(\cdot)$ some estimator for $\pi(\cdot)$ based on \tcr{$\cl_k^-$} 
$(k=1,\ldots,\kK)$. See the discussion below \eqref{ds2} for the motivation and benefit of cross fitting.

Compared to $\muhatss$, the estimators $\muhatss^\dag$ substitute $\pihatn(\cdot)$ for $\pihatN(\cdot)$, approximating the working propensity score model $\pis(\cdot)$ using $\cl$ only. We thus impose the following condition on the behavior of $\pihatn(\cdot)$, \tcr{as} a counterpart of \tcr{our earlier} Assumption \ref{api4}.

\begin{assumption}\label{apin4}
The function $\hat{D}_{n,k}(\x):=\{\pihatnk(\x)\}^{-1}-\{\pis(\x)\}^{-1}$ satisfies\tcr{:}
\bse
(\E_\X[\{\hat{D}_{n,k}(\X)\}^2])^{1/2}~=~O_p(s_n), ~~\tcr{\mbox{and}}~~ \{\E_\Z([\hat{D}_{n,k}(\X)\{Y- m^*(\X) \}]^2)\}^{1/2}~=~O_p(b_n)\tcr{,}
\ese
for some positive sequences $s_n$ and $b_n$ $(k=1,\ldots,\kK)$.
\end{assumption}

Replacing $\pihatN(\cdot)$ by $\pihatn(\cdot)$ in Corollary \ref{corate}, we immediately obtain the next corollary regarding the properties of $\muhatss^\dag$. \tcr{(This serves as the counterpart of our Corollary \ref{corate} on $\muhatss$.)}

\begin{corollary}\label{corate_dagger}
Under Assumptions \ref{ass_equally_distributed}, \ref{ahmu} and \ref{apin4} as well as the condition that $\nu=0$ \tcr{as in \eqref{disproportion}}, the SS estimator $\muhatss^\dag$ defined by \eqref{hatmuss_dag} has the stochastic expansion\tcr{:}
\bse
&&\muhatss^\dag-\mu_0~=~\nn\sl\zess(\Z_i)~+~O_p\{n^{-1/2}(w_{n,2}+b_n)+s_n\,w_{n,2}\}~+ \\
&&\phantom{\muhatss-\mu_0~=~}~I\{\pis(\X)\neq\pi(\X)\}O_p(w_{n,1})~+~I\{ m^*(\X) \neq m(\X) \}O_p(s_n), ~~\tcr{\mbox{where}}
\ese
{\cred $ 
\zess(\Z)~\equiv~\{\pis(\X)\}^{-1}T\{Y- m^*(\X) \}+\E\{ m^*(\X) \}~-~\mu_0\tcr{,}
$} 
\tcr{as in Corollary \ref{corate},} satisfying $\E\{\zess(\Z)\}=0$ given either $\pis(\X)=\pi(\X)$ or $ m^*(\X) = m(\X) $ but not necessarily both.

\vskip0.05in
\tcr{Further,} if $\pis(\X)=\pi(\X)$, $ m^*(\X) = m(\X) $ and
{\cred $ 
n^{-1/2}(w_{n,2}+b_n)+s_n\,w_{n,2}~=~o(n^{-1/2}),
$} 
\be
\tcr{\mbox{then}} ~~~ n^{1/2}\lamss^{-1}(\muhatss^\dag-\mu_0)~\xrightarrow{d}~\mn(0,1)\quad (n, \tcr{N} \to\infty)\tcr{,}
\label{ate_ss_dagger_normality}
\ee
with $\lamss^2\equiv\E[\{\zess(\Z)\}^2]=\var[\{\pi(\X)\}^{-1}T\{Y- m(\X) \}]$.
\end{corollary}

\begin{remark}[Comparison of estimators using different types of data]\label{remark_hatmuss_dag}
We can see \tcr{from} 
Corollary \ref{corate_dagger} that $\muhatss^\dag$ possesses the same robustness as the supervised estimator $\muhatsup$ in \eqref{sup_ate}. Specifically, it is consistent whenever one \tcr{among} 
$\{\pi(\cdot), m(\cdot)\}$ is correctly specified, while its $n^{1/2}$-consistency and asymptotic normality in \eqref{ate_ss_dagger_normality} require both \tcr{to be correct}. As regards 
efficiency, as long as the limiting distribution \eqref{ate_ss_dagger_normality} holds, the asymptotic variance $\lamss^2$ of $\muhatss^\dag$ equals that of $\muhatss$ in Theorem \ref{thate}, implying that $\muhatss^\dag$ outperforms $\muhatsup$ and enjoys semi-parametric optimality as discussed in Remark \ref{remark_ate_efficiency}. We summarize in Table \ref{table_ate_summary} \tcr{the} achievable properties of \tcr{all} the ATE estimators based on different types of available data. Estimation of the QTE using the data $\cl\cup\cu^\dag$ is similar in spirit while technically more laborious. We will hence omit the relevant discussion considering such a
setting is not 
\tcr{our} main interest.
\end{remark}

\vskip-0.2in
\begin{table}[H]
\def~{\hphantom{0}}
\caption{
\tcr{SS ATE estimation and its benefits: a complete picture of 
the a}chievable \tcr{robustness and efficiency} properties of the ATE estimators based on different types of available data. Here\tcr{,} the efficiency (Eff.) gain is relative to the supervised estimator \eqref{sup_ate} when $\{m^*(\cdot),\pi^*(\cdot)\}=\{m(\cdot),\pi(\cdot)\}$, \tcr{while} the optimality (Opt.) 
\tcr{refers to} attaining the \tcr{corresponding} semi-parametric efficiency bound. The abbreviation $n^{1/2}$-CAN stands for $n^{1/2}$-consistency and asymptotic normality\tcr{, while DR stands for doubly robust (in terms of consistency only).}}
{
\begin{tabular}{c||c|c|c|c|c}
\hline
\multirow{3}{*}{Data} & \multirow{3}{*}{DR} & \multicolumn{2}{c|}{$n^{1/2}$-CAN} & \multirow{3}{*}{Eff. gain} & \multirow{3}{*}{Opt.} \\ \cline{3-4}
& &$ \pis(\cdot)=\pi(\cdot)$ & $ \pis(\cdot)=\pi(\cdot)$& & \\
& & $m^*(\cdot)=m(\cdot)$ &$m^*(\cdot)\neq m(\cdot)$ & & \\
\hline
$\cl$ & \cmark & \cmark & \xmark & \xmark & \xmark \\
$\cl\cup\cu^\dag$ & \cmark & \cmark & \xmark & \cmark & \cmark \\
$\cl\cup\cu$ & \cmark & \cmark & \cmark & \cmark &\cmark \\
\hline
\end{tabular}}
\label{table_ate_summary}
\end{table}

\subsection{Final \tcr{SS} estimator for the ATE}\label{sec_ate_difference}
In 
\tcr{Sections \ref{sec_ate_ss}--\ref{sec_ate_efficiency_comparison},}
we have established the asymptotic properties of our SS estimator $\muhatss\equiv\muhatss(1)$ for $\mu_0\equiv\mu_0(1)$. We now propose  \tcr{our \emph{final SS estimator for the ATE,} i.e., the difference $\mu_0(1)-\mu_0(0)$ in \eqref{ate}, as: $\muhatss(1)-\muhatss(0)$, with} 
\bse
\muhatss(0)~:=~\E_{n+N}\{\mhatn(\X,0)\}+\E_n[\{1-\pihatN(\X)\}^{-1}(1-T)\{Y-\mhatn(\X,0)\}],
\ese
where the estimator $\mhatn(\X,0)$ is constructed by cross fitting procedures similar to \eqref{ds1}--\eqref{ds2} and has a probability limit $m^*(\X,0)$, a working outcome model for the conditional expectation $\E\{Y(0)\mid\X\}$. Adapting Theorem \ref{thate} and Corollary \ref{corate} with $\{Y,T\}$ therein replaced by $\{Y(0),1-T\}$, we can directly obtain theoretical results \tcr{for} 
$\muhatss(0)$ including its stochastic expansion and limiting distribution. By arguments analogous to those in Remarks \ref{remark_ate_robustness}--\ref{remark_ate_efficiency}, one can easily conclude the double robustness, asymptotic normality, efficiency gain compared to the supervised counterparts and semi-parametric optimality of $\muhatss(0)$. Also, it is straightforward to show these properties are possessed by the difference estimator $\muhatss(1)-\muhatss(0)$ as well. Among all the above conclusions, a particularly important one is that\tcr{:}
\be
n^{1/2}\lamate^{-1}[\{\muhatss(1)-\muhatss(0)\}-\{\mu_0(1)-\mu_0(0)\}]~\xrightarrow{d}~\mn(0,1)\quad (n, \tcr{N} \to\infty)\tcr{,}
\label{ate_difference_distribution}
\ee
under the conditions in Corollary \ref{corate} for $\muhatss(1)$ as well as their counterparts for $\muhatss(0)$, where the asymptotic variance\tcr{:}
\bse
\lamate^2~:=~\var[\{\pi(\X)\}^{-1}T\{Y- m^*(\X) \}-\{1-\pi(\X)\}^{-1}(1-T)\{Y(0)- m^*(\X,0) \}]
\ese
can be estimated by\tcr{:}
\bse
\var_n[\{\pihatN(\X)\}^{-1}T\{Y- \mhatn(\X) \}-\{1-\pihatN(\X)\}^{-1}(1-T)\{Y(0)- \mhatn(\X,0) \}].
\ese
In theory, the limiting distribution \eqref{ate_difference_distribution} provides the basis for \tcr{our SS} inference regarding the ATE\tcr{:} $\mu_0(1)-\mu_0(0)$; see the data analysis in Section \ref{sec_data_analysis} for an instance of its application.

\begin{remark}[Comparison with \citet{zhang2019high}]\label{remark_comparison_zhang2019}
It is worth mentioning \tcr{here} that our work on the ATE bears \tcr{some resemblance} 
with \tcr{the} 
recent article by \citet{zhang2019high}, who discussed SS inference for the ATE as an illustration of their SS mean estimation method and mainly focused on using a linear working model for $\E(Y\mid\X)$. We, however, treat this problem in more generality \tcr{-- both in methodology and theory}. Specifically, we allow for a wide range of methods to estimate \tcr{the} nuisance functions \tcr{in our estimators,} 
\tcr{allowing flexibility in terms of} model misspecification\tcr{, and also establish through this whole section a suit of generally applicable results -- with only high-level conditions on the nuisance estimators -- giving a complete understanding/characterization of our SS ATE estimators' properties, uncovering in the process, various interesting aspects of their robustness and efficiency benefits.}
\tcr{In Section \ref{secnf} later,} 
we \tcr{also} provide a careful study of a  \tcr{family of} outcome model estimators based on kernel smoothing, inverse probability weighting and dimension reduction, 
establishing novel results \tcr{on} 
their uniform convergence rates, which verify the high-level conditions required in Corollary \ref{corate} and ensure the efficiency superiority of our method discussed in Remark \ref{remark_ate_efficiency}; see Section \ref{sec_nf_ate} for \tcr{more}
detail\tcr{s}. \tcr{In general, we believe the SS ATE estimation problem warranted a more detailed and thorough analysis in its own right, as we attempt to do in this paper.} \tcr{Moreover,} 
we also consider, \tcr{as in the next section, the QTE estimation problem, which to our knowledge is an entirely novel contribution in the area of SS (causal) inference}.
\end{remark}

\section{SS estimation for the QTE}\label{secqte}
We now study SS estimation of the QTE \tcr{in \eqref{qte}}. As before \tcr{in Section \ref{secos}}, we will simply focus \tcr{here} on \tcr{SS estimation of the} $\tau$\tcr{-}quantile $\vt\equiv\vt(1,\tau)\in\Theta\subset\rR$ of $Y\equiv Y(1)$\tcr{, as in \eqref{generic_notation},} with some fixed and known $\tau\in(0,1)$. \tcr{This will be our goal in Sections \ref{sec_qte_general}--\ref{sec_qte_efficiency_comparison}, after which we finally address SS inference for the QTE in Section \ref{sec_qte_difference}.}

\begin{remark}[Technical difficulties \tcr{with} 
QTE estimation]\label{qte_challenges}
While the basic ideas \tcr{underlying the SS estimation of the QTE} 
are similar in spirit to those in Section \ref{secos} for the ATE, the inherent inseparability of $Y$ and $\theta$ in the quantile estimating equation \eqref{defqte} poses significantly more challenges in both implementation and theory. To overcome these difficulties, we use the strategy of one-step update in the construction of our QTE estimators, and \tcr{also}
develop technical novelties of empirical process theory in the proof of their properties; see Section \ref{sec_qte_general} as well as Lemma \ref{1v2} \tcr{(}in 
\tcr{Appendix} \ref{sm_lemmas} of the Supplementary Material\tcr{)} for \tcr{more} details.
\end{remark}

\begin{remark}[Semantic clarification for Sections \ref{sec_qte_general}--\ref{sec_qte_efficiency_comparison}]\label{remark_semantics}
\tcr{
As mentioned above, our estimand in Sections \ref{sec_qte_general}--\ref{sec_qte_efficiency_comparison} is the quantile $\vt$ of $Y(1)$, not QTE, per se. However, for semantic convenience,
we will occasionally refer to it as ``QTE'' (and the estimators as ``QTE estimators'') while presenting our results and discussions in these sections. We hope this 
slight abuse of terminology
is not a distraction, as the true estimand should be clear from context.}
\end{remark}

\subsection[SS estimators for theta0: general construction and properties]{ \tcr{SS estimators for $\vt$: g}eneral construction and properties }\label{sec_qte_general}
\tcr{Let us define} 
$\phi(\X,\theta):=\E\{\psi(Y,\theta)\mid\X\}$. Analogous to \tcr{the construction} \eqref{ate_dr_representation} for \tcr{the mean} $\mu_0$, we observe that, for arbitrary functions $\pis(\cdot)$ and $\phis(\cdot,\cdot)$, the equation \eqref{defqte} \tcr{for $\vt$} satisfies the DR type representation\tcr{:}
\be
0~=~\E\{\psi(Y,\vt)\} ~=~ \E\{  \phis(\X,\vt)\}+ \E[\{\pis(\X)\}^{-1} T\{\psi(Y,\vt) -  \phis(\X,\vt)\}]\tcr{,}
\label{qte_dr_representation}
\ee
given either $\pis(\X)=\pi(\X)$ or $\phis(\X,\theta)=\phi(\X,\theta)$ but {\it not} necessarily both.

\tcr{To} clarify the \tcr{basic} logic behind the construction of our \tcr{SS} estimators, suppose momentarily that $\{\pis(\cdot),\phis(\cdot,\cdot)\}$ are known and equal to $\{\pi(\cdot),\phi(\cdot,\cdot)\}$. One may then expect to obtain a supervised estimator of $\vt$ by solving the empirical version of \eqref{qte_dr_representation} based on $\cl$, i.e.,
\be
\E_n\{  \phi(\X,\theta)\}+ \E_n[\{\pi(\X)\}^{-1} T\{\psi(Y,\theta) -  \phi(\X,\theta)\}] ~=~0,
\label{sv}
\ee
with respect to $\theta$. However, solving \eqref{sv} directly is not a simple task due to its \tcr{inherent} non-smoothness and non-linearity \tcr{in $\theta$}. 
\tcr{A reasonable strategy to adopt instead is a} \emph{one-step update} \tcr{approach} \citep{van2000asymptotic, tsiatis2007semiparametric}\tcr{,} using 
the corresponding {influence function} 
\tcr{(a term used a bit loosely here to denote the expected influence function in the supervised case):}
\be
\{f(\vt)\}^{-1} (\E[\{\pi(\X)\}^{-1} T\{\phi(\X,\vt)-\psi(Y,\vt)\}]-\E\{  \phi(\X,\vt)\}).
\label{qte_influence_function}
\ee
Specifically, by replacing the unknown functions $\{\pi(\cdot),~\phi(\cdot,\cdot)\}$ in \eqref{qte_influence_function} with {some} estimators $\{\pihatn(\cdot),~ \phihatn(\cdot,\cdot)\}$ based on $\cl$ that \tcr{may} target possibly misspecified limits $\{\pis(\cdot),~\phis(\cdot,\cdot)\}$, 
we immediately obtain a {\it supervised estimator} \tcr{of $\vt$ \tcr{via a one-step update approach as follows:}}
\be
&&\thetahatsup~:=~ \thetahatinit +\{\hf(\thetahatinit)\}^{-1}(\E_n[\{\pihatn(\X)\}^{-1}T\{\phihatn(\X,\thetahatinit) - \psi(Y,\thetahatinit)\}]- \label{sup_qte} \\
&&\phantom{\thetahatsup~:=~ \thetahatinit +\{\hf(\thetahatinit)\}^{-1}(}\E_{n}\{\phihatn(\X,\thetahatinit)\})\tcr{,}  \nonumber
\ee
with $\thetahatinit$ an initial estimator for $\vt$ and $\hf(\cdot)$ an estimator for the density function $f(\cdot)$ of $Y$.

\paragraph*{SS estimators \tcr{of $\vt$}} \tcr{With the above motivation for a one-step update approach, and recalling the basic principles of our SS approach in Section \ref{sec_ate_ss}, 
we now formalize the details of our SS estimators of $\vt$.} 
Similar to the \tcr{rationale used in the} construction of
\tcr{\eqref{ss_ate}} for \tcr{estimating $\mu_0$ in context of} the ATE, replacing $\E_{n}\{\phihatn(\X,\thetahatinit)\}$ and $\pihatn(\X)$ in \eqref{sup_qte} by $\E_{n+N}\{\phihatn(\X,\thetahatinit)\}$ and $\pihatN(\X)$, respectively\tcr{, now} \tcg{produces a family of \emph{SS estimators} $\thetahatss$ for $\vt$, given by:} 
\be
&&\thetahatss~:=~  \thetahatinit +\{\hf(\thetahatinit)\}^{-1}(\E_n[\{\pihatN(\X)\}^{-1}T\{\phihatn(\X,\thetahatinit) - \psi(Y,\thetahatinit)\}]- \label{ss_qte}\\
&&\phantom{\thetahatss~:=~  \thetahatinit +\{\hf(\thetahatinit)\}^{-1}(}\E_{n+N}\{\phihatn(\X,\thetahatinit)\}). \nonumber
\ee
Here\tcr{, a} 
cross fitting technique \tcr{similar to \eqref{ds1}--\eqref{ds2}} is applied 
to \tcr{obtain $\phihatn(\X_i,\cdot)$:} 
\be
\phihatn(\X_i,\theta)&~:=~&\kK^{-1}\sk\phihatnk(\X_i,\theta)\quad (i=n+1,\ldots,n+N), \quad \tcr{\mbox{and}} \label{ds3}\\ \phihatn(\X_i,\theta)&~:=~&\phihatnk(\X_i,\theta)\quad (i\in\I_k\tcr{;\ k=1,\ldots,\kK}), \label{ds4}
\ee
where $\phihatnk(\cdot,\cdot)$ is an estimator for $\phis(\cdot,\cdot)$ based \tcr{only} on the data set $\cl_k^-$  $(k=1,\ldots,\kK)$.

\vskip0.05in
We now have a family of SS estimators for $\vt$ indexed by $\{\pihatN(\cdot),\phihatn(\cdot,\cdot)\}$ from \eqref{ss_qte}. \tcr{To establish their theoretical properties, we will require the following (high-level) assumptions.} 

\begin{assumption}
\label{adensity}
The quantile $\vt$ is in the interior of its parameter space $\Theta$. The density function $f(\cdot)$ of $Y$ is positive and has a bounded derivative in $\mbtv$ \tcr{for some $\varepsilon > 0$}.
\end{assumption}

\begin{assumption}
\label{ainit}
\tcr{The initial estimator $\thetahatinit$ and the density estimator $\hf(\cdot)$} 
satisfy that, for some positive sequences $u_n=o(1)$ and $v_n=o(1)$,
\be
&&\thetahatinit-\vt~=~O_p(u_{n}), ~~ \tcr{\mbox{and}} \label{hvti}\\
&&\hf(\thetahatinit)-f(\vt)~=~O_p(v_n). \label{hf}
\ee
\end{assumption}

\begin{assumption}\label{api}
Recall \tcr{that}
$\pis(\cdot)$ is some function such that $\pis(\x)\in(c,1-c)$ for any $\x\in\mx$ and some $c\in(0,1)$. Then\tcr{,}
the function $\hD(\x)\equiv\{\pihatN(\x)\}^{-1}-\{\pis(\x)\}^{-1}$ satisfies\tcr{:}
\be
&&(\E_\X[\{\hD(\X)\}^2])^{1/2}~=~O_p(s_N), ~~ \tcr{\mbox{and}} \label{d2} \\
&&\sx|\hD(\x)|~=~O_p(1)\tcr{,} \label{dsup}
\ee
for some positive sequence $s_N$ that is possibly divergent.
\end{assumption}

\begin{assumption}\label{abound}
The function $\phis(\cdot,\cdot)$ \tcr{-- the (possibly misspecified) target of $\phihatn(\cdot,\cdot)$ --} is bounded. \tcr{Further, t}he set $\mm:=\{\phis(\X,\theta):\theta\in\mbtv\}$ \tcr{for some $\varepsilon > 0$,} satisfies\tcr{:}
\be
N_{[\,]}\{\eta,\mm,L_2(\P_\X)\}~\leq~ c_1\,\eta^{-c_2},
\label{bmm}
\ee
where the symbol $N_{[\,]}(\cdot,\cdot,\cdot)$ refers to the {bracketing number}\tcr{, as} defined in \citet{van1996weak} and \citet{van2000asymptotic}. In addition, for any sequence $\tvt\to\vt$ in probability,
\be
&&\mbG_n[\{\pis(\X)\}^{-1}T\{\phis(\X,\tvt)-\phis(\X,\vt)\}]~=~o_p(1), ~~\tcr{\mbox{and}} \label{unipi1}\\ &&\mbG_{n+N}\{\phis(\X,\tvt)-\phis(\X,\vt)\}~=~o_p(1).\label{unipi2}
\ee
\end{assumption}

\begin{assumption}\label{aest}
Denote
\be
&&\hpsi(\X,\theta)~:=~\phihatnk(\X,\theta)-\phis(\X,\theta), ~~\tcr{\mbox{and}} \label{error}\\ &&\Delta_k(\cl)~:=~(\sb\E_\X[\{\hpsi(\X,\theta)\}^2])^{1/2} \quad (k=1,\ldots,\kK).\nonumber
\ee
Then\tcr{,} \tcr{for some $\varepsilon > 0$,} the set\tcr{:}
\be
\mp_{n,k}~:=~\{\hpsi(\X,\theta):\theta\in\mbtv\}
\label{pnk}
\ee
satisfies that, for any $\eta\in(0,\Delta_k(\cl)+c\,]$ \tcr{for some $c > 0$},
\be
N_{[\,]}\{\eta,\mp_{n,k}\mid\cl,L_2(\P_\X)\}~\leq~ H(\cl) \eta^{-c} \quad (k=1,\ldots,\kK)
\label{vc}
\ee
with some function $H(\cl)>0$ such that $H(\cl)=O_p(a_n)$ for some positive sequence $a_n$ that is possibly divergent. Here\tcr{,} $\mp_{n,k}$ is indexed by $\theta$ \tcr{\it only} and treats $\hpsi(\cdot,\theta)$ as a non\tcr{-}random function $(k=1,\ldots,\kK)$. Moreover, we assume \tcr{that:}
\be
&&\sb\E_\X\{|\hpsi(\X,\theta)|\}~=~O_p(d_{n,1}),~~~ \Delta_k(\cl)~=~O_p(d_{n,2}), ~~~\tcr{\mbox{and}} \nonumber\\
&& \sbx|\hpsi(\tcg{\x},\theta)|~=~O_p(d_{n,\infty}) \quad (k=1,\ldots,\kK), \nonumber
\ee
where $d_{n,1}$, $d_{n,2}$ and $d_{n,\infty}$ are some positive sequences that are possibly divergent.
\end{assumption}

\begin{remark}\label{remark_qte_assumptions}
	The basic conditions in Assumption \ref{adensity} ensure the identifiability and estimability of $\vt$. Assumption \ref{ainit} is standard for one-step estimators, regulating the behavior of $\thetahatinit$ and $\hf(\cdot)$. Assumption \ref{api} is an analogue of Assumption \ref{api4}, adapted \tcr{suitably} for the technical proofs of
	the QTE estimators. Assumption \ref{abound} outlines the features of a suitable working outcome model $\phis(\cdot,\cdot)$. According to Example 19.7 and Lemma 19.24 of \citet{van2000asymptotic}, the conditions \eqref{bmm}--\eqref{unipi2} hold as long as $\phi^{\tcr{*}}(\X,\theta)$ is Lipschitz continuous in $\theta$. Lastly, Assumption \ref{aest} imposes restrictions on the bracketing number and norms of the error term \eqref{error}. The requirements in Assumptions \ref{abound} and \ref{aest} should be expected to hold for most reasonable choices of $\{\phi^{\tcr{*}}(\cdot,\cdot),\phihatnk(\cdot,\cdot)\}$ using standard results from empirical process theory \citep{van1996weak, van2000asymptotic}. All the positive sequences in Assumptions \ref{api} and \ref{aest} are possibly divergent, so the relevant restrictions are fairly mild and weaker than \tcr{requiring} $L_\infty$ convergence. The validity of these assumptions for some choices of the nuisance functions and their estimators will be di\tcr{s}cussed in Section \ref{secnf}.
\end{remark}

\tcr{We now} 
\tcr{present the asymptotic properties of $\thetahatss$ in Theorem \ref{thqte} and Corollary \ref{corqte} below.}

\begin{theorem}\label{thqte}
Suppose that Assumptions \ref{ass_equally_distributed} and \ref{adensity}--\ref{aest} hold, and that
either $\pis(\X)=\pi(\X)$ or $\phis(\X,\theta)=\phi(\X,\theta)$ but not necessarily both. Then\tcr{, it holds that:} $\thetahatss-\vt=$
\bse
&&\{nf(\vt)\}^{-1}\sl\omega_{n,N}(\Z_i,\vt)~+~O_p\{u_n^2+u_nv_n+n^{-1/2}(r_n+z_{n,N})+s_N d_{n,2}\}~+ \\
&&~I\{\pis(\X)\neq\pi(\X)\}O_p(d_{n,1})+I\{\phis(\X,\theta)\neq\phi(\X,\theta)\}O_p(s_N)+o_p(n^{-1/2})\tcr{,}
\ese
when $\nu\geq 0$, where
\bse
&&r_n~:=~d_{n,2}\{\log\,a_n+\log(d_{n,2}^{-1})\}~+~n_{\kK}^{-1/2}d_{n,\infty}\{(\log\,a_n)^2+(\log\,d_{n,2})^2\},\\
&&z_{n,N}~:=~s_N\log\, (s_N^{-1})~+~n^{-1/2}(\log\,s_N)^2, ~~\tcr{\mbox{and}}\\
&&\omega_{n,N}(\Z,\theta)~:=~\{\pis(\X)\}^{-1}T\{\phis(\X,\theta)-\psi(Y,\theta)\}-\E_{n+N}\{\phis(\X,\theta)\}\tcr{,}
\ese
satisfying $\E\{\omega_{n,N}(\Z,\vt)\}=0$ \tcr{if either $\phis(\cdot) = \phi(\cdot)$ or $\pis(\cdot) = \pi(\cdot)$ but not necessarily both.}
\end{theorem}

\begin{corollary}\label{corqte}
Suppose that the conditions in Theorem \ref{thqte} hold true, that $\nu=0$ \tcr{as in \eqref{disproportion},} and that $\pis(\X)=\pi(\X)$. Then\tcr{,} the stochastic expansion of $\thetahatss$ is \tcr{given by:} $\thetahatss-\vt=$
\bse
&&\{nf(\vt)\}^{-1}\sl\omss(\Z_i,\vt)~+~O_p\{u_n^2+u_nv_n+n^{-1/2}(r_n+z_{n,N})+s_N d_{n,2}\}~+ \\
&&~I\{\phis(\X,\theta)\neq\phi(\X,\theta)\}O_p(s_N)~+~o_p(n^{-1/2}),
\ese
where
\bse
\omss(\Z,\theta)~:=~\{\pi(\X)\}^{-1}T\{\phis(\X,\theta)-\psi(Y,\theta)\}-\E\{\phis(\X,\theta)\}\tcr{,}
\ese
satisfying $\E\{\omss(\Z,\vt)\}=0$\tcr{, and $\phis(\X,\theta)$ is arbitrary, i.e., not necessarily equal to $\phi(\x,\theta)$.}
\vskip0.1in
Further, if either $s_N=o(n^{-1/2})$ or $\phis(\X,\theta)=\phi(\X,\theta)$ but not necessarily both, and
\be
u_n^2+u_nv_n+n^{-1/2}(r_n+z_{n,N})+s_N d_{n,2}~=~o(n^{-1/2}),
\label{srn}
\ee
\tcr{then} the limiting distribution of $\thetahatss$ is\tcr{:}
\be
n^{1/2}f(\vt)\sigss^{-1}(\thetahatss-\vt)~\xrightarrow{d}~\mn(0,1)\quad (n, \tcr{N}\to\infty)\tcr{,}
\label{qte_normality}
\ee
with $\sigss^2:=\E[\{\omss(\Z,\vt)\}^2]=\var[\{\pi(\X)\}^{-1}T\{\psi(Y,\vt)-\phis(\X,\vt)\}]$\tcr{, and t}he asymptotic variance $\{f(\vt)\}^{-2}\sigss^2$ can be estimated \tcr{as:}
\bse
\{\hf(\thetahatss)\}^{-2}\var_n[\{\pihatN(\X)\}^{-1}T\{\psi(Y,\thetahatss)-\phihatn(\X,\thetahatss)\}].
\ese
\end{corollary}

\begin{remark}[Robustness and first-order insensitivity of $\thetahatss$]\label{remark_qte_property}
\tcr{Theorem \ref{thqte} and Corollary \ref{corqte} establish the general properties of $\thetahatss$, in the same spirit as
those \tcg{of} $\muhatss$ in Section \ref{sec_ate_ss}. 
The results show, in particular, that $\thetahatss$} are always DR, while enjoying first-order insensitivity, \tcr{and} $n^{1/2}$-consistency and asymptotic normality\tcr{, {\it regardless} of whether $\phi(\cdot,\cdot)$ is misspecified,} as long as we can correctly estimate $\pi(\X)$ at a\tcr{n} $L_2$\tcr{-}rate faster than $n^{-1/2}$ \tcr{by exploiting the plentiful observations in $\cu$}. In contrast, \tcr{such} $n^{1/2}$-consistency and asymptotic normality are 
unachievable \tcr{(in general)} for supervised QTE estimators \tcr{if} 
$\phi(\cdot,\cdot)$ is misspecified. This is
analogous 
to the case of the ATE\tcr{; see} 
Remark \ref{remark_ate_robustness} for more discussions on these properties.
\end{remark}

\begin{remark}[Choice\tcr{s} of $\{\thetahatinit,\hf(\cdot)\}$]\label{remark_qte_initial_estimator}
While the general conclusions in Theorem \ref{thqte} and Corollary \ref{corqte} hold true for {any} estimators $\{\thetahatinit,\hf(\cdot)\}$ satisfying Assumption \ref{ainit}, a reasonable choice in practice \tcr{for both would be {\it IPW type estimators}.} 
Specifically, the initial estimator $\thetahatinit$ can be obtained by solving\tcr{:} $\E_n[\{\pihatN(\X)\}^{-1}T\psi(Y,\thetahatinit)]=0$, while $\hf(\cdot)$ may be defined as a kernel density estimator based on the weighted sample\tcr{:} $\{\{\pihatN(\X_i)\}^{-1}T_iY_i:i=1,\ldots,n\}$. Under the conditions in Corollary \ref{corqte}, it is not hard to show that Assumption \ref{ainit} as well as the part of \eqref{srn} related to $\{u_n,v_n\}$ \tcr{are} 
indeed satisfied by such $\{\thetahatinit,\hf(\cdot)\}$, using the basic proof techniques of quantile \tcr{method}s \citep{k2005} and kernel-based approaches \citep{hansen2008uniform}, \tcr{along} with suitable modifications \tcr{used to incorporate the IPW weights.} 
\end{remark}

\subsection{Efficiency comparison}\label{sec_qte_efficiency_comparison}

For efficiency comparison among QTE estimators, 
similar to $\muhatsup^*$ in Section \ref{secos} \tcr{for the ATE}, 
we now consider the
{\it pseudo-supervised estimator\tcr{(s)}} \tcr{of $\vt$:}
\be
&&\thetahatsup^*~:=~  \thetahatinit +\{\hf(\thetahatinit)\}^{-1}(\E_n[\{\hat{\pi}_N(\X)\}^{-1}T\{\phihatn(\X,\thetahatinit) - \psi(Y,\thetahatinit)\}]- \label{pseudo_sup_qte}\\
&&\phantom{\thetahatsup^*~:=~  \thetahatinit +\{\hf(\thetahatinit)\}^{-1}(}\E_{n}\{\phihatn(\X,\thetahatinit)\}),  \nonumber
\ee
\tcr{i.e., the version of the purely supervised estimator $\thetahatsup$ in \eqref{sup_qte} with $\pihatn(\cdot)$ therein replaced by $\pihatN(\cdot)$ from $\cu$. $\thetahatsup^*$ thus has the same robustness as $\thetahatss$ and is considered solely for efficiency comparison -- among SS and supervised estimators of $\vt$ (setting aside any robustness benefits the former already enjoys). This is based on the same motivation and rationale as those discussed in detail in Section \ref{sec_ate_efficiency_comparison} in the context of ATE estimation; so we do not repeat those here for brevity. We now present the properties of $\thetahatsup^*$ followed by the efficiency comparison.}

\begin{corollary}\label{corsup}
Under the conditions in Corollary \ref{corqte}, the pseudo-supervised estimators $\thetahatsup^*$ given by \eqref{pseudo_sup_qte} \tcr{satisfies the following expansion:} 
$\thetahatsup^*-\vt=$
\be
&&\quad\{nf(\vt)\}^{-1}\sl\oms(\Z_i,\vt)~+~O_p\{u_n^2+u_nv_n+n^{-1/2}(r_n+z_{n,N})+s_N d_{n,2}\}~+ \nonumber\\
&&\quad ~I\{\phis(\X,\theta)\neq\phi(\X,\theta)\}O_p(s_N)~+~o_p(n^{-1/2}), ~~\tcr{\mbox{and}} \nonumber\\
&&\quad n^{1/2}f(\vt)\sigsup^{-1}(\thetahatsup^*-\vt)~\xrightarrow{d}~\mn(0,1)\quad (n, \tcr{N}\to\infty),
\label{qte_sup_normality}
\ee
where
\bse
\oms(\Z,\theta)~:=~\{\pi(\X)\}^{-1}T\{\phis(\X,\theta)-\psi(Y,\theta)\}-\phis(\X,\theta)\tcr{,}
\ese
satisfying $\E\{\oms(\Z,\vt)\}=0$, and $\sigsup^2:=\E[\{\oms(\Z,\vt)\}^2]=$
{
\bse
\var[\{\pi(\X)\}^{-1}T\{\psi(Y,\vt)-\phis(\X,\vt)\}]-\var\{\phis(\X,\vt)\}+ 2\,\E\{\phis(\X,\vt)\psi(Y,\vt)\}.
\ese
}
\end{corollary}
\begin{remark}[Efficiency improvement \tcr{of $\thetahatss$ and optimality}]\label{remark_qte_efficiency}
Inspecting the asymptotic variances in Corollaries \ref{corqte} and \ref{corsup}, we see \tcr{that} $\sigss^2\leq\sigsup^2$ 
\tcr{with {\it any} choice of $\phis(\X,\theta)$ such that}
$\phis(\X,\theta)=\E\{\psi(Y,\theta)\mid \bfg(\X)\}$ for some \tcr{(possibly)} unknown function $\bfg(\cdot)$, since
\bse
\sigsup^2-\sigss^2~=~2\,\E\{\phis(\X,\vt)\psi(Y,\vt)\}-\var\{\phis(\X,\vt)\}~=~\E[\{\phis(\X,\vt)\}^2]~\geq~ 0.
\ese
Such a comparison reveals the 
{superiority} \tcr{in} efficiency of our SS estimators $\thetahatss$ over the \tcr{corresponding} ``best'' achievable ones in supervised settings \tcr{\it even if} the difference \tcr{(i.e.,  improvement)} in robustness is ignored. When $\phis(\X,\theta)=\E\{\psi(Y,\theta)\mid\X\}$, the SS variance\tcr{:}
\be
\sigss^2&~=~&\var(\{\pi(\X)\}^{-1}T[\psi(Y,\vt)-\E\{\psi(Y,\vt)\mid\X\}]) \nonumber\\
&~=~&\E(\{\pi(\X)\}^{-2}T[\psi(Y,\vt)-\E\{\psi(Y,\vt)\mid\X\}]^2) \label{qte_eff}\\
&~\leq~&\E[\{\pi(\X)\}^{-2}T\{\psi(Y,\vt)-g(\X)\}^2]\tcr{,} \nonumber
\ee
for any function $g(\cdot)$ while the equality holds only if $g(\X)=\E\{\psi(Y,\vt)\mid\X\}$ almost surely. In this sense $\thetahatss$ is asymptotically \tcr{\it optimal} among all regular and asymptotically linear estimators of $\vt$, whose influence functions have the form $\{f(\vt)\pi(\X)\}^{-1}T\{g(\X)-\psi(Y,\vt)\}$ for some function $g(\cdot)$. \tcg{Under the semi-parametric model \eqref{semiparametric_model}}, one can show \tcg{if Assumption \ref{adensity} holds true}, the representation \eqref{qte_eff} equals the efficient asymptotic variance for estimating $\vt$, that is, the \tcr{SS} estimator $\thetahatss$ achieves the \tcr{\it semi-parametric efficiency bound}. In Section \ref{sec_nf_qte}, we will \tcr{also} detail the above choices of $\phis(\cdot,\cdot)$ and some corresponding estimators $\phihatnk(\cdot,\cdot)$.
\end{remark}

\subsection{Final \tcr{SS} estimator for the QTE}\label{sec_qte_difference}
Similar to the arguments \tcr{used in \tcr{Section \ref{sec_ate_difference} for}
the case} of $\{\muhatss(1),\muhatss(0)\}$ \tcr{to obtain the ATE estimator,} 
substituting $\{Y(0),1-T\}$ for $\{Y,T\}$ in the aforementioned discussions concerning $\thetahatss\equiv\thetahatss(1)$ and $\vt\equiv\vt(1)$ immediately gives \tcg{\tcr{us} a family of \tcr{SS} estimators} $\thetahatss(0)$ for $\vt(0)$ as well as their \tcr{corresponding} properties \tcr{(as the counterparts of the properties established for $\thetahatss(1)$ so far)}. \tcr{Subsequently, we may obtain our final SS estimator(s) for the QTE, i.e., the difference $\vt(1)-\vt(0)$ in \eqref{defqte}, simply as:  $\thetahatss(1)-\thetahatss(0)$.} Then we know that, if the conditions in Corollary \ref{corqte} for $\thetahatss(1)$ and their counterparts for $\thetahatss(0)$ hold, the asymptotic distribution of \tcg{our \emph{final SS \tcr{QTE} estimators} $\thetahatss(1)-\thetahatss(0)$ 
is\tcr{:}} 
\be
n^{1/2}\sigqte^{-1}[\{\thetahatss(1)-\thetahatss(0)\}-\{\vt(1)-\vt(0)\}]~\xrightarrow{d}~\mn(0,1)\quad (n, \tcr{N}\to\infty),
\label{qte_difference_distribution}
\ee
where the asymptotic variance\tcr{:}
\bse
&&\sigqte^2~:=~\var(\{f(\vt)\pi(\X)\}^{-1}T\{\psi(Y,\vt)- \phis(\X,\vt) \}- \\
&&\phantom{\sigqte^2~:=~\var(}[f\{\vt(0),0\}\{1-\pi(\X)\}]^{-1}(1-T)[\psi\{Y(0),\vt(0)\}- \phis\{\X,\vt(0),0\} ])
\ese
can be estimated by\tcr{:}
\bse
&&\var_n(\{\hf(\thetahatss)\pihatN(\X)\}^{-1}T\{\psi(Y,\thetahatss)- \phihatn(\X,\thetahatss) \}- \\
&&\phantom{\var_n(}[\hf\{\thetahatss(0),0\}\{1-\pihatN(\X)\}]^{-1}(1-T)[\psi\{Y(0),\thetahatss(0)\}- \phihatn\{\X,\thetahatss(0),0\} ]).
\ese
In the above\tcr{,} $\hf(\cdot,0)$  and $\phihatn(\X,\theta,0)$ are \tcr{\it some} estimators for the density function $f(\cdot,0)$ of $Y(0)$ and the working model $\phis(\X,\theta,0)$ of $\E[\psi\{Y(0),\theta\}\mid\X]$, respectively. We will use \eqref{qte_difference_distribution} to construct confidence intervals for the QTE in the data analysis of Section \ref{sec_data_analysis}.

\section{Choice and estimation of \tcg{the} nuisance functions}\label{secnf}
In this section, we study some reasonable choices and estimators of the nuisance functions \tcr{involved} in the SS estimators $\muhatss$ and $\thetahatss$ from Sections \ref{secos} and \ref{secqte}, which form a critical component in the implementation of \tcr{all} our approaches. The results claimed in the last two sections, however, are completely general and allow for any choices as long as they satisfy the high-level conditions therein. \tcr{In Sections \ref{sec_PS}--\ref{sec_nf_qte} below, we discuss some choices of $\pi(\cdot)$ and the outcome models for ATE and QTE.}

\subsection{Propensity score}\label{sec_PS}

Under the assumption \eqref{disproportion}, the specification and estimation of $\pi(\cdot)$ is a relatively easier task and can be done through applying any reasonable and flexible enough regression method (parametric, semi-parametric or non-parametric) to the plentiful observations for $(T,\X\trans)\trans$ in $\cu$. For instance, one can use the {\it ``extended'' parametric families} 
$\pis(\x)\equiv h\{\bbeta_0\trans\bPsi(\x)\}$ as the working model for the propensity score $\pi(\cdot)$, where $h(\cdot)\in(0,1)$ is a {\it known} link function, the components of $\bPsi(\cdot):\rR^p\mapsto\rR^{p^*}$ are (known) basis functions of $\x$ with $p^*\equiv p^*_n$ allowed to diverge and exceed  $n$, and $\bbeta_0\in\rR^{p^*}$ is an {\it unknown} parameter vector. Such a $\pis(\x)$ can be estimated by $\pihatN(\x)\equiv h\{\bbetahat\trans\bPsi(\x)\}$ with $\bbetahat$ obtained from the corresponding parametric regression 
process of $T$ vs. $\bPsi(\X)$ using $\cu$. Regularization may be applied \tcr{here} via, for example, the $L_1$ penalty if necessary \tcr{(e.g., in high dimensional settings)}.

The families above include, as a special case, the logistic regression models with
\bse
h(x)~\equiv~\{1+\exp(-x)\}^{-1}\hbox{ and } \bPsi(\x)~\equiv~\{1,\bPsi_1\trans(\x),\bPsi_2\trans(\x),\ldots,\bPsi_M\trans(\x)\}\trans\tcr{,}
\ese
for $\bPsi_m(\x):=(\x_{[1]}^m,\x_{[2]}^m,\ldots,\x_{[p]}^m)\trans$ $(m=1,\ldots,M)$ and some positive integer $M$. Section 5.1 of \citet{chakrabortty2019high} along with Section B.1 in the supplementary material of that article provided a detailed discussion on these ``extended'' parametric families and established their (non-asymptotic) properties, sufficient for the high-level conditions on $\{\pis(\cdot),\pihatN(\cdot)\}$ in Sections \ref{secos} and \ref{secqte}. In addition, it is noteworthy that, in high dimensional scenarios \tcr{in our setup,} where $n\ll p^*\ll N$, {the parameter vector $\bbeta_0$ is totally free of sparsity} and can be estimated by unregularized methods based on $\cu$. Such a relaxation of assumptions is incurred
by the usage of massive unlabeled data and \tcr{is} generally unachievable in purely supervised settings.

\subsection{Outcome model for the ATE}\label{sec_nf_ate}
We now consider the working outcome model $m^*(\cdot)$ \tcr{involved} in our ATE estimators. As discussed in Remark \ref{remark_ate_efficiency}, one may expect to achieve semi-parametric optimality by letting $m^*(\X)\equiv\E(Y\mid\X)$. However, specifying \tcr{the} $\E(Y\mid\X)$ correctly in high dimensional scenarios is usually unrealistic while approximating it fully non-parametrically would typically bring \tcr{in}
undesirable issues such as
under-smoothing \citep{newey1998undersmoothing} even if there are only a moderate number of covariates. We therefore adopt a principled and flexible semi-parametric strategy, \tcr{via}
conducting dimension reduction followed by non-parametric calibrations and targeting $\E(Y\mid\S)$ instead of $\E(Y\mid\X)$, where $\S:=\mbP_0\trans\X\in\ms\subset\rR^r$ and $\mbP_0$ is a $r\times p$ {\it transformation matrix} with some fixed and known $r\leq p$. \tcr{(The choice $r = p$ of course leads to a trivial case with $\mbP_0 = I_p$.)} It is noteworthy that we \tcr{\it always} allow the dimension reduction to be \emph{insufficient} and do {\it not} assume anywhere \tcr{that}
\be
\E(Y\mid\S)~=~\E(Y\mid\X).
\label{sufficient_dimension_reduction}
\ee
The efficiency comparison in Remark \ref{remark_ate_efficiency} shows that, when\tcr{ever}
$\pihatN(\cdot)$ converges to $\pi(\cdot)$ fast enough, setting $m^*(\X)\equiv\E(Y\mid\mbP_0\trans\X)$ \tcr{always} guarantees our SS estimators $\muhatss$ \tcr{to} dominate any supervised competitors using the same working model $m^*(\cdot)$ \tcr{--} {no matter} whether \eqref{sufficient_dimension_reduction} holds or not. Hence\tcr{,} one is free to let $\mbP_0$ equal {any} user-defined and data-dependent matrix. If $\mbP_0$ is completely determined by the distribution of $\X$, its estimation error is very likely to be negligible owing to the large number of observations for $\X$ provided by $\cu$. An instance \tcr{of such a choice is} 
the $r$ leading principal component directions of $\X$. Nevertheless, to make the dimension reduction as ``sufficient'' as possible, one may prefer to use a transformation matrix $\mbP_0$ which depends on the joint distribution of $(Y,\X\trans)\trans$\tcr{, and thus} needs to be estimated with significant errors. We will give some examples of such $\mbP_0$ in Remark \ref{remark_choice_of_P0}.

To \tcr{estimate} 
the conditional mean $m^*(\x)\equiv\E(Y\mid \mbP_0\trans\X=\mbP_0\trans\x)$, we may employ any \tcr{suitable} smoothing technique, such as kernel smoothing, kernel machine regression or smoothing splines. For illustration, we focus  on the \tcr{\it IPW type kernel smoothing estimator(s):}
\be
\mhatnk(\x)~\equiv~\mhatnk(\x,\hmbP)~:=~\{\hlz(\x,\hmbP)\}^{-1}\hlo(\x,\hmbP)\quad (k=1,\ldots,\kK),
\label{ks_ate}
\ee
where
\bse
\hlt(\x,\mbP)~:=~ h_n^{-r} \Enk[\{\pihatN(\X)\}^{-1}T Y^tK_h\{\mbP\trans(\x-\X)\}]\quad (t=0,1),
\ese
with the notation $\E_{n,k}\{\ghat(\bfZ)\}:=n_{\kK^-}^{-1}\slk \ghat(\bfZ_i)$ for any possibly random function $\ghat(\cdot)$, \tcr{and with} $\hmbP$ \tcr{being \emph{any}} estimator of $\mbP_0$ using $\cl_k^-$, $K_h(\s):=K(h_n^{-1}\s) $, $K(\cdot)$ a kernel function \tcr{(e.g., the standard Gaussian kernel)} and $ h_n\to 0$ \tcr{denoting} a bandwidth sequence.

\begin{remark}[Subtlety and benefit\tcr{s} of the inverse probability weighting scheme]\label{remark_ks_ate_weight}
The \tcr{IPW based}
weight\tcr{s} $\{\pihatN(\X)\}^{-1}$ \tcr{involved} in $\mhatnk(\x)$ \tcr{in \eqref{ks_ate}} \tcr{play} 
a key role in \tcr{its} achieving \tcr{an {important}} {\it DR {property}}, which means $\mhatnk(\x)$ has the limit $\E(Y\mid\S=\s)$ whenever either \eqref{sufficient_dimension_reduction} is true or $\pis(\cdot)=\pi(\cdot)$\tcr{,}
but \emph{not} necessarily both. This property will be proved in Theorem \ref{theorem_ks_ate}\tcr{,}
and formally stated \tcr{and discussed} in Remark \ref{remark_ks_ate_DR}. In contrast, the (standard) complete-case version without the \tcr{IPW} weights $\{\pihatN(\X)\}^{-1}$ actually targets $\E(Y\mid\S=\s,T=1)$ that equals $\E(Y\mid\S=\s)$ \emph{only if} \eqref{ks_ate} holds. Recalling the clarification in Remark \ref{remark_ate_efficiency}, we can see that such a subtlety \tcr{(enabled by the involvement of the weights) in the construction} of $\mhatnk(\cdot)$ ensures the efficiency advantage of our SS estimators $\muhatss$ over any supervised competitors constructed with the same $\mhatnk(\cdot)$,  when $\pi(\cdot)$ is correctly specified but $m(\cdot)$ is not.

\tcr{Lastly, a}lthough $\mhatnk(\cdot)$ contains $\pihatN(\cdot)$ and thereby involves the unlabeled data $\cu$, we suppress the subscript $N$ \tcr{in $\mhatnk(\cdot)$}
for brevity considering its convergence rate mainly relies on $n$; see Theorem \ref{theorem_ks_ate}. In principle, cross fitting procedures analogous to (\ref{ds1}) and (\ref{ds2}) should be conducted for $\cu$ as well to guarantee the independence of $\mhatnk(\cdot)$ and $\X_i$ in $\mhatnk(\X_i)$ $(i=n+1,\ldots,n+N)$. However, from our experience, such extra cross fitting \tcg{procedures} bring only marginal benefits in practice while making the implementation more laborious. We hence stick to estimating $\pis(\cdot)$ using the whole $\cu$ in \tcr{our} numerical studies.
\end{remark}

\vskip-0.02in
\tcr{There is substantial literature on kernel smoothing estimators with unknown estimated covariate transformations, but mostly in low (fixed) dimensional settings  \citep{mammen2012nonparametric,mammen_rothe_schienle_2016, escanciano2014uniform}.} Considering\tcr{, however,} that \tcr{in our setting,}
the dimension $p$ of $\X$ can be \tcr{\it divergent} \tcr{(possibly exceeding $n$),}
and that the transformation matrix $\mbP_0$ as well as the weights $\{\pis(\X)\}^{-1}$ nee\tcr{d} 
to be \tcr{\it estimated} \tcr{as well},
establishing the uniform convergence property of $\mhatnk(\x,\hmbP)$ \tcr{in \eqref{ks_ate},} in fact\tcr{,} poses substantial technical challenges and has not been studied in the literature yet. \tcr{Our results here are thus {\it novel} to the best of our knowledge.} To derive \tcr{the results} 
we impose the following \tcr{conditions.}

\begin{assumption}\label{al1}
The estimator $\hmbP$ satisfies $\|\hmbP-\mbP_0\|_1=O_p(\alpha_n)$ for some \tcr{$\alpha_n \geq 0$}. 
\end{assumption}

\begin{assumption}[\tcr{Smoothness conditions}]\label{akernel}
(i) The function $K(\cdot):\rR^r\mapsto\rR$ is a symmetric kernel of order $d\geq 2$ with a finite $d$th moment. Moreover, it  is bounded, square integrable and continuously differentiable with a derivative $\nabla K(\s):=\partial K(\s)/\partial \s$ such that $\|\nabla K(\s)\|\leq c_1\,\|\s\|^{-v_1}$ for some constant $v_1>1$ and any $\|\s\|>c_2$. (ii) The support $\ms$ of $\S\equiv\mbP_0\trans\X$ is compact. The density function $f_{\S}(\cdot)$ of $\S$ is bounded and bounded away from zero on $\ms$. In addition, it is $d$ times continuously differentiable with a bounded $d$th derivative on some open set $\ms_0\supset\ms$. (iii) For some constant $u>2$, the response $Y$ satisfies $\ss\E(Y^{2u}\mid\S=\s)<\infty$. (iv) The function $\kappa_t(\s):=\E[\{\pis(\X)\}^{-1}TY^t\mid \S=\s]$ $(t=0,1)$
is $d$ times continuously differentiable and has 
\tcr{b}ounded $d$th \tcr{order} derivatives on $\ms_0$.
\end{assumption}

\begin{assumption}[\tcr{Required only when $\mbP_0$ needs to be estimated}] \label{ahbey}
(i) The support $\mx$ of $\X$ is such that $\sx\|\x\|_\infty<\infty$. (ii) The function $\nabla K(\cdot)$ has a bounded derivative satisfying $\|\partial \{\nabla K(\s)\}/\partial \s\|\leq c_1\,\|\s\|^{-v_2}$ for some constant $v_2>1$ and any $\|\s\|>c_2$. Further, it is locally Lipschitz continuous, i.e., $\|\nabla K(\s_1)-\nabla K(\s_2)\|\leq \|\s_1-\s_2\|\rho(\s_2)$ for any $\|\s_1-\s_2\|\leq c$, where $\rho(\cdot)$ is some bounded, square integrable and differentiable function with a bounded derivative $\nabla\rho(\cdot)$ such that $\|\nabla\rho(\s)\|\leq c_1\|\s\|^{-v_3}$ for some constant $v_3>1$ and any $\|\s\|>c_2$. (iii) Let $\bchi_{t[j]}(\s)$ be the $j$th component of $\bchi_{t}(\s):=\E[\X \{\pis(\X)\}^{-1}TY^t\mid \S=\s]$. Then\tcr{,} $\bchi_{t[j]}(\s)$ is continuously differentiable and has a bounded first derivative on $\ms_0$\tcr{, for each $t=0,1$ and $j=1,\ldots,p$.}
\end{assumption}

\vskip-0.02in
In the above, Assumption \ref{al1} regulates the behavior of $\hmbP$ as an estimator of the transformation matrix $\mbP_0$. Moreover, the smoothness and moment conditions in Assumption \ref{akernel} are almost adopted from \citet{hansen2008uniform} and are fairly standard in the literature of kernel-based approaches \citep{newey1994large, andrews1995nonparametric, masry1996multivariate}. Further, we require Assumption \ref{ahbey} to control \tcr{the} errors from approximating $\mbP_0$ by $\hmbP$\tcr{,} while Assumption \ref{ahbey} (ii) \tcr{in particular} is satisfied by the second-order Gaussian kernel, among others. Similar conditions were imposed by \citet{chakrabortty2018efficient} to study unweighted kernel smoothing estimators with dimension reduction  \tcr{in} 
low \tcr{(fixed)} dimensional \tcr{settings.} 
Based on these conditions, we \tcr{provide} 
the uniform convergence rate of $\mhatnk(\x,\hmbP)$ \tcr{in the following result.} 

\begin{theorem}[\tcr{Uniform consistency of $\mhatnk(\cdot)$}]\label{theorem_ks_ate}
Set $\xi_n:=\{(nh_n^r)^{-1}\log\,n\}^{1/2}$, $b_n^{(1)}:=\xi_n+h_n^d$ and $b_{n,N}^{(2)}:=h_n^{-2}\alpha_n^2+h_n^{-1}\xi_n\alpha_n+\alpha_n+h_n^{-r/2}s_N$. Suppose that Assumptions \ref{ass_equally_distributed}, \ref{api4} and \ref{al1}--\ref{ahbey} hold true and that $b_n^{(1)}+b_{n,N}^{(2)}=o(1)$. Then\tcr{,}
\bse
\sx|\mhatnk(\x,\hmbP)-\tmu(\x,\mbP_0)| ~=~O_p\{b_n^{(1)}+b_{n,N}^{(2)}\} \quad (k=1,\ldots,\kK),
\ese
where $\tmu(\x,\mbP):=\{\kappa_0(\mbP\trans\x)\}^{-1}\kappa_1(\mbP\trans\x)$\tcr{, with $\kappa_0(\cdot)$ and $\kappa_1(\cdot)$ as given in Assumption \ref{akernel}.}
\end{theorem}

\begin{remark}[Double robustness \tcr{of $\mhatnk$}]\label{remark_ks_ate_DR}
As long as either $\pis(\x)=\pi(\x)$ or $m^*(\x)\equiv \E(Y\mid \S=\s)=\E(Y\mid\X=\x)\equiv m(\x)$ but {\it not} necessarily both, we have\tcr{:}
\bse
\tmu(\x,\mbP_0)&~=~&(\E[\{\pis(\X)\}^{-1}\pi(\X)\mid \S=\s])^{-1}\E[\{\pis(\X)\}^{-1}\pi(\X)m(\X)\mid \S=\s] \\
&~=~&\E(Y\mid\S=\s)~\equiv~ m^*(\x).
\ese
Theorem \ref{theorem_ks_ate} therefore shows $\mhatnk(\x,\hmbP)$ is a \tcr{\it DR estimator} of $m^*(\x)$.
\tcr{This is an important consequence of the IPW scheme used in the construction of $\mhatnk(\cdot)$, and its benefits (in the bigger context of our final SS estimator) were discussed in detail in Remark \ref{remark_ks_ate_weight}.
}

\end{remark}

\begin{remark}[Uniform convergence \tcr{-- some examples}]\label{remark_choice_of_P0}
According to the result in Theorem \ref{theorem_ks_ate}, the uniform consistency of $\mhatnk(\x,\hmbP)$ as an estimator of $\tmu(\x,\mbP_0)$ holds at the \tcr{optimal bandwidth} \tcr{order} 
$h_{\hbox{\tiny opt}}=O\{n^{-1/(2d+r)}\}$ for any kernel order $d\geq 2$ and \tcr{a} fixed $r$, given
\be
s_N~=~o \{n^{-r/(4d+2r)}\}\quad \hbox{and} \quad \alpha_n~=~o\{n^{-1/(2d+r)}\}.
\label{s_N_alpha_n}
\ee
The first part of \eqref{s_N_alpha_n} is actually weaker than the assumption \tcr{$s_N=o(n^{-1/2})$ used}
in Corollary \ref{corate} 
and thus \tcr{should be} easy to be ensured in the SS setting \eqref{disproportion}. As regards the validity of the second part, we consider it for \tcr{some} frequently used {choices of $\mbP_0$} including, for instance, the least square regression parameter $(r=1)$ satisfying $\E\{\X(Y-\mbP_0\trans\X)\}=\bze_p$, and the $r$ leading eigenvectors of the matrix $\var\{\E(\X\mid Y)\}$, which can be estimated by 
\tcr{sliced} inverse regression \citep{li1991sliced}. When $p$ is fixed, there typically exist $n^{1/2}$-consistent estimators $\hmbP$ for $\mbP_0$\tcr{,} so the second part of \eqref{s_N_alpha_n} is satisfied by the fact that $\alpha_n=O(n^{-1/2})$. In high dimensional scenarios where $p$ is divergent and greater than $n$, one can obtain $\hmbP$ from the \tcr{$L_1$-}regularized version\tcr{(s)} of linear regression  or 
\tcr{sliced} inverse regression \citep{lin2019sparse}. The sequence $\alpha_n=O\{q(\log\,p/n)^{1/2}\}$ when the $L_1$ penalty is applied under some suitable conditions \citep{buhlmann2011statistics, negahban2012unified, wainwright2019high}, where $q:=\|\mbP_0\|_0$ represents the sparsity level of $\mbP_0$. Thus\tcr{,} the second part of \eqref{s_N_alpha_n} holds as long as
\bse
q(\log\,p)^{1/2}~=~o\{n^{(2d+r-2)/(4d+2r)}\}.
\ese
\end{remark}

\subsection{Outcome model for the QTE}\label{sec_nf_qte}

As regards the outcome model $\phis(\cdot,\cdot)$ for the QTE, we adopt the same strategy as in Section \ref{sec_nf_ate}. Specifically, \tcr{with $\mbP_0$ similar as before,} we set
\be
\phis(\x,\theta)~\equiv~\E\{\psi(Y,\theta)\mid \mbP_0\trans\X=\mbP_0\trans\x\}~\equiv~\E\{\psi(Y,\theta)\mid \S=\s\}\tcr{,}
\label{phis}
\ee
and estimate it by the IPW type kernel smoothing estimator\tcr{:}
\be
\phihatnk(\x,\theta)\equiv\phihatnk(\x,\theta,\hmbP):=\{\hatez(\x,\theta,\hmbP)\}^{-1}\hateo(\x,\theta,\hmbP)\quad (k=1,\ldots,\kK),
\label{ks_qte}
\ee
where\tcr{, with $K(\cdot)$, $h_n$ and $K_h(\cdot)$ similarly defined as in Section \ref{sec_nf_ate},}
\bse
\hatet(\x,\theta,\mbP)~:=~h_n^{-r} \Enk[\{\pihatN(\X)\}^{-1}T \{\psi(Y,\theta)\}^tK_h\{\mbP\trans(\x-\X)\}]\quad (t=0,1).
\ese
We first verify Assumption \ref{abound} for \tcr{a choice of} $\phis(\x,\theta)$ \tcr{as} in \eqref{phis}\tcr{, via the following result.}

\begin{proposition}\label{thphi}
If the conditional density $f(\cdot\mid\s)$ of $Y$ given $\S=\s$ is such that
\be
\E[\{\sb f(\theta\mid\S)\}^2]~<~\infty,
\label{conditional_density}
\ee
then Assumption \ref{abound} is satisfied by setting $\phis(\X,\theta)\equiv\E\{\psi(Y,\theta)\mid\S\}$.
\end{proposition}

We now study the uniform convergence of the estimator $\phihatnk(\x,\theta)$. It is noteworthy that establishing properties of $\phihatnk(\x,\theta)$ is \tcr{even more} technically involved compared to the case of $\mhatnk(\x)$ in Section \ref{sec_nf_ate}, since handling function class $\{\psi(Y,\theta):\theta\in\mb(\vt,\v)\}$ inevitably needs tools from empirical process theory. We itemize the relevant assumptions as follows.

\begin{assumption}[\tcr{Smoothness conditions}]\label{akernel_qte} 
(i) Assumption \ref{akernel} (i) \tcr{holds}. (ii) Assumption \ref{akernel} (ii) \tcr{holds}. (iii) 
\tcr{T}he function $\varphi_t(\s,\theta):=\E[\{\pis(\X)\}^{-1}T\{\psi(Y,\theta)\}^t\mid \S=\s]$ $(t=0,1)$ is $d$ times continuously differentiable \tcr{with respect to $\s$,} and has 
\tcr{b}ounded $d$th \tcr{order} derivatives
on $\ms_0\times\mbtv$ \tcr{for some $\varepsilon > 0$}.
\end{assumption}

\begin{assumption}[\tcr{Required only 
\tcr{if} $\mbP_0$ needs to be estimated}] \label{ahbe}
(i) Assumption \ref{ahbey} (i) \tcr{holds}. (ii) The function $\nabla K(\cdot)$ is continuously differentiable and satisfies $\|\partial \{\nabla K(\s)\}/\partial \s\|$ \tcr{$\leq c_1\,\|\s\|^{-v_2}$} for some constant $v_2>1$ and any $\|\s\|>c_2$. Further, it is locally Lipschitz continuous, i.e., $\|\nabla K(\s_1)-\nabla K(\s_2)\|\leq \|\s_1-\s_2\|\rho(\s_2)$ for any $\|\s_1-\s_2\|\leq c$, where $\rho(\cdot)$ is some bounded and square integrable function with a bounded derivative $\nabla\rho(\cdot)$. (iii) Let $\bfeta_{t[j]}(\s,\theta)$ be the $j$th component of $\bfeta_{t}(\s,\theta):=\E[\X \{\pis(\X)\}^{-1}T\{\psi(Y,\theta)\}^t\mid \S=\s]$. Then, with respect to $\s$, the function $\bfeta_{t[j]}(\s,\theta)$ is continuously differentiable and has a bounded first derivative on $\ms_0\times\mbtv$ \tcr{ for some $\varepsilon > 0$,} \tcr{for each $t=0,1$ and $j=1,\ldots p$.}
\end{assumption}

The above two assumptions can be viewed as 
\tcr{the natural} variant\tcr{s} of Assumptions \ref{akernel}--\ref{ahbey} adapted \tcr{suitably} for the case of the QTE. We now propose the following result \tcr{for $\phihatnk(\cdot,\cdot)$}.

\begin{theorem}[\tcr{Uniform convergence rate of $\phihatnk(\cdot,\cdot)$}]\label{thhd}
Set $\gamma_n:=[(nh_n^r)^{-1}\{\log(h_n^{-r})+\log(\log\,n)\}]^{1/2}$, $a_{n}^{(1)}:=\gamma_n+h_n^d$ and $a_{n,N}^{(2)}:=h_n^{-2}\alpha_n^2+h_n^{-1}\gamma_n\alpha_n+\alpha_n+h_n^{-r/2}s_N$. Suppose that Assumptions \ref{ass_equally_distributed}, \ref{api}, \ref{al1}, \ref{akernel_qte} and \ref{ahbe} hold true and that $a_{n}^{(1)}+a_{n,N}^{(2)}=o(1)$. Then
\bse
\sbx|\phihatnk(\x,\theta,\hmbP)-\tphi(\x,\theta,\mbP_0)| ~=~O_p\{a_{n}^{(1)}+a_{n,N}^{(2)}\} \quad (k=1,\ldots,\kK),
\ese
where $\tphi(\x,\theta,\mbP):=\{\varphi_0(\mbP\trans\x,\theta)\}^{-1}\varphi_1(\mbP\trans\x,\theta)$ \tcr{with $\varphi_0(\cdot)$ and $\varphi_1(\cdot)$ as in Assumption \ref{akernel_qte}.}
\end{theorem}

\begin{remark}[Double robustness and uniform convergence \tcr{of $\phihatnk(\cdot,\cdot)$}]
Whenever either $\pis(\x)=\pi(\x)$ or $\phis(\x,\theta)\equiv \E\{\psi(Y,\theta)\mid \S=\s\}=\E\{\psi(Y,\theta)\mid \X=\x\}\equiv\phi(\x,\theta)$\tcr{,} but {\it not} necessarily both, we can see \tcr{that:}
\bse
\tphi(\x,\theta,\mbP_0)&~=~&(\E[\{\pis(\X)\}^{-1}\pi(\X)\mid \S=\s])^{-1}\E[\{\pis(\X)\}^{-1}\pi(\X)\phi(\X,\theta)\mid \S=\s] \\
&~=~&\E\{\psi(Y,\theta)\mid\S=\s\}~\equiv~\phis(\x,\theta).
\ese
In this sense\tcr{,} $\phihatnk(\x,\theta,\hmbP)$ is a \tcr{\it DR estimator} of $\phis(\x,\theta)$. Moreover, it is straightforward to show $\phihatnk(\x,\theta,\hmbP)$ is uniformly consistent for $\tphi(\x,\theta,\mbP_0)$ at the optimal bandwidth rate under the same conditions on $\{s_N,\alpha_n\}$ as those in Remark \ref{remark_choice_of_P0}, while the choices of $\{\mbP_0,\hmbP\}$ therein also apply to the case of $\phihatnk(\x,\theta,\hmbP)$\tcr{; s}ee the discussion in Remark \ref{remark_choice_of_P0} for details.
\end{remark}

Theorem \ref{thhd} \tcr{therefore} has shown \tcr{(among other things) that} the sequences $\{d_{n,1}, d_{n,2}, d_{n,\infty}\}$ in \tcr{our high-level} Assumption \ref{aest} \tcr{on $\phihatnk(\cdot,\cdot)$} are all of \tcr{order} $o(1)$ when one sets\tcr{:}
\be
\hpsi(\X,\theta)~\equiv~\phihatnk(\X,\theta,\hmbP)-\phis(\X,\theta),
\label{psihat}
\ee
where $\phis(\x,\theta)$ and $\phihatnk(\x,\theta,\mbP)$ are as defined in (\ref{phis}) and (\ref{ks_qte}), respectively. 
\tcr{Furthermore, as a final verification of our high-level conditions in Assumption \ref{aest},} we validate the condition \eqref{vc} \tcr{therein} on the bracketing number \tcr{via the following proposition}.
\begin{proposition}\label{thbn}
Under the condition \eqref{conditional_density}, the function $\hpsi(\X,\theta)$ in \eqref{psihat} satisfies\tcr{:}
\bse
N_{[\,]}\{\eta,\mp_{n,k}\mid\cl,L_2(\P_\X)\}~\leq~ c\,(n+1)\eta^{-1},
\ese
where the set $\mp_{n,k}$ is as defined in \eqref{pnk}. Therefore\tcr{,} the sequence $a_n$ \tcr{characterizing the growth of} 
the function $H(\cl)$ in the condition \eqref{vc} \tcr{of Assumption \ref{aest}} is of \tcr{order} $O(n)$.
\end{proposition}

\begin{remark}[Other outcome model estimators]\label{remark_other_nuisance_functions}
\tcr{Finally, as we conclude our discussion on the nuisance functions' estimation, it is worth pointing out that i}n 
addition to the IPW type kernel smoothing estimators with necessary dimension reduction, which have been investigated thoroughly in Sections \ref{sec_nf_ate}--\ref{sec_nf_qte}, one may also employ \tcr{any} other reasonable choices of $\mhatnk(\cdot)$ and $\phihatnk(\cdot,\cdot)$ to construct $\muhatss$ and $\thetahatss$, as long as they satisfy the high-level conditions in Sections \ref{secos}--\ref{secqte}. Examples include estimators generated by parametric (\tcr{e.g,} linear$/$logistic) regression \tcr{methods, possibly with penalization in high dimensional settings \citep{farrell2015robust},} and random forest \citep{breiman2001random} without use of dimension reduction, as well as many \tcr{other} popular non-parametric machine learning approaches 
that have been advocated by some
recent works for other
related problems in analogous settings 
\citep{chernozhukov2018double, farrell2021deep}. We will consider some of these methods in \tcr{our} 
simulations and data analysis\tcr{,}
while omitting their theoretical study, which is not of our primary interest in this article\tcr{; see} 
Sections \ref{sec_simulations} and \ref{sec_data_analysis} for their implementation details and numerical performance.
\end{remark}

\section{Simulations}\label{sec_simulations}
We now investigate the numerical performance of our \tcr{SS ATE and QTE estimators} $\muhatss$ and $\thetahatss$ on simulated data 
\tcr{under a variety of data generating mechanisms}. \tcr{(We clarify here that without loss of generality we focus on $\mu_0$ and $\vt$ in \eqref{generic_notation} as our targets, though with some abuse of terminology, we occasionally refer to them as ATE and QTE respectively.)} 
We set the sample sizes $n\in\{200,500\}$ and $N=10,000$ throughout. The covariates $\X$ are drawn from a $p$-dimensional normal distribution with a zero mean and an identity covariance matrix, where $p\in\{10,200\}$ \tcr{denotes low and high dimensional choices, respectively}. For any kernel smoothing steps involved, we always use the second order Gaussian kernel and select the bandwidths 
\tcr{using}
cross validation. Regularization is applied to all regression procedures via
the $L_1$ penalty when $p=200$, while the tuning parameters are chosen \tcr{using} 
ten-fold cross validation. The number of folds in the cross fitting steps \eqref{ds1}--\eqref{ds2} and \eqref{ds3}--\eqref{ds4} is $\kK=10$. By the term ``complete-case'', we refer to conducting a process on $\{(Y_i,T_i=1,\X_i\trans)\trans:i\in\I^*\}$ without weighting, where $\I^*\equiv\I_k^-$ if cross fitting is involved while $\I^*\equiv\I$ otherwise.

\subsection{\tcr{Data generating mechanisms and nuisance estimator choices}} \tcr{We use the following choices as the \emph{true} data generating models for $T \mid \X$ and $Y \mid \X $.} Let $\X_q:=(\X_{[1]},\ldots,\X_{[q]})\trans$  where $q=p$ when $p=10$, and $q\in\{5,\ceil{p^{1/2}}\}$ when $p=200$, \tcr{representing the (effective) \emph{sparsity} (fully dense for $p = 10$, and sparse or moderately dense for $p = 200$, respectively) of the true data generating models for the nuisance functions, as described below}.

\vskip0.05in
\tcr{For the \emph{propensity score} $\pi(\X)$}, \tcr{and with $T \mid \X \sim \mbox{Bernoulli} \{\pi(\X)\}$,} we set \tcr{the choices:}
\begin{enumerate}[(i)]
\item $\pi(\X)\equiv h(\bon_q\trans\X_q/q^{1/2})$, a {\it linear }model;

\item $\pi(\X)\equiv h\{\bon_q\trans\X_q/q^{1/2}+(\bon_q\trans\X_q)^2/(2q)\}$, a {\it single index} model;

\item $\pi(\X)\equiv h\{\bon_q\trans\X_q/q^{1/2}+\|\X_q\|^2/(2q)\}$, a {\it quadratic} model.
\end{enumerate}
In the above $h(x)\equiv\{1+\exp(-x)\}^{-1}$ \tcr{denotes the usual ``expit'' link function for a logistic model}. To approximate $\pi(\X)$ using the data $\cu$, we obtain the \emph{estimator} $\pihatN(\x)$ from\tcr{:}
\begin{enumerate}[I.]
\item unregularized or regularized \tcr{(linear)} logistic regression of $T$ vs. $\X$ (Lin), \tcg{which correctly specifies the propensity score (i) but misspecifies (ii) and (iii)}; ~~\tcr{or}

\item unregularized or regularized \tcr{(quadratic)} logistic regression of $T$ vs. $(\X\trans,\X_{[1]}^2,\ldots,\X_{[p]}^2)\trans$ (Quad), \tcg{which correctly specifies the propensity scores (i) and (iii) but misspecifies (ii)}.
\end{enumerate}

\tcr{T}he \emph{conditional outcome model} is $Y\mid\X\sim\mn\{m(\X),1\}$ with \tcr{
choices of $m(\cdot)$ as follows:}
\begin{enumerate}[(a)]
\item $m(\X)\equiv \bon_q\trans\X_q$, a {\it linear} model;

\item $m(\X)\equiv \bon_q\trans\X_q+(\bon_q\trans\X_q)^2/q$, a {\it single index} model;

\item $m(\X)\equiv \bon_q\trans\X_q+\|\X_q\|^2/3$, a {\it quadratic} model;

\item $m(\X)\equiv 0$, a {\it null} model;

\item $m(\X)\equiv \bon_p\trans\X\{1+2(\bze_{p/2}\trans,\bon_{p/2}\trans)\X/p\}$, a {\it double index} model.
\end{enumerate}
The outcome models (d) and (e) are considered for cases with $p=10$ only and their results are summarized in 
\tcr{Appendix \ref{sm_simulations}} of the Supplementary Material. The following discussions mainly focus on the outcome models (a)--(c).

\tcr{T}he \tcr{\emph{estimators}} $\mhatnk(\x)$ and $\phihatnk(\x,\thetahatinit)$ are constructed based on the data $\cl_k^-$ through\tcr{:}

\begin{enumerate}[I.]
\item kernel smoothing (KS), \tcr{in} \eqref{ks_ate} and \eqref{ks_qte}, where $\hmbP\in\rR^{p\times r}$ is \tcr{chosen as:}

\vskip0.04in
\begin{enumerate}[1.]
\item the slope vector ($r=1$) from the complete-case version of unregularized or regularized linear regression of $Y$ vs. $\X$ (KS$_1$), \tcg{which correctly specifies the outcome models (a), (b) and (d) but misspecifies (c) and (e)}; ~~\tcr{or}

\item the first two directions ($r=2$) selected by the complete-case version of the unregularized (with $\ceil{n/5}$ slices of equal width) or regularized (with $4$ slices of equal size) sliced inverse regression \citep{li1991sliced, lin2019sparse} of $Y$ vs. $\X$ (KS$_2$), \tcg{which correctly specifies the outcome models (a), (b), (d) and (e) but misspecifies (c)}; ~~\tcr{or}
\end{enumerate}

\vskip0.04in
\item parametric regression (PR), giving
\bse
\mhatnk(\x)~\equiv~(1,\x\trans)\trans\bxihat_k \tcr{\quad \hbox{and} \quad} \phihatnk(\x,\thetahatinit)~\equiv~ h\{(1,\x\trans)\trans\bgammahat_k\}-\tau\tcr{,}
\ese
with $\bxihat_k/\bgammahat_k$ \tcr{respectively being} the slope vector from the complete-case version of unregularized or regularized linear$/$logistic regression of $Y/I(Y<\thetahatinit)$ vs. $\X$ using $\cl_k^-$, \tcg{which correctly specifies the outcome models \{(a), (d)\}  and (d) for the ATE and QTE estimation, respectively, while misspecifying the others.}
\end{enumerate}
\tcg{In general, our choices of $\{\pi(\x),m(\x)\}$ incorporate \tcr{both} linear \tcr{and non-linear effects, including} quadratic and interaction effects, that are commonly encountered in practice. Also, our approaches to constructing $\{\pihatN(\x), \mhatnk(\x), \phihatnk(\x,\theta)\}$ represent a broad class of flexible and user-friendly (parametric or semi-parametric) strategies \tcr{often adopted} for modeling the relation between a continuous or binary response and a set of (possibly high dimensional) covariates.} \tcr{They also allow for a variety of scenarios in terms of correct/incorrect specifications of the (working) nuisance models.} 
\tcr{B}ased on the various $\mhatnk(\cdot)$ and $\phihatnk(\cdot,\cdot)$ described above, we obtain $\mhatn(\cdot)$ and $\phihatn(\cdot,\cdot)$ via the cross fitting procedures \eqref{ds1}--\eqref{ds2} and \eqref{ds3}--\eqref{ds4}. In addition, for the QTE estimation, we plug $\thetahatinit$ and $\hf(\cdot)$ from Remark \ref{remark_qte_initial_estimator} into $\thetahatss$ defined by \eqref{ss_qte}, while obtaining the initial estimator and estimated density for $\thetahatsup$ in \eqref{sup_qte} through the same IPW approach but with $\pihatn(\cdot)$ instead of $\pihatN(\cdot)$ \tcr{(i.e., the version based on $\cl$ instead of $\cu$). The same $\pihatn(\cdot)$ is also used for constructing the supervised ATE estimator $\muhatsup$ in \eqref{sup_ate}.}

\tcr{For all combinations of the true data generating models, and for {any} of the choices of the nuisance function estimators as listed above, we implement our SS ATE and QTE estimators, evaluate their performances for both estimation \tcr{(see Section \ref{sec_sim_estimation})} and inference \tcr{(see Section \ref{sec_sim_inference})}, and also compare their estimation efficiency with respect to a variety of corresponding supervised estimators, \eqref{sup_ate} and \eqref{sup_qte}, as well as their oracle versions} \tcg{(see the\tcr{ir formal} descriptions \tcr{in Section \ref{sec_sim_estimation}})}.  All the results are summarized from 500 replications.

\begin{table}
\def~{\hphantom{0}}
\caption{Efficiencies of the ATE estimators relative to the corresponding oracle supervised estimators; \tcg{see Remark \ref{remark_interpretation_RE} for interpretations of these relative efficiencies.} Here\tcr{,} $n$ denotes the labeled data size, $p$ the number of covariates, $q$ the model sparsity, $m(\X)\equiv\E(Y\mid\X)$, $\pi(\X)\equiv\E(T\mid\X)$, $\hat{\pi}(\X)$ \tcr{--} the estimated propensity score, Lin \tcr{--} logistic regression of $T$ vs. $\X$\tcr{,} and Quad \tcr{--} logistic regression of $T$ vs. $(\X\trans,\X_{[1]}^2,\ldots,\X_{[p]}^2)\trans$; KS$_1/$KS$_2$ represents kernel smoothing on the one$/$two direction(s) selected by linear regression$/$
\tcr{sliced} inverse regression; PR \tcr{denotes} parametric regression\tcr{,} and ORE oracle relative efficiency. The \textbf{\tcn{blue}} color implies
the best efficiency in each case.}{
\resizebox{\textwidth}{!}{
\begin{tabular}{ccc||ccc|ccc||ccc|ccc||c}
\hline
\multicolumn{3}{c||}{\multirow{2}{*}{$p=10$}}  & \multicolumn{6}{c||}{$n=200$}                      & \multicolumn{6}{c||}{$n=500$}                      & \multirow{3}{*}{ORE} \\
\cline{4-15}
&  &  & \multicolumn{3}{c|}{Supervised} & \multicolumn{3}{c||}{\textbf{SS}}& \multicolumn{3}{c|}{Supervised} & \multicolumn{3}{c||}{\textbf{SS}}&                      \\
$m(\X)$              & $\pi(\X)$            & $\hat{\pi}(\X)$      & KS$_1$  & KS$_2$ & PR   & KS$_1$ & KS$_2$ & PR   & KS$_1$  & KS$_2$ & PR   & KS$_1$ & KS$_2$ & PR   &                      \\ \hline
\multirow{6}{*}{(a)} & (i)   & Lin  & 0.87 & 0.86 & 0.96 & 2.99 & 2.74 & \tcn{\bf 3.72} & 0.99 & 0.98 & 0.99 & 3.35 & 3.19 & \tcn{\bf 3.70} & 4.37 \\
&       & Quad & 0.79 & 0.63 & 0.91 & 3.00 & 2.74 & \tcn{\bf 3.74} & 0.97 & 0.96 & 0.98 & 3.34 & 3.20 & \tcn{\bf 3.69} & 4.37 \\
& (ii)  & Lin  & 0.93 & 0.91 & 0.99 & 3.37 & 3.10 & \tcn{\bf 4.05} & 1.00 & 1.00 & 0.99 & 3.64 & 3.55 & \tcn{\bf 3.93} & 4.78 \\
&       & Quad & 0.88 & 0.85 & 0.91 & 3.43 & 3.19 & \tcn{\bf 4.07} & 0.99 & 1.00 & 0.98 & 3.68 & 3.59 & \tcn{\bf 3.96} & 4.78 \\
& (iii) & Lin  & 0.87 & 0.84 & 0.95 & 2.89 & 2.53 & \tcn{\bf 4.05} & 0.96 & 0.95 & 0.99 & 3.21 & 3.08 & \tcn{\bf 3.88} & 4.99 \\
&       & Quad & 0.86 & 0.81 & 0.91 & 3.08 & 2.70 & \tcn{\bf 4.13} & 0.98 & 0.98 & 1.00 & 3.44 & 3.31 & \tcn{\bf 3.92} & 4.99 \\ \hline
\multirow{6}{*}{(b)} & (i)   & Lin  & 0.93 & 0.92 & 0.51 & \tcn{\bf 3.62} & 3.42 & 1.03 & 0.99 & 0.98 & 0.67 & \tcn{\bf 3.73} & 3.61 & 1.17 & 5.07 \\
&       & Quad & 0.92 & 0.77 & 0.40 & \tcn{\bf 3.64} & 3.49 & 1.02 & 0.98 & 0.98 & 0.61 & \tcn{\bf 3.74} & 3.59 & 1.16 & 5.07 \\
& (ii)  & Lin  & 0.94 & 0.86 & 0.26 & \tcn{\bf 2.29} & 1.69 & 0.36 & 0.92 & 0.91 & 0.15 & \tcn{\bf 2.29} & 2.16 & 0.18 & 3.55 \\
&       & Quad & 0.85 & 0.81 & 0.28 & \tcn{\bf 2.35} & 1.76 & 0.41 & 0.91 & 0.90 & 0.17 & \tcn{\bf 2.34} & 2.20 & 0.21 & 3.55 \\
& (iii) & Lin  & 0.90 & 0.89 & 0.51 & \tcn{\bf 3.10} & 2.83 & 0.88 & 0.97 & 0.97 & 0.60 & \tcn{\bf 3.05} & 3.00 & 0.84 & 4.39 \\
&       & Quad & 0.87 & 0.84 & 0.56 & \tcn{\bf 3.20} & 2.90 & 1.08 & 0.98 & 0.96 & 0.63 & \tcn{\bf 3.11} & 3.04 & 1.07 & 4.39 \\ \hline
\multirow{6}{*}{(c)} & (i)   & Lin  & 0.62 & 0.61 & 0.67 & \tcn{\bf 1.23} & 1.21 & 1.17 & 0.78 & 0.79 & 0.74 & 1.52 & \tcn{\bf 1.58} & 1.45 & 9.52 \\
&       & Quad & 0.61 & 0.54 & 0.60 & \tcn{\bf 1.21} & 1.21 & 1.15 & 0.84 & 0.85 & 0.80 & 1.50 & \tcn{\bf 1.56} & 1.41 & 9.52 \\
& (ii)  & Lin  & 0.70 & 0.66 & 0.56 & \tcn{\bf 1.32} & 1.17 & 1.01 & 0.85 & 0.84 & 0.55 & \tcn{\bf 1.58} & 1.52 & 0.96 & 8.71 \\
&       & Quad & 0.79 & 0.75 & 0.83 & \tcn{\bf 1.35} & 1.19 & 1.32 & 0.90 & 0.89 & 0.83 & 1.47 & 1.46 & \tcn{\bf 1.49} & 8.71 \\
& (iii) & Lin  & 0.57 & 0.58 & 0.53 & 0.92 & \tcn{\bf 0.95} & 0.87 & 0.48 & 0.49 & 0.43 & 0.70 &\tcn{\bf  0.72} & 0.61 & 9.42 \\
&       & Quad & 0.78 & 0.74 & 0.83 & \tcn{\bf 1.42} & 1.40 & 1.51 & 0.94 & 0.92 & 0.92 & 1.59 & \tcn{\bf 1.60} & 1.55 & 9.42\\
\hline
\multicolumn{16}{c}{ } \\\hline
\multicolumn{3}{c||}{\multirow{2}{*}{$p=200,q=5$}} & \multicolumn{6}{c||}{$n=200$}                      & \multicolumn{6}{c||}{$n=500$}                      & \multirow{3}{*}{ORE} \\
\cline{4-15}
\multicolumn{3}{c||}{} & \multicolumn{3}{c|}{Supervised} & \multicolumn{3}{c||}{\textbf{SS}}& \multicolumn{3}{c|}{Supervised} & \multicolumn{3}{c||}{\textbf{SS}}&                      \\
$m(\X)$              & $\pi(\X)$            & $\hat{\pi}(\X)$      & KS$_1$  & KS$_2$ & PR   & KS$_1$ & KS$_2$ & PR   & KS$_1$  & KS$_2$ & PR   & KS$_1$ & KS$_2$ & PR   &                      \\ \hline
\multirow{6}{*}{(a)}   & (i)         & Lin              & 0.72    & 0.22   & 0.46 & \tcn{\bf 1.60}   & 0.67   & 1.43 & 0.94    & 0.85   & 0.73 & \tcn{\bf 1.88}   & 1.62   & 1.73 & 2.68                 \\
&             & Quad             & 0.70    & 0.20   & 0.43 & \tcn{\bf 1.61}   & 0.67   & 1.42 & 0.94    & 0.83   & 0.68 & \tcn{\bf 1.89}   & 1.62   & 1.72 & 2.68                 \\
& (ii)        & Lin              & 0.87    & 0.45   & 0.70 & \tcn{\bf 1.89}   & 0.91   & 1.73 & 0.97    & 0.88   & 0.80 & \tcn{\bf 2.15}   & 2.00   & 2.05 & 2.89                 \\
&             & Quad             & 0.86    & 0.44   & 0.69 & \tcn{\bf 1.91}   & 0.92   & 1.75 & 0.97    & 0.88   & 0.78 & \tcn{\bf 2.15}   & 1.99   & 2.07 & 2.89                 \\
& (iii)       & Lin              & 0.82    & 0.34   & 0.57 & \tcn{\bf 1.74}   & 0.79   & 1.64 & 0.95    & 0.89   & 0.76 & \tcn{\bf 2.35}   & 2.06   & 2.17 & 3.00                 \\
&             & Quad             & 0.80    & 0.32   & 0.55 & \tcn{\bf 1.79}   & 0.84   & 1.68 & 0.95    & 0.86   & 0.72 & \tcn{\bf 2.45}   & 2.13   & 2.19 & 3.00                 \\ \hline
\multirow{6}{*}{(b)}   & (i)         & Lin              & 0.86    & 0.35   & 0.76 & \tcn{\bf 1.60}   & 0.94   & 1.06 & 0.95    & 0.95   & 0.65 & \tcn{\bf 2.04}   & 1.97   & 1.04 & 3.37                 \\
&             & Quad             & 0.83    & 0.31   & 0.74 & \tcn{\bf 1.61}   & 0.93   & 1.08 & 0.95    & 0.95   & 0.65 & \tcn{\bf 2.04}   & 1.97   & 1.03 & 3.37                 \\
& (ii)        & Lin              & 0.35    & 0.23   & 0.22 & \tcn{\bf 0.44}   & 0.40   & 0.35 & 0.55    & 0.35   & 0.14 & \tcn{\bf 0.73}   & 0.49   & 0.15 & 2.29                 \\
&             & Quad             & 0.35    & 0.22   & 0.22 & \tcn{\bf 0.45}   & 0.42   & 0.37 & 0.54    & 0.34   & 0.14 & \tcn{\bf 0.75}   & 0.51   & 0.16 & 2.29                 \\
& (iii)       & Lin              & 0.82    & 0.49   & 0.66 & \tcn{\bf 0.99}   & 0.72   & 0.68 & 0.88    & 0.85   & 0.68 & \tcn{\bf 1.48}   & 1.35   & 0.60 & 2.74                 \\
&             & Quad             & 0.80    & 0.45   & 0.64 & \tcn{\bf 1.13}   & 0.78   & 0.80 & 0.90    & 0.86   & 0.71 & \tcn{\bf 1.66}   & 1.55   & 0.84 & 2.74                 \\ \hline
\multirow{6}{*}{(c)}   & (i)         & Lin              & 0.59    & 0.23   & 0.39 & \tcn{\bf 1.00}   & 0.65   & 0.93 & 0.75    & 0.71   & 0.72 & 1.16   & 1.10   & \tcn{\bf 1.20} & 4.13                 \\
&             & Quad             & 0.57    & 0.20   & 0.36 & \tcn{\bf 1.00}   & 0.64   & 0.92 & 0.76    & 0.70   & 0.71 & 1.17   & 1.10   & \tcn{\bf 1.20} & 4.13                 \\
& (ii)        & Lin              & 0.64    & 0.35   & 0.43 & \tcn{\bf 0.99}   & 0.63   & 0.90 & 0.74    & 0.64   & 0.38 & \tcn{\bf 1.14}   & 1.05   & 0.79 & 3.63                 \\
&             & Quad             & 0.64    & 0.34   & 0.42 & \tcn{\bf 1.02}   & 0.64   & 0.94 & 0.74    & 0.64   & 0.37 & \tcn{\bf 1.21}   & 1.12   & 0.91 & 3.63                 \\
& (iii)       & Lin              & 0.39    & 0.19   & 0.25 & \tcn{\bf 0.68}   & 0.47   & 0.60 & 0.38    & 0.32   & 0.26 & \tcn{\bf 0.50}   & 0.47   & 0.43 & 3.78                 \\
&             & Quad             & 0.39    & 0.18   & 0.24 & \tcn{\bf 0.95}   & 0.59   & 0.82 & 0.40    & 0.33   & 0.26 & \tcn{\bf 1.33}   & 1.15   & 1.04 & 3.78                 \\
\hline
\multicolumn{16}{c}{ } \\ \hline
\multicolumn{3}{c||}{\multirow{2}{*}{$p=200,q=\ceil{p^{1/2}}$}} & \multicolumn{6}{c||}{$n=200$}                      & \multicolumn{6}{c||}{$n=500$}                      & \multirow{3}{*}{ORE} \\
\cline{4-15}
\multicolumn{3}{c||}{} & \multicolumn{3}{c|}{Supervised} & \multicolumn{3}{c||}{\textbf{SS}}& \multicolumn{3}{c|}{Supervised} & \multicolumn{3}{c||}{\textbf{SS}}&                      \\
$m(\X)$              & $\pi(\X)$            & $\hat{\pi}(\X)$      & KS$_1$  & KS$_2$ & PR   & KS$_1$ & KS$_2$ & PR   & KS$_1$  & KS$_2$ & PR   & KS$_1$ & KS$_2$ & PR   &                      \\ \hline
\multirow{6}{*}{(a)}   & (i)         & Lin              & 0.35    & 0.09   & 0.29 & \tcn{\bf 1.38}   & 0.46   & 1.20 & 0.83    & 0.60   & 0.60 & \tcn{\bf 3.59}   & 2.04   & 2.96 & 6.05                 \\
&             & Quad             & 0.34    & 0.09   & 0.28 & \tcn{\bf 1.36}   & 0.43   & 1.17 & 0.81    & 0.55   & 0.55 & \tcn{\bf 3.57}   & 2.01   & 2.87 & 6.05                 \\
& (ii)        & Lin              & 0.68    & 0.23   & 0.61 & \tcn{\bf 1.74}   & 0.51   & 1.64 & 0.97    & 0.73   & 0.80 & \tcn{\bf 3.90}   & 2.55   & 3.71 & 6.65                 \\
&             & Quad             & 0.67    & 0.23   & 0.60 & \tcn{\bf 1.78}   & 0.52   & 1.66 & 0.97    & 0.72   & 0.79 & \tcn{\bf 3.91}   & 2.51   & 3.72 & 6.65                 \\
& (iii)       & Lin              & 0.62    & 0.14   & 0.49 & \tcn{\bf 2.07}   & 0.60   & 1.91 & 0.91    & 0.74   & 0.70 & \tcn{\bf 3.77}   & 2.65   & 3.54 & 6.99                 \\
&             & Quad             & 0.60    & 0.13   & 0.48 & \tcn{\bf 2.13}   & 0.60   & 1.94 & 0.90    & 0.69   & 0.66 & \tcn{\bf 3.80}   & 2.67   & 3.50 & 6.99                 \\ \hline
\multirow{6}{*}{(b)}   & (i)         & Lin              & 0.40    & 0.11   & 0.34 & \tcn{\bf 1.29}   & 0.55   & 1.16 & 0.91    & 0.77   & 0.89 & \tcn{\bf 3.89}   & 2.96   & 2.27 & 6.78                 \\
&             & Quad             & 0.38    & 0.11   & 0.33 & \tcn{\bf 1.29}   & 0.52   & 1.16 & 0.88    & 0.70   & 0.89 & \tcn{\bf 3.91}   & 2.92   & 2.29 & 6.78                 \\
& (ii)        & Lin              & 0.31    & 0.18   & 0.24 & \tcn{\bf 0.68}   & 0.44   & 0.56 & 0.60    & 0.53   & 0.21 & \tcn{\bf 1.55}   & 1.43   & 0.34 & 4.97                 \\
&             & Quad             & 0.31    & 0.17   & 0.23 & \tcn{\bf 0.65}   & 0.42   & 0.54 & 0.59    & 0.52   & 0.21 & \tcn{\bf 1.52}   & 1.39   & 0.34 & 4.97                 \\
& (iii)       & Lin              & 0.63    & 0.18   & 0.54 & \tcn{\bf 1.64}   & 0.75   & 1.33 & 0.96    & 0.82   & 0.93 & \tcn{\bf 3.43}   & 2.71   & 2.09 & 6.14                 \\
&             & Quad             & 0.61    & 0.17   & 0.53 & \tcn{\bf 1.68}   & 0.77   & 1.36 & 0.94    & 0.78   & 0.93 & \tcn{\bf 3.45}   & 2.72   & 2.15 & 6.14                 \\ \hline
\multirow{6}{*}{(c)}   & (i)         & Lin              & 0.16    & 0.10   & 0.13 & \tcn{\bf 0.56}   & 0.41   & 0.52 & 0.61    & 0.36   & 0.38 & \tcn{\bf 1.27}   & 0.93   & 1.15 & 17.23                \\
&             & Quad             & 0.16    & 0.09   & 0.12 & \tcn{\bf 0.56}   & 0.39   & 0.51 & 0.59    & 0.32   & 0.34 & \tcn{\bf 1.26}   & 0.91   & 1.13 & 17.23                \\
& (ii)        & Lin              & 0.31    & 0.22   & 0.26 & 0.65   & 0.49   & \tcn{\bf 0.67} & 0.63    & 0.48   & 0.36 & \tcn{\bf 1.23}   & 1.07   & 1.06 & 16.30                \\
&             & Quad             & 0.30    & 0.22   & 0.25 & 0.65   & 0.48   & \tcn{\bf 0.65} & 0.63    & 0.49   & 0.35 & \tcn{\bf 1.24}   & 1.07   & 1.05 & 16.30                \\
& (iii)       & Lin              & 0.16    & 0.10   & 0.13 & \tcn{\bf 0.54}   & 0.40   & 0.48 & 0.39    & 0.26   & 0.22 & \tcn{\bf 0.72}   & 0.59   & 0.59 & 17.82                \\
&             & Quad             & 0.16    & 0.10   & 0.12 & \tcn{\bf 0.68}   & 0.52   & 0.53 & 0.38    & 0.24   & 0.21 & \tcn{\bf 1.27}   & 0.94   & 0.96 & 17.82 \\ \hline
\end{tabular}
}}
\label{table_ate_efficiency}
\end{table}

\begin{table}
\def~{\hphantom{0}}
\caption{\tcr{Efficiencies of QTE estimators.}
We consider the same scenario\tcr{(s)} as \tcr{in} Table \ref{table_ate_efficiency}, but now the estimand is the QTE.} 
{
\resizebox{\textwidth}{!}{
\begin{tabular}{ccc||ccc|ccc||ccc|ccc||c}
\hline
\multicolumn{3}{c||}{\multirow{2}{*}{$p=10$}}  & \multicolumn{6}{c||}{$n=200$}                      & \multicolumn{6}{c||}{$n=500$}                      & \multirow{3}{*}{ORE} \\
\cline{4-15}
&  &  & \multicolumn{3}{c|}{Supervised} & \multicolumn{3}{c||}{\textbf{SS}}& \multicolumn{3}{c|}{Supervised} & \multicolumn{3}{c||}{\textbf{SS}}&                      \\
$m(\X)$              & $\pi(\X)$            & $\hat{\pi}(\X)$      & KS$_1$  & KS$_2$ & PR   & KS$_1$ & KS$_2$ & PR   & KS$_1$  & KS$_2$ & PR   & KS$_1$ & KS$_2$ & PR   &                      \\ \hline
\multirow{6}{*}{(a)} & (i)                  & Lin                  & 0.96    & 0.90   & 0.79 & \tcn{\bf 1.98}   & 1.88   & 1.34 & 0.99    & 0.98   & 0.93 & 1.85   & 1.80   & \tcn{\bf 1.90} & 2.24                 \\
&                      & Quad                 & 0.74    & 0.69   & 0.65 & \tcn{\bf 2.05}   & 1.93   & 1.36 & 0.99    & 0.98   & 0.91 & 1.86   & 1.82   & \tcn{\bf 1.89} & 2.24                 \\
& (ii)                 & Lin                  & 0.86    & 0.85   & 0.82 & \tcn{\bf 1.56}   & 1.44   & 0.98 & 0.99    & 0.97   & 0.97 & 1.55   & 1.51   & \tcn{\bf 1.59} & 2.12                 \\
&                      & Quad                 & 0.79    & 0.77   & 0.73 & \tcn{\bf 1.56}   & 1.48   & 1.00 & 0.99    & 0.97   & 0.95 & 1.57   & 1.50   & \tcn{\bf 1.61} & 2.12                 \\
& (iii)                & Lin                  & 0.94    & 0.90   & 0.93 & 1.77   & 1.61   & \tcn{\bf 1.96} & 1.01    & 1.01   & 1.02 & \tcn{\bf 2.26}   & 2.24   & 2.18 & 2.42                 \\
&                      & Quad                 & 0.88    & 0.80   & 0.93 & 1.85   & 1.69   & \tcn{\bf 1.89} & 0.96    & 0.97   & 0.99 & \tcn{\bf 2.29}   & 2.27   & 2.15 & 2.42                 \\ \hline
\multirow{6}{*}{(b)} & (i)                  & Lin                  & 0.93    & 0.90   & 0.85 & \tcn{\bf 1.82}   & 1.70   & 1.42 & 0.95    & 0.93   & 0.92 & 1.78   & 1.73   & \tcn{\bf 1.84} & 2.13                 \\
&                      & Quad                 & 0.77    & 0.74   & 0.72 & \tcn{\bf 1.86}   & 1.73   & 1.45 & 0.96    & 0.95   & 0.91 & 1.78   & 1.72   & \tcn{\bf 1.81} & 2.13                 \\
& (ii)                 & Lin                  & 0.78    & 0.73   & 0.80 & \tcn{\bf 1.22}   & 1.10   & 1.08 & 0.82    & 0.75   & 0.78 & \tcn{\bf 1.38}   & 1.19   & 1.19 & 1.92                 \\
&                      & Quad                 & 0.66    & 0.65   & 0.74 & \tcn{\bf 1.28}   & 1.15   & 1.11 & 0.84    & 0.78   & 0.80 & \tcn{\bf 1.44}   & 1.26   & 1.24 & 1.92                 \\
& (iii)                & Lin                  & 0.90    & 0.88   & 0.89 & 1.57   & 1.45   & \tcn{\bf 1.79} & 0.93    & 0.93   & 0.95 & 1.82   & 1.84   & \tcn{\bf 1.92} & 2.16                 \\
&                      & Quad                 & 0.85    & 0.83   & 0.90 & 1.74   & 1.60   & \tcn{\bf 1.89} & 0.92    & 0.91   & 0.96 & 1.89   & 1.93   & \tcn{\bf 1.97} & 2.16                 \\ \hline
\multirow{6}{*}{(c)} & (i)                  & Lin                  & 0.71    & 0.70   & 0.69 & \tcn{\bf 1.12}   & 1.06   & 1.02 & 0.77    & 0.77   & 0.83 & 1.22   & 1.19   & \tcn{\bf 1.33} & 2.35                 \\
&                      & Quad                 & 0.69    & 0.69   & 0.60 & \tcn{\bf 1.11}   & 1.05   & 1.01 & 0.83    & 0.83   & 0.87 & 1.18   & 1.15   & \tcn{\bf 1.26} & 2.35                 \\
& (ii)                 & Lin                  & 0.70    & 0.70   & 0.66 & \tcn{\bf 0.99}   & 0.93   & 0.87 & 0.74    & 0.74   & 0.78 & 1.00   & 1.02   & \tcn{\bf 1.02} & 2.25                 \\
&                      & Quad                 & 0.82    & 0.79   & 0.74 & \tcn{\bf 1.08}   & 1.02   & 0.94 & 0.84    & 0.84   & 0.87 & 1.16   & \tcn{\bf 1.19}   & 1.09 & 2.25                 \\
& (iii)                & Lin                  & 0.61    & 0.63   & 0.65 & 0.82   & 0.80   & \tcn{\bf 0.96} & 0.58    & 0.58   & 0.63 & 0.77   & 0.77   & \tcn{\bf 0.88} & 2.55                 \\
&                      & Quad                 & 0.86    & 0.85   & 0.86 & 1.16   & 1.12   & \tcn{\bf 1.25} & 0.95    & 0.93   & 0.92 & \tcn{\bf 1.28}   & 1.25   & 1.26 & 2.55 \\\hline
\multicolumn{16}{c}{ } \\
\hline
\multicolumn{3}{c||}{\multirow{2}{*}{$p=200,q=5$}} & \multicolumn{6}{c||}{$n=200$}                      & \multicolumn{6}{c||}{$n=500$}                      & \multirow{3}{*}{ORE} \\
\cline{4-15}
\multicolumn{3}{c||}{} & \multicolumn{3}{c|}{Supervised} & \multicolumn{3}{c||}{\textbf{SS}}& \multicolumn{3}{c|}{Supervised} & \multicolumn{3}{c||}{\textbf{SS}}&                      \\
$m(\X)$              & $\pi(\X)$            & $\hat{\pi}(\X)$      & KS$_1$  & KS$_2$ & PR   & KS$_1$ & KS$_2$ & PR   & KS$_1$  & KS$_2$ & PR   & KS$_1$ & KS$_2$ & PR   &                      \\ \hline
\multirow{6}{*}{(a)}   & (i)         & Lin              & 0.73    & 0.39   & 0.35 & \tcn{\bf 1.29}   & 0.72   & 0.81 & 0.92    & 0.93   & 0.71 & \tcn{\bf 1.45}   & 1.40   & 1.22 & 1.78                 \\
&             & Quad             & 0.71    & 0.36   & 0.32 & \tcn{\bf 1.28}   & 0.70   & 0.80 & 0.90    & 0.91   & 0.69 & \tcn{\bf 1.45}   & 1.40   & 1.21 & 1.78                 \\
& (ii)        & Lin              & 0.88    & 0.44   & 0.35 & \tcn{\bf 1.03}   & 0.67   & 0.70 & 0.96    & 0.92   & 0.60 & \tcn{\bf 1.45}   & 1.35   & 1.05 & 1.69                 \\
&             & Quad             & 0.87    & 0.44   & 0.35 & \tcn{\bf 1.04}   & 0.69   & 0.69 & 0.95    & 0.91   & 0.57 & \tcn{\bf 1.46}   & 1.37   & 1.07 & 1.69                 \\
& (iii)       & Lin              & 0.91    & 0.47   & 0.43 & \tcn{\bf 1.31}   & 0.81   & 0.96 & 0.94    & 0.94   & 0.72 & \tcn{\bf 1.57}   & 1.55   & 1.33 & 1.86                 \\
&             & Quad             & 0.88    & 0.43   & 0.39 & \tcn{\bf 1.41}   & 0.83   & 1.00 & 0.96    & 0.95   & 0.71 & \tcn{\bf 1.61}   & 1.59   & 1.36 & 1.86                 \\ \hline
\multirow{6}{*}{(b)}   & (i)         & Lin              & 0.59    & 0.38   & 0.42 & \tcn{\bf 1.05}   & 0.73   & 0.79 & 0.89    & 0.90   & 0.96 & \tcn{\bf 1.29}   & 1.24   & 1.17 & 1.50                 \\
&             & Quad             & 0.55    & 0.36   & 0.39 & \tcn{\bf 1.06}   & 0.73   & 0.78 & 0.81    & 0.80   & 0.91 & \tcn{\bf 1.30}   & 1.26   & 1.19 & 1.50                 \\
& (ii)        & Lin              & 0.38    & 0.21   & 0.20 & \tcn{\bf 0.41}   & 0.33   & 0.35 & 0.77    & 0.70   & 0.22 & \tcn{\bf 0.81}   & 0.67   & 0.25 & 1.45                 \\
&             & Quad             & 0.38    & 0.21   & 0.20 & \tcn{\bf 0.43}   & 0.34   & 0.35 & 0.75    & 0.68   & 0.21 & \tcn{\bf 0.81}   & 0.69   & 0.26 & 1.45                 \\
& (iii)       & Lin              & 0.69    & 0.45   & 0.41 & \tcn{\bf 0.76}   & 0.64   & 0.67 & 0.95    & 0.93   & 0.88 & \tcn{\bf 1.08}   & 1.04   & 0.82 & 1.50                 \\
&             & Quad             & 0.67    & 0.40   & 0.38 & \tcn{\bf 0.83}   & 0.69   & 0.74 & 0.90    & 0.89   & 0.87 & \tcn{\bf 1.14}   & 1.11   & 0.95 & 1.50                 \\ \hline
\multirow{6}{*}{(c)}   & (i)         & Lin              & 0.67    & 0.35   & 0.30 & \tcn{\bf 0.91}   & 0.66   & 0.72 & 0.81    & 0.77   & 0.56 & \tcn{\bf 1.09}   & 1.05   & 0.91 & 1.81                 \\
&             & Quad             & 0.63    & 0.33   & 0.28 & \tcn{\bf 0.91}   & 0.67   & 0.71 & 0.81    & 0.77   & 0.55 & \tcn{\bf 1.08}   & 1.03   & 0.87 & 1.81                 \\
& (ii)        & Lin              & 0.66    & 0.34   & 0.30 & \tcn{\bf 0.77}   & 0.51   & 0.61 & 0.77    & 0.75   & 0.44 & 1.03   & \tcn{\bf 1.03}   & 0.75 & 1.74                 \\
&             & Quad             & 0.67    & 0.34   & 0.30 & \tcn{\bf 0.79}   & 0.52   & 0.62 & 0.75    & 0.73   & 0.42 & 1.08   & \tcn{\bf 1.09}   & 0.82 & 1.74                 \\
& (iii)       & Lin              & 0.55    & 0.24   & 0.22 & \tcn{\bf 0.62}   & 0.46   & 0.52 & 0.51    & 0.50   & 0.29 & \tcn{\bf 0.59}   & 0.57   & 0.49 & 1.91                 \\
&             & Quad             & 0.54    & 0.23   & 0.21 & \tcn{\bf 0.86}   & 0.55   & 0.68 & 0.55    & 0.53   & 0.29 & \tcn{\bf 0.97}   & 0.93   & 0.80 & 1.91                 \\ \hline
\multicolumn{16}{c}{ } \\ \hline
\multicolumn{3}{c||}{\multirow{2}{*}{$p=200,q=\ceil{p^{1/2}}$}} & \multicolumn{6}{c||}{$n=200$}                      & \multicolumn{6}{c||}{$n=500$}                      & \multirow{3}{*}{ORE} \\
\cline{4-15}
\multicolumn{3}{c||}{} & \multicolumn{3}{c|}{Supervised} & \multicolumn{3}{c||}{\textbf{SS}}& \multicolumn{3}{c|}{Supervised} & \multicolumn{3}{c||}{\textbf{SS}}&                      \\
$m(\X)$              & $\pi(\X)$            & $\hat{\pi}(\X)$      & KS$_1$  & KS$_2$ & PR   & KS$_1$ & KS$_2$ & PR   & KS$_1$  & KS$_2$ & PR   & KS$_1$ & KS$_2$ & PR   &                      \\ \hline
\multirow{6}{*}{(a)}   & (i)         & Lin              & 0.53    & 0.14   & 0.09 & \tcn{\bf 0.89}   & 0.44   & 0.43 & 0.85    & 0.80   & 0.45 & \tcn{\bf 2.06}   & 1.74   & 1.16 & 2.62                 \\
&             & Quad             & 0.53    & 0.14   & 0.09 & \tcn{\bf 0.92}   & 0.42   & 0.42 & 0.80    & 0.73   & 0.37 & \tcn{\bf 2.05}   & 1.73   & 1.12 & 2.62                 \\
& (ii)        & Lin              & 0.68    & 0.21   & 0.15 & \tcn{\bf 0.99}   & 0.40   & 0.41 & 0.79    & 0.71   & 0.33 & \tcn{\bf 1.63}   & 1.40   & 0.79 & 2.45                 \\
&             & Quad             & 0.67    & 0.21   & 0.15 & \tcn{\bf 1.01}   & 0.39   & 0.39 & 0.80    & 0.71   & 0.32 & \tcn{\bf 1.66}   & 1.43   & 0.75 & 2.45                 \\
& (iii)       & Lin              & 0.77    & 0.21   & 0.14 & \tcn{\bf 1.42}   & 0.58   & 0.62 & 0.85    & 0.80   & 0.50 & \tcn{\bf 2.21}   & 1.69   & 1.31 & 2.87                 \\
&             & Quad             & 0.76    & 0.20   & 0.14 & \tcn{\bf 1.40}   & 0.58   & 0.61 & 0.81    & 0.74   & 0.43 & \tcn{\bf 2.14}   & 1.68   & 1.32 & 2.87                 \\ \hline
\multirow{6}{*}{(b)}   & (i)         & Lin              & 0.46    & 0.12   & 0.08 & \tcn{\bf 0.73}   & 0.43   & 0.42 & 0.76    & 0.77   & 0.48 & \tcn{\bf 1.85}   & 1.62   & 1.10 & 2.59                 \\
&             & Quad             & 0.45    & 0.12   & 0.08 & \tcn{\bf 0.73}   & 0.41   & 0.39 & 0.70    & 0.70   & 0.40 & \tcn{\bf 1.82}   & 1.61   & 1.07 & 2.59                 \\
& (ii)        & Lin              & 0.38    & 0.18   & 0.13 & \tcn{\bf 0.56}   & 0.38   & 0.40 & 0.67    & 0.63   & 0.33 & \tcn{\bf 1.21}   & 1.16   & 0.72 & 2.29                 \\
&             & Quad             & 0.37    & 0.17   & 0.13 & \tcn{\bf 0.56}   & 0.35   & 0.37 & 0.69    & 0.64   & 0.32 & \tcn{\bf 1.15}   & 1.14   & 0.70 & 2.29                 \\
& (iii)       & Lin              & 0.68    & 0.19   & 0.13 & \tcn{\bf 0.97}   & 0.62   & 0.61 & 0.82    & 0.74   & 0.50 & \tcn{\bf 2.06}   & 1.66   & 1.37 & 2.73                 \\
&             & Quad             & 0.66    & 0.18   & 0.12 & \tcn{\bf 0.98}   & 0.63   & 0.61 & 0.80    & 0.72   & 0.46 & \tcn{\bf 1.99}   & 1.60   & 1.35 & 2.73                 \\ \hline
\multirow{6}{*}{(c)}   & (i)         & Lin              & 0.27    & 0.13   & 0.10 & \tcn{\bf 0.55}   & 0.42   & 0.45 & 0.72    & 0.67   & 0.27 & \tcn{\bf 1.11}   & 0.97   & 0.73 & 2.72                 \\
&             & Quad             & 0.27    & 0.13   & 0.09 & \tcn{\bf 0.53}   & 0.41   & 0.43 & 0.67    & 0.61   & 0.23 & \tcn{\bf 1.09}   & 0.95   & 0.69 & 2.72                 \\
& (ii)        & Lin              & 0.37    & 0.22   & 0.17 & \tcn{\bf 0.54}   & 0.42   & 0.47 & 0.67    & 0.57   & 0.21 & \tcn{\bf 0.94}   & 0.80   & 0.51 & 2.58                 \\
&             & Quad             & 0.37    & 0.22   & 0.17 & \tcn{\bf 0.54}   & 0.41   & 0.46 & 0.67    & 0.56   & 0.21 & \tcn{\bf 0.94}   & 0.81   & 0.49 & 2.58                 \\
& (iii)       & Lin              & 0.26    & 0.14   & 0.12 & \tcn{\bf 0.56}   & 0.42   & 0.45 & 0.62    & 0.49   & 0.23 & \tcn{\bf 0.87}   & 0.75   & 0.60 & 3.04                 \\
&             & Quad             & 0.26    & 0.14   & 0.11 & \tcn{\bf 0.59}   & 0.46   & 0.47 & 0.59    & 0.46   & 0.21 & \tcn{\bf 1.06}   & 0.89   & 0.71 & 3.04 \\ \hline
\end{tabular}
}}
\label{table_qte_efficiency}
\end{table}

\begin{table}
\def~{\hphantom{0}}
\caption{Inference based on the SS estimators 
\underline{\tcr{using} kernel smoothing on the direction selected by linear regression \tcr{(KS$_1$)}} \tcr{as the choice of the working outcome model, for the ATE and the QTE,} when $n=500$. 
Here\tcr{,} ESE is the empirical standard error, \tcr{Bias is the empirical bias,} ASE \tcr{is} the average of the estimated standard errors\tcr{,} and CR \tcr{is} the \tcr{empirical} coverage rate of the 95\% confidence intervals. \tcr{All o}ther notations are the same as in Table \ref{table_ate_efficiency}. The \textbf{{\color{navyblue} blue}} color \tcr{highlights settings where} 
the propensity scores and the outcome models are \tcr{both} correctly specified, while the \textbf{boldfaces} \tcr{indicate ones where} 
the propensity scores are correctly specified but the outcome models are not.}{
\resizebox{\textwidth}{!}{
\begin{tabular}{ccc|cccc|cccc|cccc}
\hline
\multicolumn{3}{c|}{ATE} & \multicolumn{4}{c|}{$p=10$} & \multicolumn{4}{c|}{$p=200,q=5$} & \multicolumn{4}{c}{$p=200,q=\ceil{p^{1/2}}$} \\
$m(\X)$              & $\pi(\X)$            & $\hat{\pi}(\X)$      & ESE  & Bias  & ASE  & CR   & ESE    & Bias   & ASE   & CR    & ESE       & Bias       & ASE      & CR       \\ \hline
& (i)   & {\color{navyblue} \textbf{Lin}}  & {\color{navyblue} \textbf{0.08}} & {\color{navyblue} \textbf{0.00}} & {\color{navyblue} \textbf{0.08}} & {\color{navyblue} \textbf{0.93}} & {\color{navyblue} \textbf{0.08}} & {\color{navyblue} \textbf{0.01}} & {\color{navyblue} \textbf{0.08}} & {\color{navyblue} \textbf{0.93}} & {\color{navyblue} \textbf{0.09}} & {\color{navyblue} \textbf{0.01}} & {\color{navyblue} \textbf{0.09}} & {\color{navyblue} \textbf{0.93}} \\
&       & {\color{navyblue} \textbf{Quad}} & {\color{navyblue} \textbf{0.08}} & {\color{navyblue} \textbf{0.00}} & {\color{navyblue} \textbf{0.08}} & {\color{navyblue} \textbf{0.93}} & {\color{navyblue} \textbf{0.08}} & {\color{navyblue} \textbf{0.01}} & {\color{navyblue} \textbf{0.07}} & {\color{navyblue} \textbf{0.95}} & {\color{navyblue} \textbf{0.09}} & {\color{navyblue} \textbf{0.02}} & {\color{navyblue} \textbf{0.09}} & {\color{navyblue} \textbf{0.93}} \\
& (ii)  & Lin                                  & 0.07                                 & 0.00                                 & 0.08                                 & 0.95                                 & 0.07                                 & 0.00                                 & 0.07                                 & 0.97                                 & 0.08                                 & 0.00                                 & 0.08                                 & 0.95                                 \\
&       & Quad                                 & 0.07                                 & 0.00                                 & 0.07                                 & 0.96                                 & 0.07                                 & 0.00                                 & 0.07                                 & 0.96                                 & 0.08                                 & 0.00                                 & 0.08                                 & 0.95                                 \\
& (iii) & Lin                                  & 0.08                                 & 0.00                                 & 0.08                                 & 0.93                                 & 0.07                                 & 0.01                                 & 0.07                                 & 0.94                                 & 0.08                                 & 0.01                                 & 0.08                                 & 0.94                                 \\
\multirow{-6}{*}{(a)} &       & {\color{navyblue} \textbf{Quad}} & {\color{navyblue} \textbf{0.08}} & {\color{navyblue} \textbf{0.00}} & {\color{navyblue} \textbf{0.07}} & {\color{navyblue} \textbf{0.93}} & {\color{navyblue} \textbf{0.07}} & {\color{navyblue} \textbf{0.01}} & {\color{navyblue} \textbf{0.07}} & {\color{navyblue} \textbf{0.94}} & {\color{navyblue} \textbf{0.08}} & {\color{navyblue} \textbf{0.01}} & {\color{navyblue} \textbf{0.08}} & {\color{navyblue} \textbf{0.94}} \\ \hline
& (i)   & {\color{navyblue} \textbf{Lin}}  & {\color{navyblue} \textbf{0.08}} & {\color{navyblue} \textbf{0.00}} & {\color{navyblue} \textbf{0.08}} & {\color{navyblue} \textbf{0.93}} & {\color{navyblue} \textbf{0.08}} & {\color{navyblue} \textbf{0.00}} & {\color{navyblue} \textbf{0.08}} & {\color{navyblue} \textbf{0.95}} & {\color{navyblue} \textbf{0.09}} & {\color{navyblue} \textbf{0.00}} & {\color{navyblue} \textbf{0.09}} & {\color{navyblue} \textbf{0.94}} \\
&       & {\color{navyblue} \textbf{Quad}} & {\color{navyblue} \textbf{0.08}} & {\color{navyblue} \textbf{0.00}} & {\color{navyblue} \textbf{0.08}} & {\color{navyblue} \textbf{0.94}} & {\color{navyblue} \textbf{0.08}} & {\color{navyblue} \textbf{0.00}} & {\color{navyblue} \textbf{0.08}} & {\color{navyblue} \textbf{0.94}} & {\color{navyblue} \textbf{0.09}} & {\color{navyblue} \textbf{0.01}} & {\color{navyblue} \textbf{0.09}} & {\color{navyblue} \textbf{0.94}} \\
& (ii)  & Lin                                  & 0.07                                 & 0.02                                 & 0.08                                 & 0.94                                 & 0.08                                 & 0.06                                 & 0.08                                 & 0.87                                 & 0.09                                 & 0.07                                 & 0.09                                 & 0.90                                 \\
&       & Quad                                 & 0.07                                 & 0.02                                 & 0.07                                 & 0.95                                 & 0.08                                 & 0.06                                 & 0.08                                 & 0.87                                 & 0.09                                 & 0.07                                 & 0.09                                 & 0.89                                 \\
& (iii) & Lin                                  & 0.08                                 & 0.00                                 & 0.07                                 & 0.93                                 & 0.08                                 & 0.01                                 & 0.08                                 & 0.96                                 & 0.08                                 & 0.01                                 & 0.08                                 & 0.95                                 \\
\multirow{-6}{*}{(b)} &       & {\color{navyblue} \textbf{Quad}} & {\color{navyblue} \textbf{0.08}} & {\color{navyblue} \textbf{0.00}} & {\color{navyblue} \textbf{0.07}} & {\color{navyblue} \textbf{0.93}} & {\color{navyblue} \textbf{0.08}} & {\color{navyblue} \textbf{0.00}} & {\color{navyblue} \textbf{0.07}} & {\color{navyblue} \textbf{0.96}} & {\color{navyblue} \textbf{0.08}} & {\color{navyblue} \textbf{0.00}} & {\color{navyblue} \textbf{0.08}} & {\color{navyblue} \textbf{0.95}} \\ \hline
& (i)   & \textbf{Lin}                         & \textbf{0.13}                        & \textbf{0.00}                        & \textbf{0.13}                        & \textbf{0.96}                        & \textbf{0.11}                        & \textbf{0.01}                        & \textbf{0.10}                        & \textbf{0.92}                        & \textbf{0.17}                        & \textbf{0.02}                        & \textbf{0.16}                        & \textbf{0.93}                        \\
&       & \textbf{Quad}                        & \textbf{0.13}                        & \textbf{0.00}                        & \textbf{0.13}                        & \textbf{0.95}                        & \textbf{0.11}                        & \textbf{0.01}                        & \textbf{0.10}                        & \textbf{0.92}                        & \textbf{0.17}                        & \textbf{0.03}                        & \textbf{0.16}                        & \textbf{0.92}                        \\
& (ii)  & Lin                                  & 0.11                                 & 0.01                                 & 0.12                                 & 0.97                                 & 0.09                                 & 0.02                                 & 0.09                                 & 0.95                                 & 0.15                                 & 0.04                                 & 0.15                                 & 0.94                                 \\
&       & Quad                                 & 0.11                                 & -0.04                                & 0.12                                 & 0.96                                 & 0.09                                 & 0.01                                 & 0.09                                 & 0.96                                 & 0.15                                 & 0.04                                 & 0.15                                 & 0.94                                 \\
& (iii) & Lin                                  & 0.12                                 & 0.13                                 & 0.12                                 & 0.83                                 & 0.09                                 & 0.11                                 & 0.09                                 & 0.78                                 & 0.15                                 & 0.15                                 & 0.15                                 & 0.83                                 \\
\multirow{-6}{*}{(c)} &       & \textbf{Quad}                        & \textbf{0.12}                        & \textbf{0.01}                        & \textbf{0.12}                        & \textbf{0.95}                        & \textbf{0.09}                        & \textbf{-0.01}                       & \textbf{0.10}                        & \textbf{0.97}                        & \textbf{0.16}                        & \textbf{-0.02}                       & \textbf{0.17}                        & \textbf{0.96}
\\ \hline
\multicolumn{15}{c}{ } \\ \hline
\multicolumn{3}{c|}{QTE}& \multicolumn{4}{c|}{$p=10$} & \multicolumn{4}{c|}{$p=200,q=5$} & \multicolumn{4}{c}{$p=200,q=\ceil{p^{1/2}}$} \\
$m(\X)$              & $\pi(\X)$            & $\hat{\pi}(\X)$      & ESE  & Bias  & ASE  & CR   & ESE    & Bias   & ASE   & CR    & ESE       & Bias       & ASE      & CR       \\ \hline
& (i)   & {\color{navyblue} \textbf{Lin}}  & {\color{navyblue} \textbf{0.15}} & {\color{navyblue} \textbf{0.04}} & {\color{navyblue} \textbf{0.15}} & {\color{navyblue} \textbf{0.92}} & {\color{navyblue} \textbf{0.13}} & {\color{navyblue} \textbf{0.01}} & {\color{navyblue} \textbf{0.13}} & {\color{navyblue} \textbf{0.95}} & {\color{navyblue} \textbf{0.17}} & {\color{navyblue} \textbf{-0.01}} & {\color{navyblue} \textbf{0.17}} & {\color{navyblue} \textbf{0.94}} \\
&       & {\color{navyblue} \textbf{Quad}} & {\color{navyblue} \textbf{0.15}} & {\color{navyblue} \textbf{0.04}} & {\color{navyblue} \textbf{0.15}} & {\color{navyblue} \textbf{0.93}} & {\color{navyblue} \textbf{0.13}} & {\color{navyblue} \textbf{0.01}} & {\color{navyblue} \textbf{0.13}} & {\color{navyblue} \textbf{0.95}} & {\color{navyblue} \textbf{0.17}} & {\color{navyblue} \textbf{-0.01}} & {\color{navyblue} \textbf{0.17}} & {\color{navyblue} \textbf{0.94}} \\
& (ii)  & Lin                                  & 0.15                                 & 0.04                                 & 0.14                                 & 0.91                                 & 0.13                                 & 0.01                                 & 0.12                                 & 0.94                                 & 0.18                                 & -0.01                                 & 0.16                                 & 0.92                                 \\
&       & Quad                                 & 0.15                                 & 0.04                                 & 0.14                                 & 0.91                                 & 0.13                                 & 0.01                                 & 0.12                                 & 0.94                                 & 0.18                                 & -0.01                                 & 0.16                                 & 0.93                                 \\
& (iii) & Lin                                  & 0.13                                 & 0.02                                 & 0.13                                 & 0.94                                 & 0.11                                 & 0.01                                 & 0.12                                 & 0.96                                 & 0.15                                 & 0.01                                  & 0.15                                 & 0.95                                 \\
\multirow{-6}{*}{(a)} &       & {\color{navyblue} \textbf{Quad}} & {\color{navyblue} \textbf{0.13}} & {\color{navyblue} \textbf{0.02}} & {\color{navyblue} \textbf{0.13}} & {\color{navyblue} \textbf{0.94}} & {\color{navyblue} \textbf{0.11}} & {\color{navyblue} \textbf{0.01}} & {\color{navyblue} \textbf{0.12}} & {\color{navyblue} \textbf{0.96}} & {\color{navyblue} \textbf{0.15}} & {\color{navyblue} \textbf{0.01}}  & {\color{navyblue} \textbf{0.15}} & {\color{navyblue} \textbf{0.95}} \\ \hline
& (i)   & {\color{navyblue} \textbf{Lin}}  & {\color{navyblue} \textbf{0.15}} & {\color{navyblue} \textbf{0.02}} & {\color{navyblue} \textbf{0.14}} & {\color{navyblue} \textbf{0.92}} & {\color{navyblue} \textbf{0.13}} & {\color{navyblue} \textbf{0.01}} & {\color{navyblue} \textbf{0.13}} & {\color{navyblue} \textbf{0.95}} & {\color{navyblue} \textbf{0.18}} & {\color{navyblue} \textbf{0.00}}  & {\color{navyblue} \textbf{0.17}} & {\color{navyblue} \textbf{0.93}} \\
&       & {\color{navyblue} \textbf{Quad}} & {\color{navyblue} \textbf{0.15}} & {\color{navyblue} \textbf{0.02}} & {\color{navyblue} \textbf{0.14}} & {\color{navyblue} \textbf{0.93}} & {\color{navyblue} \textbf{0.13}} & {\color{navyblue} \textbf{0.01}} & {\color{navyblue} \textbf{0.13}} & {\color{navyblue} \textbf{0.95}} & {\color{navyblue} \textbf{0.18}} & {\color{navyblue} \textbf{0.00}}  & {\color{navyblue} \textbf{0.17}} & {\color{navyblue} \textbf{0.94}} \\
& (ii)  & Lin                                  & 0.14                                 & 0.05                                 & 0.14                                 & 0.94                                 & 0.12                                 & 0.07                                 & 0.12                                 & 0.94                                 & 0.19                                 & 0.05                                  & 0.17                                 & 0.92                                 \\
&       & Quad                                 & 0.14                                 & 0.05                                 & 0.14                                 & 0.95                                 & 0.12                                 & 0.07                                 & 0.12                                 & 0.93                                 & 0.19                                 & 0.04                                  & 0.17                                 & 0.92                                 \\
& (iii) & Lin                                  & 0.13                                 & 0.02                                 & 0.13                                 & 0.95                                 & 0.12                                 & 0.02                                 & 0.12                                 & 0.94                                 & 0.15                                 & 0.00                                  & 0.15                                 & 0.95                                 \\
\multirow{-6}{*}{(b)} &       & {\color{navyblue} \textbf{Quad}} & {\color{navyblue} \textbf{0.13}} & {\color{navyblue} \textbf{0.02}} & {\color{navyblue} \textbf{0.13}} & {\color{navyblue} \textbf{0.95}} & {\color{navyblue} \textbf{0.12}} & {\color{navyblue} \textbf{0.01}} & {\color{navyblue} \textbf{0.12}} & {\color{navyblue} \textbf{0.95}} & {\color{navyblue} \textbf{0.15}} & {\color{navyblue} \textbf{0.00}}  & {\color{navyblue} \textbf{0.15}} & {\color{navyblue} \textbf{0.95}} \\ \hline
& (i)   & \textbf{Lin}                         & \textbf{0.19}                        & \textbf{0.01}                        & \textbf{0.21}                        & \textbf{0.96}                        & \textbf{0.16}                        & \textbf{0.02}                        & \textbf{0.16}                        & \textbf{0.97}                        & \textbf{0.26}                        & \textbf{0.00}                         & \textbf{0.27}                        & \textbf{0.95}                        \\
&       & \textbf{Quad}                        & \textbf{0.20}                        & \textbf{0.01}                        & \textbf{0.21}                        & \textbf{0.95}                        & \textbf{0.16}                        & \textbf{0.03}                        & \textbf{0.16}                        & \textbf{0.97}                        & \textbf{0.26}                        & \textbf{0.00}                         & \textbf{0.27}                        & \textbf{0.95}                        \\
& (ii)  & Lin                                  & 0.20                                 & 0.07                                 & 0.19                                 & 0.92                                 & 0.14                                 & 0.04                                 & 0.15                                 & 0.94                                 & 0.24                                 & 0.05                                  & 0.24                                 & 0.95                                 \\
&       & Quad                                 & 0.19                                 & 0.01                                 & 0.19                                 & 0.95                                 & 0.14                                 & 0.02                                 & 0.15                                 & 0.95                                 & 0.24                                 & 0.04                                  & 0.24                                 & 0.96                                 \\
& (iii) & Lin                                  & 0.18                                 & 0.15                                 & 0.18                                 & 0.88                                 & 0.15                                 & 0.13                                 & 0.15                                 & 0.86                                 & 0.22                                 & 0.15                                  & 0.23                                 & 0.91                                 \\
\multirow{-6}{*}{(c)} &       & \textbf{Quad}                        & \textbf{0.18}                        & \textbf{0.01}                        & \textbf{0.18}                        & \textbf{0.95}                        & \textbf{0.14}                        & \textbf{0.05}                        & \textbf{0.14}                        & \textbf{0.93}                        & \textbf{0.22}                        & \textbf{0.11}                         & \textbf{0.23}                        & \textbf{0.93}
\\ \hline
\end{tabular}
}}
\label{table_inferece}
\end{table}

\subsection{\tcr{Results on estimation efficiency} 
}\label{sec_sim_estimation}
In Tables \ref{table_ate_efficiency}--\ref{table_qte_efficiency}, we report the efficiencies, measured by mean squared errors, of various supervised and SS estimators relative to the corresponding ``oracle'' supervised estimators $\muhatora$ and $\thetahatora$, constructed via substituting $\{\pi(\cdot),m(\cdot),\phi(\cdot,\cdot)\}$ for $\{\pihatn(\cdot),\mhatn(\cdot),\phihatn(\cdot,\cdot)\}$ in \eqref{sup_ate} and \eqref{sup_qte}. The supervised ``oracle'' estimators of the QTE use the initial estimators and estimated densities from the IPW approach described in Remark \ref{remark_qte_initial_estimator} with $\pihatN(\cdot)$ replaced by $\pi(\cdot)$. \tcg{\tcr{We clarify here that s}uch ``oracle'' estimators \tcr{(for both the ATE and the QTE)} are obviously \emph{unrealistic}\tcr{,}
and \tcr{are used here} just \tcr{to} serve as suitable benchmarks that are always consistent. Specifically, the relative efficiencies in Table \ref{table_ate_efficiency} are calculated by\tcr{:}}
\bse
\E\{(\muhatora-\mu_0)^2\}/\E\{(\muhatsup-\mu_0)^2\}\hbox{ and } \E\{(\muhatora-\mu_0)^2\}/\E\{(\muhatss-\mu_0)^2\},
\ese
\tcg{while those in Table \ref{table_qte_efficiency} are given by\tcr{:}}
\bse
\E\{(\thetahatora-\theta_0)^2\}/\E\{(\thetahatsup-\theta_0)^2\}\hbox{ and } \E\{(\thetahatora-\theta_0)^2\}/\E\{(\thetahatss-\theta_0)^2\}.
\ese
For reference, we provide the ``oracle'' relative efficiencies \tcr{(denoted \tcr{as} ``ORE'' in the tables)} given by\tcr{:} $\lams^2/\lamss^2$ and $\sigsup^2/\sigss^2$ with $\{m^*(\cdot),\phis(\cdot,\cdot)\}=\{m(\cdot),\phi(\cdot,\cdot)\}$ as well, where $\lams^2$, $\lamss^2$, $\sigsup^2$ and $\sigss^2$  are the asymptotic variances in \eqref{ate_normality}, \eqref{ate_sup_normality}, \eqref{qte_normality} and \eqref{qte_sup_normality}. The unknown quantities therein as well as the true values of $\mu_0$ and $\vt$ are approximated by Monte Carlo based on $100,000$ realizations of $(Y,T,\X\trans)\trans$ independent of $\cl\cup\cu$. It is noteworthy \tcr{here} that these ``oracle'' relative efficiencies can be achieved only asymptotically, \tcr{and that too \emph{only}} when $\{\pi(\cdot),m(\cdot),\phi(\cdot,\cdot)\}$ are all correctly specified and estimated at fast enough rates.

\vskip0.08in
\tcg{Generally speaking, the results in Tables \ref{table_ate_efficiency}--\ref{table_qte_efficiency} clearly show that our SS estimators uniformly outperform their supervised competitors\tcr{,} and even yield better efficiency than the supervised ``oracle'' estimators in most of the cases, indicated by numbers greater than one in the tables. Specifically, inspecting the two tables reveals that, among all the settings, our SS estimators make the most significant efficiency improvement when all the nuisance models are correctly specified. For instance, when $\{m(\X),\pi(\X)\}=\{(a),(i)\}$, the combination of Lin and PR correctly estimate the nuisance functions and give fairly impressive results for the ATE case.}

\tcg{\tcr{Moreover,} when both correctly approximating $\pi(\X)$, Lin and Quad yields similar results. However, under the setups with \tcr{$\{m(\X), \pi(\X)\}=\{(c),(iii)\}$}, for example, where Quad produces estimators converging to the true $\pi(\X)$ but Lin does not, and all the working outcome models misspecify the underlying relation between $Y/I(Y<\vt)$ vs. $\X$, Quad shows notable advantages over Lin. This substantiates the importance of the propensity score estimators $\pihatN(\X)$ in our methods, which has been
stated in Corollaries \ref{corate} and \ref{corqte}. As regards the choices of $\mhatnk(\X)$ and $\phihatnk(\X, \theta)$, KS$_1$ gives the best efficiency for most of the cases, justifying the approach combining kernel smoothing and dimension reduction to estimating the outcome models, as demonstrated in Sections \ref{sec_nf_ate}--\ref{sec_nf_qte}. Further, we observe that, as the labeled data size increases, the relative efficiencies of our SS estimators rise substantially, except for a few cases\tcr{,} such as the ATE 
\tcr{estimator} with the PR outcome model estimators when $p=10$.} 
\tcg{\tcr{The} improvement verifies the asymptotic properties claimed in Section \ref{sec_ate_ss} and \ref{sec_qte_general}, while \tcr{any of} the
exceptions could be explained by the fact that the performance of the benchmarks for calculating the relative efficiencies, i.e., the ``oracle'' supervised estimators, are improved by more labeled data as well. Considering \tcr{that} the ``oracle'' supervised estimators are always constructed with the true nuisance functions without \emph{any} estimation errors, the positive effect of increasing $n$ on them is very likely to be more significant than that on our SS estimators. }

\tcg{\tcr{In} addition, another interesting finding is that, in the scenario $(n,p,q)=(200,200,\ceil{p^{1/2}})$ where $q=O(n^{1/2})$, our SS estimators still beat their supervised counterparts under all the settings, and possess efficiencies close to or even \emph{better} than those of the supervised ``oracle'' estimators, which use the knowledge of the true data generating mechanisms, when all the nuisance models are correctly specified. This \tcr{(pleasantly)} surprising fact implies the performance of our methods is \tcr{somewhat}
\emph{insensitive} to the sparsity condition $q=o(n^{1/2})$, which is often required in the high dimensional \tcr{inference} literature \citep{buhlmann2011statistics, negahban2012unified, wainwright2019high} to ensure the $L_1$\tcr{--}consistency assumed in Assumption \ref{al1} for the nuisance estimators; see the relevant discussion in Remark \ref{remark_choice_of_P0} also.}

\begin{remark}[Interpretations of the relative efficiencies in Tables \ref{table_ate_efficiency}--\ref{table_qte_efficiency}]\label{remark_interpretation_RE}
\tcg{One may notice that the relative efficiencies of our SS estimators are sometimes quite different from the corresponding oracle quantities (ORE) in the tables. We attribute the differences to two reasons: (a) possible misspecification of the nuisance models, which obviously makes the oracle efficiencies unachievable, and (b) finite sample errors, from which \emph{any} practical methods have to suffer, especially in high dimensional scenarios. In contrast, the oracle relative efficiencies are calculated presuming all the nuisance models are known and the sample sizes are infinite.} \tcg{\tcr{Lastly,} it is also \tcr{worth} 
point\tcr{ing} out that the quantities in Tables \ref{table_ate_efficiency}--\ref{table_qte_efficiency} somewhat ``understate'' the efficiency gain of our methods in the sense that the benchmarks, i.e, the ``oracle'' supervised estimators, are \emph{unrealistic} due to requiring the knowledge of the underlying data generating mechanisms. When compared with the \emph{feasible} supervised estimators, the advantage of our methods is even \emph{more significant}. For example, when $(n,p,q)=(200,200,\ceil{p^{1/2}})$, $\{m(\X),\pi(\X)\}=\{(c),(i)\}$ and the nuisance functions are estimated by the combination of Lin and KS$_1$, the efficiencies of our SS estimators relative to the supervised competitors are $0.56/0.16=3.50$ and $0.55/0.27=2.04$ for the cases of the ATE and the QTE, respectively. Relative to the original numbers $0.56$ and $0.55$ in the tables, the ratios $3.50$ and $2.04$ \tcr{indeed} 
provide \tcr{a} more \tcr{direct and overwhelming} 
evidence \tcr{of} 
the efficiency superiority of our methods, while we choose the ``oracle'' supervised estimators as suitable \tcr{(common)} benchmarks \tcr{(for comparing all estimators -- supervised and semi-supervised)} just because they are always consistent, \tcr{and more importantly, are the \emph{best} achievable supervised estimators (and yet \tcr{are} idealized/infeasible, with both nuisance functions $\pi(\cdot)$ and $m(\cdot)/\phi(\cdot,\cdot)$ presumed known)}.}
\end{remark}

\subsection{\tcr{Results on inference}}\label{sec_sim_inference}
Next, Table \ref{table_inferece} presents the results of inference based on our SS estimators using KS$_1$ \tcr{(as a representative case)} to calculate $\mhatn(\cdot)$ and $\phihatn(\cdot,\cdot)$ when $n=500$. We report the bias, the empirical standard error (ESE), the average of the estimated standard errors (ASE), and the coverage rate (CR) of the 95\% confidence intervals. As expected, the biases are negligible as long as either the propensity score or the outcome model  is correctly specified, which verifies the DR property of our methods. Moreover, we can see \tcr{that} whenever $\pis(\cdot)=\pi(\cdot)$, the ASEs are fairly close to the corresponding ESEs and the CRs are all around the nominal level \tcr{of} 0.95\tcr{,} \tcr{{even if}} $m^*(\cdot)\neq m(\cdot)$ and $\phis(\cdot,\cdot)\neq\phi(\cdot,\cdot)$. See, for example, the results of the configurations marked in bold, where $\pis(\cdot)=\pi(\cdot)$ but the outcome model estimators based on KS$_1$ do {not} converge to $m(\cdot)$ (for the ATE) or $\phi(\cdot,\cdot)$ (for the QTE). Such an observation confirms that, owing to the use of the massive unlabeled data, the \tcr{{$n^{1/2}$-consistency and asymptotic normality}} of our \tcr{SS} ATE and QTE estimators \tcr{{only}} require correct specifications of $\pi(\cdot)$ as claimed in Corollaries \ref{corate} and \ref{corqte}. Also, it justifies the limiting distributions and variance estimations proposed in the two corollaries. \tcr{Lastly, as mentioned before, we only present results of inference for one case as an illustration.}  When we set $n=200$ or take other choices of $\{\mhatn(\cdot),\phihatn(\cdot,\cdot)\}$, our estimators still give satisfactory inference results similar \tcr{in flavor} to those in Table \ref{table_inferece}. We 
\tcr{therefore} skip them \tcr{here} for the sake of brevity.

\section{Real \tcr{d}ata \tcr{a}nalysis}\label{sec_data_analysis}
In this section, we apply our proposed methods to a data set from \citet{baxter2006genotypic} \tcr{that is} available at the Stanford University HIV Drug Resistance
Database \citep{rhee2003human} (https://hivdb.stanford.edu/pages/genopheno.dataset.html). \tcr{This data was also considered in \citet{zhang2019high} for illustration of their SS mean estimator\footnote{We are grateful to Yuqian Zhang for sharing details on data pre-processing in \citet{zhang2019high}.}.} In the data set, there is an observed outcome\tcr{,} $\mathbb{Y}$\tcr{,} representing the drug resistance to 
\tcr{lamivudine} (3TC), a nucleoside reverse transcriptase inhibitor, along with the indicators of mutations on $240$ positions of the HIV reverse transcriptase. \tcr{Our goal was to investigate the causal effect(s) (ATE$/$QTE) of these mutations on drug resistance.} We set the treatment indicator $T$ to be the existence of mutations on the $m$th position while regarding the other $p=239$ indicators  as the covariates $\X$. In the interest of space, we only take $m\in\{39,69,75,98,123,162,184,203\}$, a randomly selected subset of $\{1,\ldots,240\}$, for illustration. Analysis with other choices of $m$ can be conducted analogously. As regards \tcr{the} sample sizes, the labeled and unlabeled data contain $n=423$ and $N=2458$ observations, respectively. To test if the labeled and unlabeled data are equally distributed \tcg{and satisfy Assumption \ref{ass_equally_distributed}}, we calculate the Pearson \tcr{test} statistic and obtain the corresponding \tcr{$p$}-value \tcr{as} $0.18$ using a permutation distribution \citep{agresti2005multivariate}, 
implying that the labeling is indeed independent of $(T,\X\trans)\trans$. In the following,  we will estimate  the ATE \eqref{ate} and the QTE \eqref{qte} (with $\tau=0.5$) \tcr{with this data,}
based on the limiting distributions \eqref{ate_difference_distribution} and \eqref{qte_difference_distribution}\tcr{,} rather than focusing on $\mu_0(1)$ and $\vt(1)$ only.

\begin{figure}
	\centering
	\caption{\footnotesize \tcr{Data analysis:} $95\%$ confidence intervals for the ATE of \tcr{the mutations on} the drug resistance to 
		\tcr{3TC} based on the supervised estimator \eqref{sup_ate} (\underline{\tcr{undashed bars}}) and the SS estimator \eqref{ss_ate} (\underline{\tcr{dashed bars}}). Here\tcr{,} $m$ is the position of mutation regarded as the treatment indicator. We consider three different combinations to estimate the ``propensity score \& outcome model'': $(\mathrm{i})$ regularized logistic regression \& kernel smoothing on the first two directions selected by the regularized sliced inverse regression ({\color{red} 
			\textbf{red}} fill); $(\mathrm{ii})$ regularized logistic regression \& regularized parametric regression ({\color{darkpastelgreen} \textbf{green}} fill); $(\mathrm{iii})$ random forest \& random forest ({\color{bleudefrance} \textbf{blue}} fill).}
	\includegraphics[scale=0.52]{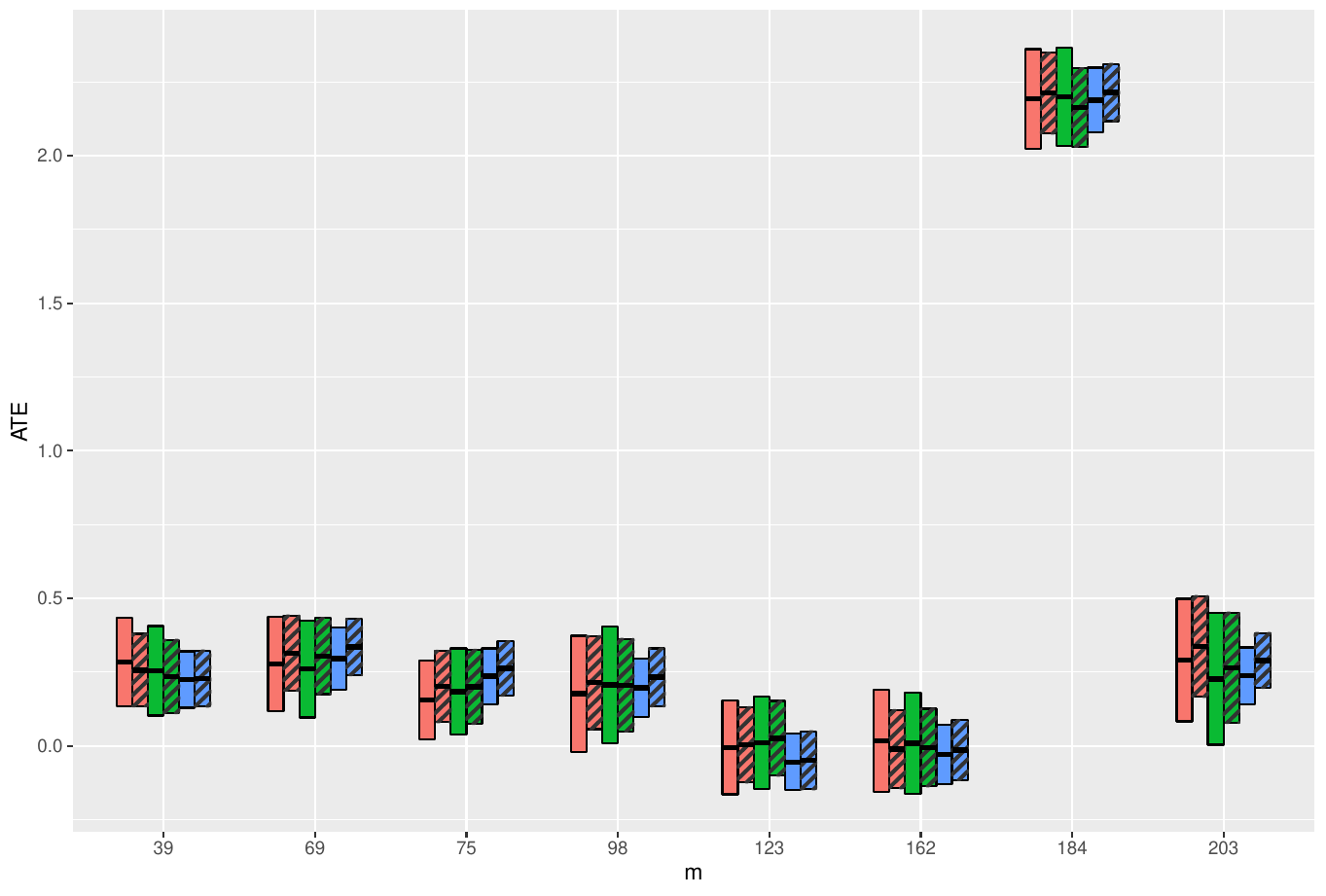}  
	\label{figure_ate}
\end{figure}

\begin{figure}
	\centering
	\caption{\footnotesize We consider the same scenario as \tcr{in} Figure \ref{figure_ate}, but now the estimand is the QTE ($\tau=0.5$). }
	\includegraphics[scale=0.52]{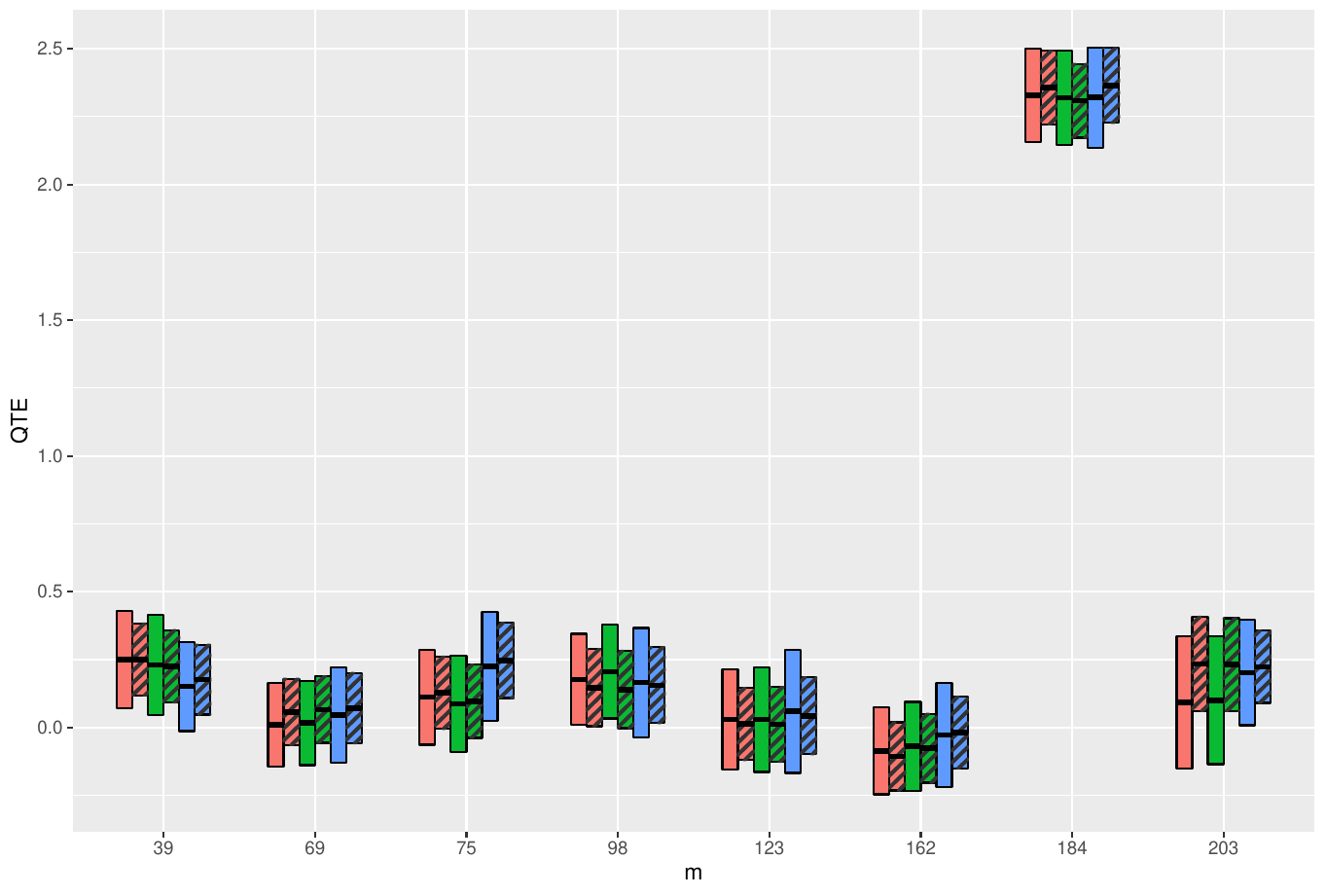} 
	\label{figure_qte}
\end{figure}

\tcr{For implementing our estimators, i}n addition to the \tcr{nuisance estimation}
approaches leveraged in 
Section \ref{sec_simulations}, we also estimate the propensity score and outcome models using random forest \tcr{here}, treating $T$, $Y$ or $I(Y<\thetahatinit)$ as the response,  growing $500$ trees and randomly sampling $\ceil{p^{1/2}}$ covariates as candidates at each split. In Figures \ref{figure_ate} and \ref{figure_qte}, we display the 95\% confidence intervals of the ATE and the QTE, respectively, averaging over 10 replications to remove 
potential randomness from cross fitting. 
\tcr{(The} confidence intervals are also presented numerically in 
\tcr{Appendix} \ref{sm_data_analysis}
of the Supplementary Material.\tcr{)} \tcr{From the plots, w}e observe that our SS approaches generally yield \tcr{\it shorter} confidence intervals than their supervised counterparts, confirming again the efficiency gain from the usage of unlabeled data. Moreover, we notice that, when $m=203$, all the SS confidence intervals of the QTE are strictly above zero, indicating significantly positive median treatment effect. This finding is, however, very
likely to be ignored in the supervised setting since zero is included by the confidence intervals constructed based on the
labeled data only. Such a contrast
\tcr{reinforces the fact that}
our SS methods \tcr{in comparison} are notably more powerful in detecting significant treatment effects.

\section{\tcr{Concluding discussion}} 
\label{sec_conclusion_discussion}
We have developed \tcr{here} a family of SS estimators for (a) the ATE and (b) the QTE\tcr{, in possibly high dimensional settings,}  
\tcr{and more importantly, we have developed a unified understanding of SS causal inference and its benefits -- {\it both} in robustness and efficiency -- something we feel has been missing in the literature}. In addition to the DR property in consistency that can be attained by purely supervised methods as well, we have proved our estimators also possess $n^{1/2}$-consistency and asymptotic normality whenever the propensity score $\pi(\cdot)$ is correctly specified. This property is useful for inference while generally unachievable in supervised settings. Even if \tcr{this} 
difference in robustness is ignored, our estimators are still guaranteed to be more efficient than their supervised counterparts. Further, as long as all the nuisance functions are correctly specified, our approaches have been shown to attain semi-parametric optimality \tcr{as well.} 
\tcr{All our theoretical claims above have also been validated numerically via extensive simulation studies and an empirical data analysis.} 

\tcr{Further, a}s a principled and flexible choice for estimating the outcome models in our methods, \tcr{we have studied thoroughly} IPW type kernel smoothing estimators \tcr{in high dimensional settings} with \tcr{possible use of} dimension reduction \tcr{techniques}.
We have shown they uniformly converge in probability to $\E(Y\mid\mbP_0\trans\X)$ (for the case of the ATE) or $\E\{\psi(Y,\theta)\mid\mbP_0\trans\X\}$ (for the case of the QTE) with some transformation matrix $\mbP_0$, given either the propensity score or the outcome model is correctly specified but {\it not} necessarily both. The precise convergence rates have been derived as well. This DR property guarantees the efficiency advantage of our SS methods over their supervised competitors. We view these results \tcr{also} as one of our major contributions. \tcr{To the best of our knowledge, results of this flavor (especially, in high dimensions, with $p$ diverging) have not been established in the relevant existing literature.} 
\tcr{They} can be applicable to many other problems \tcr{as well and should therefore be of independent interest}.


\paragraph*{\tcr{Extensions}}
\tcg{
\tcr{As mentioned in Section \ref{sec:psetup}, while we focus on the ATE and QTE
for simplicity and clarity of 
the main messages,}
our \tcr{SS} methods \tcr{\it can} be easily extended to \tcr{other causal estimands, including} the
\tcr{\it general $Z$-estimation problem} 
\tcmAC{\citep{van2000asymptotic,van1996weak}},
targeting a parameter defined as the solution to an estimating equation. 
As long as the estimand has a close form like $\mu_0\equiv\E(Y)$, one can construct a family of SS estimators in the same \tcmAC{spirit} 
as our ATE estimators \eqref{ss_ate}. An example is the \emph{linear regression parameter} $\bbeta_0^{\mbox{\tiny LIN}}:=\{\E(\Xarrow\Xarrow\trans)\}^{-1}\E(\Xarrow Y)$\tcmAC{,} that solves the equation\tcmAC{:} $\E\{\Xarrow(Y-\Xarrow\trans\bbeta_0^{\mbox{\tiny LIN}})\}=\bzero_d$, where $\Xarrow:=(1,\X\trans)\trans$. On the other hand, for estimating equations that cannot be solved straightforwardly, the one-step update strategy, used for our QTE estimators \eqref{ss_qte}, allows for simple and flexible implementations of SS estimation and inference with various choices of nuisance estimators. For instance, our approach to constructing the SS QTE estimators can be adapted for the \emph{quantile regression parameter} $\bbeta_0^{\mbox{\tiny QUAN}}$, defined by the equation $\E[\Xarrow\{I(Y<\Xarrow\trans\bbeta_0^{\mbox{\tiny QUAN}})-\tau\}]=\bzero_d$, with extra technical effort. These SS estimators for the general estimating equation problems are expected to possess desirable properties, such as improved robustness and efficiency relative to their supervised counterparts, which are similar in spirit to those stated in Sections \ref{secos} and \ref{secqte} for our SS ATE and QTE estimators. 
\tcm{We will briefly
discuss in Appendix \ref{sm_Z_estimation} \tcmAC{the methodological details of these}
possible extensions of our SS inference methods to the general $Z$-estimation problem under the potential outcome framework.}
\tcr{However, a detailed \tcmAC{theoretical}
analysis  is beyond the scope (and 
\tcmAC{primary} goals) of the current work, and therefore, we choose not to delve any further into these aspects here.}} 

\vskip0.05in
\tcr{Lastly, i}n this article, we have only considered cases where the labeled and unlabeled data are equally distributed \tcg{and thereby satisfy Assumption \ref{ass_equally_distributed}}. However, the labeling mechanisms in some practical problems are in fact not determined by design and \tcr{hence,} \tcr{\it labeling bias} \tcr{can exist} 
between $\cl$ and $\cu$. It is 
\tcr{important to note} that, due to the disproportion assumption \eqref{disproportion}, one \tcr{\it cannot} simply analyze such settings by \tcr{using} classical missing data theory \citep{tsiatis2007semiparametric, little2019statistical}, which requires the proportion of complete observations is bounded away from zero in the sample. Some recent attention has been paid to SS inference with labeling bias in the context of linear regression \tcr{\citep[Section II]{chakrabortty2018efficient}} 
and mean 
\tcr{estimation} \citep{zhang2021double_robust}. For treatment effect estimation, which is more technically complicated owing to the potential outcome framework, a primary challenge is that there exists no consistent supervised method when the labeled and unlabeled data follow different distributions\tcr{; so}  the 
goal of using unlabeled data to \tcr{`improve'} estimation accuracy compared to supervised approaches becomes somewhat ambiguous. With biased labeling mechanisms, we believe SS inference for treatment effect needs to be studied under a novel framework and thus poses an interesting problem for future research.




\begin{appendix}

{\cmag

\par\smallskip

\section{Extension to general $Z$-estimation problems}\label{sm_Z_estimation}
In this section, we briefly discuss the SS \tcmAC{inference} strategy for the \emph{general $Z$-estimation problem} \citep{van1996weak, van2000asymptotic} under the potential outcome framework, based on a natural extension of our methods for the ATE and the QTE in Sections \ref{secos} and \ref{secqte}. Specifically, for some \tcmAC{\it fixed} $d\geq 1$, we are interested in a $d$-dimensional parameter $\btheta_0\in\Lambda\subset\R^d$, \tcmAC{for some $\Lambda$,} defined as the solution to the \tcmAC{\it estimating equation}:
\be
\E\{\bpsi(Y,\X,\btheta_0)\}~=~\bzero_d,
\label{Z_estimation}
\ee
where 
$\bpsi(\cdot,\cdot,\cdot)\in\R^d$ \tcmAC{is some known function that satisfies:} $\E\{\|\bpsi(Y,\X,\btheta)\|^2\}<\infty$ for any $\btheta\in\Lambda$,  and that $\bH(\btheta):=\partial\E\{\bpsi(Y,\X,\btheta)\}/\partial\btheta$ exists and is non-singular in a neighborhood $\mb(\btheta_0,\varepsilon)$ of $\btheta_0$ for some $\varepsilon>0$. The special cases \tcmAC{with} $\bpsi(Y,\X,\btheta)\equiv Y-\btheta$ and $\bpsi(Y,\X,\btheta)\equiv I(Y<\btheta)-\tau$, with $d=1$, \tcmAC{correspond to the earlier cases of} 
the ATE and the QTE, respectively.  This type of SS $Z$-estimation problems \eqref{Z_estimation} \tcmAC{-- but} \emph{without} the missingness of the potential outcome $Y$ in the labeled data, which can be viewed as a special case of the following discussion with $T\equiv 1$, has been studied in Chapter 2 of \citet{Chakrabortty_Thesis_2016}.

\paragraph*{\tcmAC{SS estimators}} Similar in spirit to \eqref{qte_dr_representation}, we know the following \tcmAC{\it DR type representation}:
\be
\bzero_d&~=~&\E\{\bpsi(Y,\X,\btheta_0)\} \label{EE_DR_representation}\\
&~=~&\E\{  \bphi^*(\X,\btheta_0)\}+ \E[\{\pis(\X)\}^{-1} T\{\bpsi(Y,\X,\btheta_0) -  \bphi^*(\X,\btheta_0)\}], \nonumber
\ee
with arbitrary functions $\{\pi^*(\cdot),\bphi^*(\cdot,\cdot)\}$, holds true for the estimating equation \eqref{Z_estimation}, as long as either $\pi^*(\X)=\pi(\X)$ or $\bphi^*(\X,\btheta)=\bphi(\X,\btheta):=\E\{\bpsi(Y,\X,\btheta)\mid\X\}$, but \tcmAC{\it not} necessarily both. \tcmAC{The {\it empirical version} of \eqref{EE_DR_representation} constructed based on $\cl\cup\cu$} is \tcmAC{then} given by:  
\be
\E_{n+N}\{  \bphihat_n(\X,\btheta)\}+ \E_n[\{\pihatN(\X)\}^{-1} T\{\bpsi(Y,\X,\btheta) -  \bphihat_n(\X,\btheta)\}]~=~\bzero_d,
\label{sample_DR_representation}
\ee
where $\bphihat_n(\cdot,\cdot)$ is some estimator of $\bphi^*(\cdot,\cdot)$ from $\cl$, constructed via the cross-fitting procedures similar to \eqref{ds3}--\eqref{ds4} so that $\X_i$ and $\bphihat_n(\cdot,\cdot)$ are independent in $\bphihat_n(\X_i,\btheta)$ $(i=1,\ldots,n)$\tcmAC{, and $\pihatN(\cdot)$ is some estimator of $\pi(\cdot)$ based on $\cu$, same as in Sections \ref{secos}--\ref{secqte}.}
\tcmAC{Then,} following derivations analogous to those at the beginning of Section \ref{sec_qte_general}, which yielded our SS QTE estimators \eqref{ss_qte}, we can implement the one-step update approach based on the influence function corresponding to \eqref{sample_DR_representation}, and obtain \emph{a family of semi-supervised \tcmAC{$Z$-}estimators} for $\btheta_0$: \be
&&\quad\bthetahatss~:=~  \bthetahatinit +\{\bHhat_n(\bthetahatinit)\}^{-1}(\E_n[\{\pihatN(\X)\}^{-1}T\{\bphihat_n(\X,\bthetahatinit) - \bpsi(Y,\X,\bthetahatinit)\}]- \label{EE_one_step}\\
&&\phantom{\quad\bthetahatss~:=~  \bthetahatinit +\{\bHhat_n(\bthetahatinit)\}^{-1}(}\E_{n+N}\{\bphihat_n(\X,\bthetahatinit)\}),  \nonumber
\ee
indexed by $\{\pihatN(\cdot),\bphihat_n(\cdot,\cdot),\bthetahatinit,\bHhat_n(\cdot)\}$, where $\bthetahatinit$ is an initial estimator of $\btheta_0$ and $\bHhat_n(\cdot)$ is an estimator of $\bH(\cdot)$, both based on $\cl$. Of course, if the analytical solution, with respect to $\btheta$, of \eqref{sample_DR_representation} exists, one can directly take it as the SS estimator 
\tcmAC{$\bthetahatss$ itself.} Our SS ATE estimators $\muhatss$, given in \eqref{ss_ate}, are examples of this type. However, the one-step update \eqref{EE_one_step} is obviously a more general strategy that is  implementation-friendly and is broadly applicable to estimating  equations of various forms, regardless of whether their analytical solutions exist or not.

\paragraph*{\tcmAC{Properties of $\bthetahatss$ (brief sketch)}} To derive properties of our SS estimators $\bthetahatss$, we need the following restrictions on the complexity of the class of the estimating functions:
\be
&&\hbox{For some $\varepsilon>0$, the (random) function class $\{\bpsi(Y,\X,\btheta):\btheta\in\mb(\btheta_0,\varepsilon)\}$} \label{EE_P_Donsker}\\
&&\hbox{lies in a $\P$-Donsker class with square integrable envelope functions, \tcmAC{~and}}\nonumber \\
&&\hbox{$\E_\Z\{\|\bpsi(Y,\X,\tilde{\btheta})-\bpsi(Y,\X,\btheta_0)\|^2\}\tcmAC{~\xrightarrow{p}~}0$ \tcmAC{~for} any (random) \tcmAC{sequence~} $\tilde{\btheta}\xrightarrow{p}\btheta_0$.} \nonumber
\ee
\tcmAC{Further,} we require the function $\bpsi_0(\btheta):=\E\{\bpsi(Y,\X,\btheta)\}$ to be smooth enough so that, in $\mb(\btheta_0,\varepsilon)$ for some $\varepsilon>0$, it satisfies the Taylor expansion:
\be
&&\bpsi_0(\btheta)~=~\bpsi_0(\btheta_0)+\bH(\btheta_0)(\btheta-\btheta_0)+\bfr(\btheta,\btheta_0) \hbox{ \tcmAC{~for} some $\bfr(\btheta,\btheta_0)$\tcmAC{,}} \label{EE_Taylor}\\
&&\hbox{such that $\|\bfr(\btheta,\btheta_0)\|\tcmAC{~=~} O(\|\btheta-\btheta_0\|^2)$ \tcmAC{~as~} $\btheta\to\btheta_0$.} \nonumber
\ee
These conditions \eqref{EE_P_Donsker}--\eqref{EE_Taylor} are fairly mild and standard for estimating equation problems, while their analogues can be found in the \tcmAC{(supervised)} $Z$-estimation literature such as
\citet{van2000asymptotic}. It is also noteworthy that, under the basic Assumption \ref{adensity}, \eqref{EE_P_Donsker}--\eqref{EE_Taylor} are in fact satisfied by the special case $\bpsi(Y,\X,\btheta)\equiv I(Y<\btheta)-\tau$ with $d=1$, which is the estimating function \tcmAC{corresponding to the QTE;} 
see the proof of Theorem \ref{thqte} in Section \ref{proof_theorem_qte} for details.

\tcmAC{Further}, we need to regulate the behavior of the components $\{\pihatN(\cdot),\bphihat_n(\cdot,\cdot),\bthetahatinit,\bHhat_n(\cdot)\}$ in \eqref{EE_one_step} and the possibly misspecified limits $\{\pi^*(\cdot),\bphi^*(\cdot,\cdot)\}$ of $\{\pihatN^*(\cdot),\bphihat_n^*(\cdot,\cdot)\}$. Noticing that the \emph{high-level} conditions on $\{\pihatN(\cdot),\phihatn(\cdot,\cdot),\thetahatinit,\hf(\cdot),\pi^*(\cdot),\phi^*(\cdot,\cdot)\}$ \tcmAC{that were} 
enlisted in Assumptions \ref{ainit}--\ref{aest}, do \emph{not} require \emph{any} specific forms of these components, we can easily adapt them for the case of the general estimating equation \eqref{Z_estimation}, with appropriate modifications for the (fixed-dimensional) vector/matrix-valued (random) functions involved, e.g., taking the \emph{column-wise $L_2$-norms} $\|\cdot\|$ of these functions and their moments; see the definition of $\|\cdot\|$ in the Notation paragraph at the beginning of Section \ref{secos}.

Under the above assumptions on the estimating functions and the nuisance components, as well as some necessary (and \tcmAC{fairly} reasonable) convergence rate conditions, we can show the following results for our SS estimators $\bthetahatss$, which are similar in flavor to those established for our SS ATE and QTE estimators in Sections \ref{secos}--\ref{secqte}.
\begin{enumerate}[(i)]
	\item \emph{Double robustness:} Whenever either $\pi^*(\cdot)=\pi(\cdot)$ or $\bphi^*(\cdot,\cdot)=\bphi(\cdot,\cdot)$ holds, but not necessarily both, our SS estimators $\bthetahatss$ is consistent for $\btheta_0$.
	
	\item \emph{$n^{1/2}$-consistency and asymptotic normality}: Suppose $\pi^*(\cdot)=\pi(\cdot)$. Then, if either 
	$\bphi^*(\cdot,\cdot)=\bphi(\cdot,\cdot)$ or we can use the massive unlabeled data to estimate $\pi(\cdot)$ at a rate faster than $n^{-1/2}$, but \tcmAC{\it not} necessarily both, $\bthetahatss$ has the following expansion:
	\be
	&&\bthetahatss-\btheta_0~=~n^{-1}\sumi\bomegass(\Z_i,\btheta_0)+o_p(n^{-1/2}), \hbox{ with }  \bomegass(\Z,\btheta_0):= \label{EE_expansion}\\
	&&\{\bH(\btheta_0)\}^{-1}[\{\pi(\X)\}^{-1}T\{\bphi^*(\X,\btheta_0)-\bpsi(Y,\X,\btheta_0)\}-\E\{\bphi^*(\X,\btheta_0)\}], \nonumber
	\ee
	for an \emph{arbitrary} $\bphi^*(\cdot,\cdot)$, \emph{not} necessarily equal to $\bphi(\cdot,\cdot)$. This property is generally \emph{unachievable} in purely supervised settings \tcmAC{(similar in spirit to our discussions in Remarks \ref{remark_ate_robustness} and \ref{remark_qte_property})}. Further, the expansion \eqref{EE_expansion} implies the limiting distribution of $\bthetahatss$:
	\bse
	n^{1/2}(\bthetahatss-\btheta_0)~\xrightarrow{d}~\mn_d[\,\bzero_d,\cov\{\bomegass(\Z,\btheta_0)\}\,]\quad (n,N\to\infty).
	\ese
	
	\item \emph{Efficiency improvement and optimality}: Setting aside the robustness difference from our SS estimators, as stated in (ii), the \emph{best achievable influence function} of supervised estimators for $\btheta_0$, with the same outcome model estimator $\bphihat_n(\cdot,\cdot)$, is given by:
	\bse
	\bomegasup(\Z,\btheta_0)~:=~\{\bH(\btheta_0)\}^{-1}[\{\pi(\X)\}^{-1}T\{\bphi^*(\X,\btheta_0)-\bpsi(Y,\X,\btheta_0)\}-\bphi^*(\X,\btheta_0)].
	\ese
	Comparing the supervised and semi-supervised asymptotic covariance matrices, when $\bphi^*(\X,\btheta)\equiv\E\{\bpsi(Y,\X,\btheta)\mid\bfg(\X)\}$ for some function $\bfg(\cdot)$, we notice that
	\bse
	\cov\{\bomegasup(\Z,\btheta_0)\}-\cov\{\bomegass(\Z,\btheta_0)\} ~=~\{\bH(\btheta_0)\}^{-1}\cov\{\bphi^*(\X,\btheta_0)\}\{\bH(\btheta_0)\}^{-1},
	\ese
	which is positive semi-definite. This indicates the efficiency superiority of our SS estimators over their supervised counterparts. Moreover, if both the propensity score $\pi(\cdot)$ and the outcome model $\bphi(\cdot,\cdot)$ are correctly specified, the SS \tcmAC{estimator's} influence function $\bomegass(\Z,\btheta_0)$, given in \eqref{EE_expansion}, equals the \emph{efficient influence function} for estimating $\btheta_0$ under the semi-parametric model \eqref{semiparametric_model}, 
	\tcmAC{thus implying}
	$\bthetahatss$ attains the \tcmAC{corresponding} \emph{semi-parametric efficiency bound} and is \emph{(locally) semi-parametric efficient}.
\end{enumerate}
}

\section{Technical details}\label{sm_technical}

\subsection{Preliminary lemmas}\label{sm_lemmas}
The following Lemma \ref{1v2} would be useful in the proofs of the main theorems\tcr{, in particular, the results in Section \ref{secqte} regarding QTE estimation}.
\begin{lemma}\label{1v2}
Suppose there are two independent samples, $\ms_1$ and $\ms_2$, consisting of $n$ and $m$ independent copies of $(\X\trans,Y)\trans$, respectively. For $\bgamma\in\rR^d$ with some fixed $d$, let $\hg_{n}(\x,\bgamma)$ be an estimator of a measurable function $g(\x,\bgamma)\in\rR$ based on $\ms_1$ and \tcr{define:}
\bse
\mbG_{m}\{\hg_{n}(\X,\bgamma)\}~:=~ m^{1/2}[m^{-1}\hbox{$\sum_{(\X_i\trans,Y_i)\trans\in\ms_2}$}\hg_{n}(\X_i,\bgamma)-\E_\X\{\hg_{n}(\X,\bgamma)\}].
\ese
For some set $\ct\subset\rR^d$, denote
\bse
\Delta(\ms_1)~:=~(\sg\E_\X[\{\hg_n(\X,\bgamma)\}^2])^{1/2},\  M(\ms_1):=\sgx|\hg_n(\x,\bgamma)|.
\ese
For any $\eta\in(0,\Delta(\ms_1)+c\,]$, suppose $\G_{n}:=\{\hg_{n}(\X,\bgamma):\bgamma\in\ct\}$ satisfies that
\be
N_{[\,]}\{\eta,\G_{n}\mid\ms_1,L_2(\P_\X)\}~\leq~ H(\ms_1)\eta^{-c}\tcr{,}
\label{bracket2}
\ee
with some function $H(\ms_1)>0$. Here $\G_n$ is indexed by $\bgamma$ only and treats $\hg_n(\cdot,\bgamma)$ as a nonrandom function. Assume $H(\ms_1)=O_p(a_n)$, $\Delta(\ms_1)=O_p(d_{n,2})$ and $M(\ms_1)=O_p(d_{n,\infty})$  with some positive sequences $a_n$, $d_{n,2}$ and $d_{n,\infty}$ allowed to diverge, then we have\tcr{:}
\bse
\sg|\mbG_m\{\hg_n(\X,\bgamma)\}|~=~O_p(r_{n,m}),
\ese
where $r_{n,m}=d_{n,2}\{\log\,a_n+\log\,(d_{n,2}^{-1})\}+m^{-1/2}d_{n,\infty}\{(\log\,a_n)^2+(\log\,d_{n,2})^2\}$.
\end{lemma}

\subsection{Proof of Lemma \ref{1v2}} For any $\delta\in(0,\Delta(\ms_1)+c\,]$, we have that the bracketing integral
\bse
J_{[\,]}\{\delta,\G_n\mid\ms_1,L_2(\P_\X)\}&~\equiv~&\hbox{$\int_0^\delta$}[1+\log\,N_{[\,]}\{\eta,\G_n\mid\ms_1,L_2(\P_\X)\}]^{1/2}d\eta \\
&~\leq~&\hbox{$\int_0^\delta$}1+\log \,N_{[\,]}\{\eta,\G_n\mid\ms_1,L_2(\P_\X)\}d\eta  \\
&~\leq~&\hbox{$\int_0^\delta$}1+\log\,H(\ms_1)-c\,\log\,\eta\, d\eta \\
&~=~&\delta\{1+\log\,H(\ms_1)\}+c\,(\delta-\delta\,\log\,\delta),
\ese
where the third step is due to (\ref{bracket2}). This, combined with Lemma 19.36 of \citet{van2000asymptotic}, implies\tcr{:}
\bse
&&\phantom{~=~}\E_\X[\sg|\mbG_m\{\hg_n(\X,\bgamma)\}|] \\
&&~\leq~ J_{[\,]}\{\delta,\G_n\mid\ms_1,L_2(\P_\X)\}+[J_{[\,]}\{\delta,\G_n\mid\ms_1,L_2(\P_\X)\}]^2M(\ms_1)\delta^{-2}m^{-1/2} \\
&&~\leq~ \delta\{1+\log\,H(\ms_1)\}+c\,(\delta-\delta\,\log\,\delta)+\{1+\log\,H(\ms_1)+c\,(1-\log\,\delta)\}^2M(\ms_1)m^{-1/2}
\ese
for any $\delta\in(\Delta(\ms_1),\Delta(\ms_1)+c\,]$. Therefore\tcr{,}
\bse
\E_\X[\sg|\mbG_m\{\hg_n(\X,\bgamma)\}|] &~\leq~& \Delta(\ms_1)\{1+\log\,H(\ms_1)\}+c\,\{\Delta(\ms_1)-\Delta(\ms_1)\,\log\,\Delta(\ms_1)\}+\\
&&~~[1+\log\,H(\ms_1)+c\,\{1-\log\,\Delta(\ms_1)\}]^2M(\ms_1)m^{-1/2}.
\ese
Since the right hand side in the above is $O_p(r_{n,m})$, it gives that
\be
\E_\X[\sg|\mbG_m\{\hg_n(\X,\bgamma)\}|] ~=~O_p(r_{n,m}).
\label{ex}
\ee
Then, for any positive sequence $t_n\to\infty$, we have
\bse
&&\phantom{=}\P_{\ms_2}[\sg|\mbG_m\{\hg_n(\X,\bgamma)\}|>t_n r_{n,m}\mid\ms_1] \\
&&~\leq~ (t_n r_{n,m})^{-1}\E_\X[\sg|\mbG_m\{\hg_n(\X,\bgamma)\}|] ~=~o_p(1),
\ese
where the first step holds by Markov's inequality and the last step is due to (\ref{ex}). This, combined with Lemma 6.1 of \citet{chernozhukov2018double}, gives that
\bse
\P[\sg|\mbG_m\{\hg_n(\X,\bgamma)\}|>t_n r_{n,m}]~\to~ 0,
\ese
which completes the proof.

\subsection{Proof of Theorem \ref{thate}}
Denote $\Enk^*\{\hg(\Z)\}:=n_{\kK}^{-1}\sum_{i\in\I_k}\hg(\Z_i)$ for any random function $\hg(\cdot)$ $(k=1,\ldots,\kK)$. Write
\be
\muhatss-\mu_0~=~S_1+S_2+S_3+S_4+S_5,
\label{date}
\ee
where
\be
S_1&~:=~&\E_n[\{\pis(\X)\}^{-1}T\{Y- m^*(\X) \}]+\E_{n+N}\{ m^*(\X) \}-\mu_0, \label{s1}\\
S_2&~:=~&\E_n([\nu_{n,N}-\{\pis(\X)\}^{-1}T]\{\mhatn(\X)- m^*(\X) \})=\kK^{-1}\sk S_{2,k} \nonumber\\
&~:=~&\kK^{-1}\sk\Enk^*([\nu_{n,N}-\{\pis(\X)\}^{-1}T]\{\hat{m}_{n,k}(\X)- m^*(\X) \}), \nonumber\\
S_3&~:=~&(1-\nu_{n,N})\E_{N}\{\mhatn(\X)- m^*(\X) \}=\kK^{-1}\sk S_{3,k} \nonumber\\
&~:=~&\kK^{-1}\sk[(1-\nu_{n,N})\E_N\{\hat{m}_{n,k}(\X)- m^*(\X) \}], \nonumber\\
S_4&~:=~&\E_n[\hD(\X) T\{Y- m^*(\X) \}],\ S_5:=\E_n[\hD(\X) T\{ m^*(\X) -\mhatn(\X)\}].
\nonumber
\ee

We first handle $S_2$ and $S_3$. \tcr{To this end, w}e have\tcr{:}
\bse
&&\phantom{~=~}\E_\Z\{([\nu_{n,N}-\{\pis(\X)\}^{-1}T]\{\hat{m}_{n,k}(\X)- m^*(\X) \})^2\} \\
&&~\leq~ c\,\E_\X[\{\hat{m}_{n,k}(\X)- m^*(\X) \}^2] ~=~O_p(w_{n,2}^2),
\ese
where the first step uses the boundedness of $\{\pis(\X)\}^{-1}$ from Assumption \ref{api4} and the last step is due to (\ref{wn2}) of Assumption \ref{ahmu}. It now follows that
\bse
\var(S_{2,k}\mid\cl_k^-)~=~O_p(n^{-1}w_{n,2}^2),\ \var(S_{3,k}\mid\cl_k^-)~=~O_p(N^{-1}w_{n,2}^2).
\ese
Thus\tcr{,} Chebyshev's inequality gives that, for any positive sequence $t_n\to\infty$,
\bse
&&\P_{\cl_k}(|S_{2,k}-\E_\Z(S_{2,k})|\geq t_n n^{-1/2}w_{n,2}\mid\cl_k^-)~\leq~ n(t_nw_{n,2})^{-2}\var(S_{2,k}\mid\cl_k^-) ~=~o_p(1), \\
&&\P_{\cu}(|S_{3,k}-\E_\Z(S_{3,k})|\geq t_n n^{-1/2}w_{n,2}\mid\cl_k^-)~\leq~ n(t_nw_{n,2})^{-2}\var(S_{3,k}\mid\cl_k^-) ~=~o_p(1).
\ese
Then\tcr{,} Lemma 6.1 of \citet{chernozhukov2018double} implies
\bse
|S_{2,k}-\E_\Z(S_{2,k})|~=~O_p(n^{-1/2}w_{n,2}),\ |S_{3,k}-\E_\Z(S_{3,k})|~=~O_p(N^{-1/2}w_{n,2}),
\ese
which gives that
\be
|S_{2,k}+S_{3,k}-\E_\Z(S_{2,k}+S_{3,k})|~=~O_p(n^{-1/2}w_{n,2}).
\label{s23e}
\ee
In addition, we know that
\bse
|\E_\Z(S_{2,k}+S_{3,k})|&~=~&|\E_\Z([1-\{\pis(\X)\}^{-1}T]\{\hat{m}_{n,k}(\X)- m^*(\X) \})| \\
&~\leq~&c\,I\{\pis(\X)\neq\pi(\X)\}\E\{|\hat{m}_{n,k}(\X)- m^*(\X) |\} \\
&~=~&I\{\pis(\X)\neq\pi(\X)\}O_p(w_{n,1}),
\ese
where the second step uses the boundedness of $\{\pis(\X)\}^{-1}$ from Assumption \ref{api4} as well as the fact that
\bse
\E_\Z([1-\{\pi(\X)\}^{-1}T]\{\hat{m}_{n,k}(\X)- m^*(\X) \})~=~0,
\ese
and the last step holds by (\ref{wn1}) of Assumption \ref{ahmu}. This, combined with (\ref{s23e}), gives
\bse
|S_{2,k}+S_{3,k}|~=~O_p(n^{-1/2}w_{n,2})+I\{\pis(\X)\neq\pi(\X)\}O_p(w_{n,1}),
\ese
which implies\tcr{:}
\be
|S_2+S_3|&~\leq~&\kK^{-1}\sk|S_{2,k}+S_{3,k}| \nonumber\\
&~=~&O_p(n^{-1/2}w_{n,2})+I\{\pis(\X)\neq\pi(\X)\}O_p(w_{n,1}).
\label{s23}
\ee

Next, we control $S_4$. We know that
\bse
\E_\Z([\hD(\X)T\{Y- m^*(\X) \}]^2)~\leq~\E_\Z([\hD(\X)\{Y- m^*(\X) \}]^2)~=~O_p(b_N^2),
\ese
where the last step holds by (\ref{sn4}) of Assumption  \ref{api4}. This implies\tcr{:}
\bse
\var(S_{4}\mid\cu)~=~O_p(n^{-1}b_N^2).
\ese
Thus Chebyshev's inequality gives that, for any positive sequence $t_n\to\infty$,
\bse
\P_\cl(|S_{4}-\E_\Z(S_4)|\geq t_n n^{-1/2}b_N\mid\cu)~\leq~ n(t_nb_N)^{-2}\var(S_{4}\mid\cu) ~=~o_p(1).
\ese
Then, by Lemma 6.1 of \citet{chernozhukov2018double}, we have
\be
|S_{4}-\E_\Z(S_4)|~=~O_p(n^{-1/2}b_N).
\label{s41}
\ee
In addition, if $ m^*(\X) = m(\X) $, then
\bse
\E_\Z(S_4)~=~\E(\E[\hD(\X) T\{Y- m(\X) \}\mid\cu,\X]\mid \cu)~=~0.
\ese
Otherwise, we have
\bse
|\E_\Z(S_{4})|~\leq~(\E_\X[\{\hD(\X)\}^2]\E[\{Y- m^*(\X) \}^2])^{1/2} ~=~O_p(s_N),
\ese
where the first step uses H\"older's inequality and the last step is due to (\ref{sn2}) of Assumption \ref{api4}. Therefore $|\E_\Z(S_4)|=I\{ m^*(\X) \neq m(\X) \}O_p(s_N)$. This, combined with (\ref{s41}), implies\tcr{:}
\be
|S_4|~=~O_p(n^{-1/2}b_N)+I\{ m(\X) \neq m^*(\X) \}O_p(s_N).
\label{s4}
\ee

Now, we consider $S_5$. Markov's inequality gives that, for any positive sequence $t_n\to\infty$,
\be
&&\phantom{~=~}\P_\cl(\Enk^* [\{\hD(\X)\}^2]\geq t_ns_N^2\mid\cu)~\leq~ t_n^{-1}s_N^{-2}\E_\X [\{\hD(\X)\}^2]~=~o_p(1), \label{pdn}\\
&&\phantom{~=~}\P_{\cl_k}(\Enk^*[\{ m^*(\X) -\hat{m}_{n,k}(\X)\}^2]\geq t_nw_{n,2}^2\mid\cl_k^-) \nonumber\\
&&~\leq~ t_n^{-1}w_{n,2}^{-2}\E_\X[\{ m^*(\X) -\hat{m}_{n,k}(\X)\}^2]=o_p(1)\quad (k=1,\ldots,\kK),
\label{pmun}
\ee
where (\ref{pdn}) uses (\ref{sn2}) of Assumption \ref{api4} and (\ref{pmun}) holds by (\ref{wn2}) of Assumption \ref{ahmu}. Then, by Lemma 6.1 of \citet{chernozhukov2018double}, we have
\be
&&\Enk^* [\{\hD(\X)\}^2]~=~O_p(s_N^2), \label{sn}\\
&&\Enk^* [\{ m^*(\X) -\hat{m}_{n,k}(\X)\}^2]~=~O_p(w_{n,2}^2)\quad (k=1,\ldots,\kK).
\label{en}
\ee
Hence\tcr{,} H\"older's inequality implies\tcr{:}
\be
|S_5|&~\leq~&\kK^{-1}\sk\Enk^*[|\hD(\X)\{ m^*(\X) -\hat{m}_{n,k}(\X)\}|]  \nonumber\\
&~\leq~&\kK^{-1}\sk(\Enk^*[\{\hD(\X)\}^2]\Enk^*[\{ m^*(\X) -\hat{m}_{n,k}(\X)\}^2])^{1/2} =O_p(s_N\,w_{n,2}),
\label{s5}
\ee
where the last step holds by (\ref{sn}) and(\ref{en}).

Summing up, the equations (\ref{date}), (\ref{s1}), (\ref{s23}), (\ref{s4}) and (\ref{s5}) conclude the result.

\subsection{Proof of Corollary \ref{corate}}
Since $\nu=0$, we have
\bse
\E_{n+N}\{ m^*(\X) \}~=~\E\{ m^*(\X) \}+O_p\{(n+N)^{-1/2}\}~=~\E\{ m^*(\X) \}+ o_p(n^{-1/2}).
\ese
by the central limit theorem. Then the stochastic expansion directly follows from Theorem \ref{thate} and the asymptotic normality is obvious.

\subsection{Proof of Corollary \ref{coratesup}}
With $\E_{n+N}\{\mhatn(\X)\}$ substituted by $\E_n\{\mhatn(\X)\}$, the proof of Theorem \ref{thate} directly gives the stochastic expansion followed by the asymptotic normality. Then\tcr{,} we have
\bse
&&\phantom{~=~}\cov[\{\pi(\X)\}^{-1}T\{Y- m^*(\X) \}, m^*(\X) ] \\
&&~=~\E\{ m^*(\X) Y\}-\E[\{ m^*(\X) \}^2]-\E\{Y- m^*(\X) \}\E\{ m^*(\X) \} \\
&&~=~\E\{ m^*(\X) Y\}-\var\{ m^*(\X) \}.
\ese
Therefore\tcr{,}
\bse
&&\lams^2~=~\var[\{\pi(\X)\}^{-1}T\{Y- m^*(\X) \}]+\var\{ m^*(\X) \}+ \\
&&\phantom{\lams^2=~~}2\,\cov[\{\pi(\X)\}^{-1}T\{Y- m^*(\X) \}, m^*(\X) ] \\
&&\phantom{\lams^2}~=~\var[\{\pi(\X)\}^{-1}T\{Y- m^*(\X) \}]-\var\{ m^*(\X) \}+2\,\E\{ m^*(\X) (Y-\mu_0)\}.
\ese

\subsection{Proof of Corollary \ref{corate_dagger}}
The stochastic expansion can be obtained from the proof of Theorem \ref{thate} with $\pihatN(\cdot)$ replaced by $\pihatn(\cdot)$. The asymptotic normality directly follows.

\subsection{Proof of Theorem \ref{thqte}}\label{proof_theorem_qte} Write
\be
\thetahatss-\vt~=~\{T_1(\thetahatinit)-\vt\}+\{\hf(\thetahatinit)\}^{-1}\{T_2(\thetahatinit)+T_3(\thetahatinit)+T_4(\thetahatinit)\},
\label{dde}
\ee
where
\bse
T_1(\theta)&~:=~&\theta+\{\hf(\theta)\}^{-1}(\E_n[\{\pis(\X)\}^{-1}T\{\phis(\X,\theta)-\psi(Y,\theta)\}]-\E_{n+N}\{\phis(\X,\theta)\}), \\ T_2(\theta)&~:=~&\E_n([\{\pis(\X)\}^{-1}T-\nu_{n,N}]\{\phihatn(\X,\theta)-\phis(\X,\theta)\})-\\
&&~~(1-\nu_{n,N})\E_{N}\{\phihatn(\X,\theta)-\phis(\X,\theta)\}, \\
T_3(\theta)&~:=~&\E_n[\hD(\X) T\{\phis(\X,\theta)-\psi(Y,\theta)\}],\\
T_4(\theta)&~:=~&\E_n[\hD(\X) T\{\phihatn(\X,\theta)-\phis(\X,\theta)\}].
\ese

First, the conditions (\ref{hvti}) and (\ref{hf}) of Assumption \ref{ainit} give
\be
&&\P\{\thetahatinit\in\mbtv\}~\to~ 1, \label{belong}\\
&&\hL~:=~\{\hf(\thetahatinit)\}^{-1}-\{f(\vt)\}^{-1}~=~O_p(v_n)~=~o_p(1).
\label{hl}
\ee
Also, we have
\be
\hf(\thetahatinit)~=~O_p(1)\tcr{,}
\label{hfo}
\ee
due to (\ref{hf}) of Assumption \ref{ainit} and the fact that $f(\vt)>0$ from Assumption \ref{adensity}.

Now\tcr{,} we consider $T_1(\thetahatinit)$. According to (\ref{hvti}) of Assumption \ref{ainit} and (\ref{unipi1}) of Assumption \ref{abound}, we have
\bse
n^{-1/2}\mbG_n[\{\pis(\X)\}^{-1}T\phis(\X,\thetahatinit)]~=~n^{-1/2}\mbG_n[\{\pis(\X)\}^{-1}T\phis(\X,\vt)]+o_p(n^{-1/2}),
\ese
which implies that
\be
&&\phantom{~=~}\E_n[\{\pis(\X)\}^{-1}T\phis(\X,\thetahatinit)]\nonumber\\
&&~=~\E_\Z[\{\pis(\X)\}^{-1}T\phis(\X,\thetahatinit)] +\E_n[\{\pis(\X)\}^{-1}T\phis(\X,\vt)]- \nonumber\\
&&\phantom{~=~}\E_\Z[\{\pis(\X)\}^{-1}T\phis(\X,\vt)]+o_p(n^{-1/2}).
\label{t11}
\ee
Considering that $\{\psi(Y,\theta):\theta\in\mbtv\}$ is a $\P$-Donsker class from Theorem 19.3 of \citet{van2000asymptotic} and the permanence properties of $\P$-Donsker classes \citet{van1996weak}, Theorem 2.10.6 of \citet{van1996weak} gives that $\md^*=\{\{\pis(\X)\}^{-1}T\psi(Y,\theta):\theta\in\mbtv\}$ is $\P$-Donsker since $\{\pis(\X)\}^{-1}T$ and $\psi(Y,\theta)$ are bounded. Moreover, the convergence (\ref{belong}) implies that $\{\pis(\X)\}^{-1}T\psi(Y,\thetahatinit)$ is in $\md^*$ with probability tending to one. In addition, we have
\bse
&&\phantom{~=~}\E_\Z[\{\pis(\X)\}^{-2}T\{\psi(Y,\thetahatinit)-\psi(Y,\vt)\}^2] \\
&&~\leq~ c\, \E_\bfZ[\{I(Y<\thetahatinit)-I(Y<\vt)\}^2] =c\,F(\thetahatinit)+F(\vt)-2F\{\min(\thetahatinit,\vt)\}\to 0
\ese
in probability\tcr{,} because of the boundedness of $\{\pis(\X)\}^{-2}T$, the continuity of $F(\cdot)$ from Assumption \ref{adensity} and the consistency of $\thetahatinit$ from Assumption \ref{ainit}. Hence Lemma 19.24 of \citet{van2000asymptotic}
gives that
\bse
\mbG_n[\{\pis(\X)\}^{-1}T\{\psi(Y,\thetahatinit)-\psi(Y,\vt)\}]~=~o_p(1),
\ese
which implies\tcr{:}
\be
\E_n[\{\pis(\X)\}^{-1}T\psi(Y,\thetahatinit)]&~=~&\E_\Z[\{\pis(\X)\}^{-1}T\psi(Y,\thetahatinit)] +\E_n[\{\pis(\X)\}^{-1}T\psi(Y,\vt)]- \nonumber\\
&&\E_\Z[\{\pis(\X)\}^{-1}T\psi(Y,\vt)]+o_p(n^{-1/2}).\label{t12}
\ee
Further, the condition (\ref{unipi2}) gives
\be
\E_{n+N}\{\phis(\X,\thetahatinit)\}&~=~&\E_\X\{\phis(\X,\thetahatinit)\}+\E_{n+N}\{\phis(\X,\vt)\}- \nonumber\\
&&\E_\X\{\phis(\X,\vt)\}+o_p(n^{-1/2}).
\label{t13}
\ee
Since either $\phis(\cdot,\cdot)=\phi(\cdot,\cdot)$ or $\pis(\cdot)=\pi(\cdot)$, we know that
\be
\E_\Z[\{\pis(\X)\}^{-1}T\{\phis(\X,\vt)-\psi(Y,\vt)\}]-\E_\X\{\phis(\X,\vt)\}~=~0,
\label{t14}
\ee
and that
\be
&&\phantom{~=~}\E_\Z[\{\pis(\X)\}^{-1}T\{\phis(\X,\thetahatinit)-\psi(Y,\thetahatinit)\}]-\E_\X\{\phis(\X,\thetahatinit)\} \nonumber\\ &&~=~ -\E_\Z\{\psi(Y,\thetahatinit)\}.
\label{t15}
\ee
In addition, Taylor's expansion gives that
\be
\E_\bfZ\{\psi(Y,\thetahatinit)\}&~=~&f(\vt)(\thetahatinit-\vt)+O_p(|\thetahatinit-\vt|^2)\nonumber\\
&~=~&f(\vt)(\thetahatinit-\vt)+O_p(u_n^2) \label{df12} \\
&~=~&O_p(u_n),\label{df122}
\ee
where the residual term in the first step is due to (\ref{belong}) and the fact that $f(\cdot)$ has a bounded derivative in $\mbtv$ from Assumption \ref{adensity}, the second step uses (\ref{hvti}) in Assumption \ref{ainit} and the last step holds by the fact that $u_n=o(1)$ from Assumption \ref{ainit}. Therefore\tcr{,}
\be
\E_n\{\omega_{n,N}(\Z,\thetahatinit)\}&~=~&\E_n\{\omega_{n,N}(\Z,\vt)\}- \E_\Z\{\psi(Y,\thetahatinit)\}+o_p(n^{-1/2}) \nonumber\\
&~=~&\E_n\{\omega_{n,N}(\Z,\vt)\}-f(\vt)(\thetahatinit-\vt)+O_p(u_n^2)+o_p(n^{-1/2}) \label{taylor}\\
&~=~&\E_n\{\omega_{n,N}(\Z,\vt)\}+O_p(u_n)+o_p(n^{-1/2}), \nonumber
\ee
where the first step uses (\ref{t11})--(\ref{t15}), the second step is due to (\ref{df12}) and the last step holds by (\ref{df122}). It now follows that
\be
\hL\E_n\{\omega_{n,N}(\Z,\thetahatinit)\}~=~O_p(u_nv_n)+o_p(n^{-1/2})\tcr{,}\label{diffl}
\ee
from (\ref{hl}) and the fact that $\E_n\{\omega_{n,N}(\Z,\vt)\}=O_p(n^{-1/2})$ from the central limit theorem. Hence\tcr{,} we have
\be
&&T_1(\thetahatinit)-\vt~=~\thetahatinit-\vt+\{\hf(\thetahatinit)\}^{-1}\E_n\{\omega_{n,N}(\Z,\thetahatinit)\}\nonumber\\
&&\phantom{T_1(\thetahatinit)-\vt}~=~\thetahatinit-\vt+\{f(\vt)\}^{-1}\E_n\{\omega_{n,N}(\Z,\thetahatinit)\}+O_p(u_nv_n)+o_p(n^{-1/2}) \nonumber\\
&&\phantom{T_1(\thetahatinit)-\vt}~=~\thetahatinit-\vt+\{f(\vt)\}^{-1}[\E_n\{\omega_{n,N}(\Z,\vt)\}-f(\vt)(\thetahatinit-\vt)]+ \nonumber\\
&&\phantom{T_1(\thetahatinit)-\vt=}O_p(u_n^2+u_nv_n)+o_p(n^{-1/2}) \nonumber\\
&&\phantom{T_1(\thetahatinit)-\vt}~=~\{f(\vt)\}^{-1}\E_n\{\omega_{n,N}(\Z,\vt)\}+O_p(u_nv_n+u_n^2)+o_p(n^{-1/2}),
\label{t1}
\ee
where the second step uses (\ref{diffl}) and the third step is due to (\ref{taylor}).

Next, we control $T_2(\thetahatinit)$. Denote
\bse
\mp_{n,k}^*~:=~\{[\{\pis(\X)\}^{-1}T-\nu_{n,N}]\hpsi(\X,\theta):\theta\in\mbtv\}.
\ese
Due to the boundedness of $[\{\pis(\X)\}^{-1}T-\nu_{n,N}]$ from Assumption \ref{api}, we have
\be
&&\phantom{=}N_{[\,]} \{c_1\,\eta,\mp_{n,k}^*\mid\cl,L_2(\P_\X)\}~\leq~ N_{[\,]} \{\eta,\mp_{n,k}\mid\cl,L_2(\P_\X)\}~\leq~ H(\cl)\eta^{-c}, \label{bracket}\\
&&\phantom{~=~}\sbx|[\{\pis(\X)\}^{-1}T-\nu_{n,N}]\hpsi(\X,\theta)| \nonumber\\
&&~\leq~ c\,\sbx|\hpsi(\X,\theta)| =O_p(d_{n,\infty}), \label{bn}\\
&&\phantom{~~}[\sb\E_\Z\{([\{\pis(\X)\}^{-1}T-\nu_{n,N}]\hpsi(\X,\theta))\}^2]^{1/2} \nonumber\\
&&~\leq~ c\,\Delta_{k}(\cl)=O_p(d_{n,2}) \quad (k=1,\ldots,\kK)\tcr{,} \label{dn}
\ee
from Assumption \ref{aest}. Then\tcr{,} (\ref{bracket}) implies\tcr{:}
\be
N_{[\,]} \{\eta,\mp_{n,k}^*\mid\cl,L_2(\P_\X)\}~\leq~ c_1^{c_2} H(\cl)\eta^{-c_2}. \label{an}
\ee
Since $c_1^{c_2} H(\cl)=O_p(a_n)$ from Assumption \ref{aest}, combining (\ref{bn})--(\ref{an})  and applying Lemma \ref{1v2} yield that
\be
\sb|\mbG_{n_\kK,k}([\{\pis(\X)\}^{-1}T-\nu_{n,N}]\hpsi(\X,\theta))|~=~ O_p(r_n)\tcr{,}
\label{mbgpi}
\ee
with the notation
\bse
\mbG_{n_\kK,k}\{\ghat(\Z)\}~:=~n_\kK^{1/2}[n_\kK^{-1}\hbox{$\sum_{i\in\I_k}$}\ghat(\Z_i)-\E_\X\{\ghat(\Z)\}]\quad (k=1,\ldots,\kK)\tcr{,}
\ese
for any random function $\ghat(\cdot)$. In addition, we have
\be
&&\phantom{=}\sb|\E_\Z([\{\pis(\X)\}^{-1}T-1]\hpsi(\X,\theta))| \nonumber\\
&&~\leq~ c\, I\{\pis(\X)\neq\pi(\X)\} \sb\E_\Z\{|\hpsi(\X,\theta)|\}\nonumber\\
&&~=~I\{\pis(\X)\neq\pi(\X)\}O_p(d_{n,1}),
\label{idn}
\ee
where the first step holds by the boundedness of $\{\pis(\X)\}^{-1}$ from Assumption \ref{api} and the fact that
\bse
\E_\Z([\{\pi(\X)\}^{-1}T-1]\hpsi(\X,\theta))~=~0,
\ese
and the last step is due to Assumption \ref{aest}. Moreover, under Assumption \ref{aest}, Lemma \ref{1v2} implies that
\be
&&\phantom{~=~}\sb|\mbG_N\{\hpsi(\X,\theta)\}| \nonumber\\
&&~=~O_p[d_{n,2}\{\log\,a_n+\log\,(d_{n,2}^{-1})\}+N^{-1/2}d_{n,\infty}\{(\log\,a_n)^2+(\log\,d_{n,2})^2\}] \nonumber\\
&&~=~O_p(r_n)\quad (k=1,\ldots,\kK). \label{rr2}
\ee
Considering (\ref{mbgpi})--(\ref{rr2}), we know that
\bse
&&T_2(\thetahatinit)~=~\kK^{-1}\sk \{n_\kK^{-1/2}\mbG_{n_\kK,k}([\{\pis(\X)\}^{-1}T-\nu_{n,N}]\hpsi(\X,\thetahatinit))- \\
&&\phantom{T_2(\thetahatinit)=\kK^{-1}\sk\{}N^{-1/2}(1-\nu_{n,N})\mbG_N\{\hpsi(\X,\thetahatinit)\}+ \\
&&\phantom{T_2(\thetahatinit)=\kK^{-1}\sk\{}\E_\Z([\{\pis(\X)\}^{-1}T-1]\hpsi(\X,\thetahatinit))\} \\
&&\phantom{T_2(\thetahatinit)}~=~O_p(n^{-1/2}r_n)+I\{\pis(\X)\neq\pi(\X)\}O_p(d_{n,1}),
\ese
which, combined with (\ref{hfo}), implies that
\be
\{\hf(\thetahatinit)\}^{-1}T_2(\thetahatinit)~=~O_p(n^{-1/2}r_n)+I\{\pis(\X)\neq\pi(\X)\}O_p(d_{n,2}).
\label{t2}
\ee

Further, we \tcr{now} handle $T_3(\thetahatinit)$. Let $\mh:= \{\hD(\X)T\phis(\X,\theta):\theta\in\mbtv\}$ and recall $\mm= \{\phis(\X,\theta):\theta\in\mbtv\}$. We have
\be
&&\phantom{=}N_{[\,]} \{\sx|\hD(\x)|\eta,\mh\mid\cu,L_2(\P_\X)\} ~\leq~ N_{[\,]} \{\eta,\mm,L_2(\P_\X)\}\leq c_1\,\eta^{-c_2}, \label{bracketh}\\
&&\phantom{=}\sbx|\hD(\X)T\phis(\X,\theta)|~=~O_p(1), \label{bnh}\\
&&\phantom{=}(\sb\E_\Z[\{\hD(\X)T\phis(Y,\theta)\}^2])^{1/2}~=~O_p(s_N), \label{dnh}
\ee
where (\ref{bracketh}) uses (\ref{bmm}) of Assumption \ref{abound}, (\ref{bnh}) holds by (\ref{dsup}) of Assumption \ref{api} and the boundedness of $\phis(\X,\theta)$ from Assumption \ref{abound}, and (\ref{dnh}) is due to (\ref{d2}) of Assumption \ref{api} and the boundedness of $\phis(\X,\theta)$ from Assumption \ref{abound}. Then\tcr{,} (\ref{bracketh}) gives
\be
N_{[\,]} \{\eta,\mh\mid\cu,L_2(\P_\X)\} ~\leq~ c_1\,\{\sx|\hD(\x)|\}^{c_2}\eta^{-c_2}.
\label{anh}
\ee
Since $c_1\,\{\sx|\hD(\x)|\}^{c_2}=O_p(1)$ from Assumption \ref{api}, combining (\ref{bnh})--(\ref{anh})  and applying Lemma \ref{1v2} yield that
\bse
\sb|\mbG_{n}\{\hD(\X)T\phis(Y,\theta)\}|~=~O_p(z_{n,N}),
\ese
which gives that
\be
|\E_n\{\hD(\X)T\phis(Y,\thetahatinit)\}-\E_\Z\{\hD(\X)T\phis(Y,\thetahatinit)\}|~=~O_p(n^{-1/2}z_{n,N}).
\label{t311}
\ee
Analogously, by Example19.6 of \citet{van2000asymptotic} and the boundedness of $\psi(Y,\theta)$, we know that
\be
|\E_n\{\hD(\X)T\psi(Y,\thetahatinit)\}-\E_\Z\{\hD(\X)T\psi(Y,\thetahatinit)\}|~=~O_p(n^{-1/2}z_{n,N}).
\label{t312}
\ee
Combining (\ref{t311}) and (\ref{t312}) yields\tcr{:}
\be
|T_3(\thetahatinit)-\E_\Z\{T_3(\thetahatinit)\}|~=~O_p(n^{-1/2}z_{n,N}).
\label{et3}
\ee
In addition, if $\phis(\X,\theta)=\phi(\X,\theta)$, then
\bse
\E_\Z\{T_3(\thetahatinit)\}~=~\E_\Z(\E_\Z[\hD(\X)T\{\phis(\X,\thetahatinit)-\psi(Y,\thetahatinit)\}\mid\X])~=~0.
\ese
Otherwise, we have
\bse
|\E_\Z\{T_{3}(\thetahatinit)\}|~\leq~(\E_\X[\{\hD(\X)\}^2]\E[\{\phis(\X,\thetahatinit)-\psi(Y,\thetahatinit)\}^2])^{1/2} ~=~O_p(s_N),
\ese
where the last step uses the boundedness of $\phis(\X,\theta)$ from Assumption \ref{abound}. Hence\tcr{,}
\bse
|\E_\Z\{T_{3}(\thetahatinit)\}|~=~I\{\phis(\X,\theta)\neq\phi(\X,\theta)\}O_p(s_N).
\ese
This, combined with (\ref{hfo}) and (\ref{et3}), implies\tcr{:}
\be
\{\hf(\thetahatinit)\}^{-1}T_3(\thetahatinit)~=~O_p(n^{-1/2}z_{n,N})+I\{\phis(\X,\theta)\neq\phi(\X,\theta)\}O_p(s_N).
\label{t3}
\ee

Eventually, we deal with $T_4(\thetahatinit)$. Denote
\bse
\mq~:=~\{\hD (\X)T\hpsi(\X,\theta):\theta\in\mbtv\}.
\ese
Due to (\ref{dsup})  of Assumption \ref{api}, we have
\be
&&\phantom{~=~}N_{[\,]} \{\sx|\hD(\x)|\eta,\mq\mid\cl\cup\cu,L_2(\P_\X)\} \nonumber\\
&&~\leq~ N_{[\,]} \{\eta,\mp_{n,k}\mid\cl,L_2(\P_\X)\}\leq H(\cl)\eta^{-c}, \label{bracket1}\\
&&\phantom{~=~}\sbx|\hD(\X)\hpsi(\X,\theta)| \nonumber\\
&&~\leq~ \sx|\hD(\x)|\sbx|\hpsi(\X,\theta)| =O_p(d_{n,\infty}), \label{bn1}\\
&&\phantom{~=~}(\sb\E_\X[\{\hD(\X)\hpsi(\X,\theta)\}^2])^{1/2} \nonumber\\
&&~\leq~ \sx|\hD(\x)|\Delta_{k}(\cl)=O_p(d_{n,2}) \quad (k=1,\ldots,\kK)\tcr{,} \label{dn1}
\ee
from Assumption \ref{aest}. Then\tcr{,} (\ref{bracket1}) implies\tcr{:}
\be
N_{[\,]} \{\eta,\mq\mid\cl\cup\cu,L_2(\P_\X)\}~\leq~ \{\sx|\hD(\x)|\}^c H(\cl)\eta^{-c}. \label{an1}
\ee
Since $\{\sx|\hD(\x)|\}^c H(\cl)=O_p(a_n)$
from Assumptions \ref{aest} and \ref{api}, combining (\ref{bn1})--(\ref{an1})  and applying Lemma \ref{1v2} yield that
\be
\sb|\mbG_{n_\kK,k}\{\hD(\X)\hpsi(\X,\theta)\}|~=~O_p(r_n).
\label{mbgpi1}
\ee
In addition, we have
\be
&&\phantom{~=~}\sb|\E_\X\{\hD(\X)\hpsi(\X,\theta)\}|\nonumber\\
&&~\leq~ (\E_\X[\{\hD(\X)\}^2]\sb\E_\X[\{\hpsi(\X,\theta)\}^2])^{1/2}=O_p(s_N d_{n,2}),
\label{idn1}
\ee
where the first step holds by H\"older's inequality and the last step is due to Assumptions \ref{api} and \ref{aest}. Considering (\ref{mbgpi1}) and (\ref{idn1}), we know that
\bse
T_4(\thetahatinit)&~=~&\kK^{-1}\sk [n_\kK^{-1/2}\mbG_{n_\kK,k}\{\hD(\X)\hpsi(\X,\thetahatinit)\}+\E_\X\{\hD(\X)\hpsi(\X,\thetahatinit)\}] \\
&~=~&O_p(n^{-1/2}r_n+s_N d_{n,2}),
\ese
which, combined with (\ref{hfo}), implies that
\be
\{\hf(\thetahatinit)\}^{-1}T_4(\thetahatinit)~=~O_p(n^{-1/2}r_n+s_N d_{n,2}). \label{t4}
\ee

Summing up,  the equations (\ref{t1}), (\ref{t2}), (\ref{t3}) and (\ref{t4}) conclude the result.

\subsection{Proof of Corollary \ref{corqte}}
Since $\nu=0$, we have
\bse
\E_{n+N}\{\phis(\X,\vt)\}=\E\{\phis(\X,\vt)\}+O_p\{(n+N)^{-1/2}\}~=~\E\{\phis(\X,\vt)\}+ o_p(n^{-1/2})\tcr{,}
\ese
by the central limit theorem. Then\tcr{,} the stochastic expansion directly follows from Theorem \ref{thqte} and the asymptotic normality is obvious.

\subsection{Proof of Corollary \ref{corsup}}
With $\E_{n+N}\{\phihatn(\X,\thetahatinit)\}$ substituted by $\E_n\{\phihatn(\X,\thetahatinit)\}$, the proof of Theorem \ref{thqte} directly gives the stochastic expansion followed by the asymptotic normality. Then\tcr{,} we have
\bse
&&\phantom{~=~}\cov[\{\pi(\X)\}^{-1}T\{\phis(\X,\vt)-\psi(Y,\vt)\},\phis(\X,\vt)] \\
&&~=~\E[\{\phis(\X,\vt)\}^2]-\E\{\phis(\X,\vt)\psi(Y,\vt)\}-\E\{\phis(\X,\vt)-\psi(Y,\vt)\}\E\{\phis(\X,\vt)\} \\
&&~=~\var\{\phis(\X,\vt)\}-\E\{\phis(\X,\vt)\psi(Y,\vt)\}.
\ese
Therefore\tcr{,}
\bse
&&\sigsup^2~=~\var[\{\pi(\X)\}^{-1}T\{\psi(Y,\vt)-\phis(\X,\vt)\}]+\var\{\phis(\X,\vt)\}- \\
&&\phantom{\sigsup^2~=~}2\,\cov[\{\pi(\X)\}^{-1}T\{\phis(\X,\vt)-\psi(Y,\vt)\},\phis(\X,\vt)] \\
&&\phantom{\sigsup^2}~=~\var[\{\pi(\X)\}^{-1}T\{\psi(Y,\vt)-\phis(\X,\vt)\}]-\var\{\phis(\X,\vt)\}+2\,\E\{\phis(\X,\vt)\psi(Y,\vt)\}.
\ese

\subsection{Proof of Theorem \ref{theorem_ks_ate}}
Denote $\ell^{(t)}(\x,\mbP)=\kappa_t(\mbP\trans\x)f_\S(\mbP\trans\x)$  $(t=0,1)$. We now derive the convergence rate of $\hlo(\x,\hmbP)-\ell^{(1)}(\x,\mbP)$. The case of $\hlz(\x,\hmbP)-\ell^{(0)}(\x,\mbP)$ is similar.

We first deal with the error from estimating $\mbP_0$ by $\hmbP$, i.e., $\hlo(\x,\hmbP)-\hlo(\x,\mbP_0)$. Taylor's expansion gives that, for
\be
\bar{\s}_n~:=~h_n^{-1}\{\mbP_0\trans+\bmu(\hmbP-\mbP_0)\trans\}(\x-\X)\tcr{,}
\label{ysbar}
\ee
with some $\bmu:=\diag(\mu_1,\ldots,\mu_r)$ and $\mu_j\in(0,1)$ $(j=1,\ldots,r)$,
\be
&&\phantom{~=~}\hlo(\x,\hmbP)-\hlo(\x,\mbP_0) \nonumber \\
&&~=~h_n^{-(r+1)}\Enk[\{\nabla K(\bar{\s})\}\trans(\hmbP-\mbP_0)\trans(\x-\X)\{\pihatN(\X)\}^{-1}TY] \nonumber\\
&&~=~U_n(\x)+V_{n,N}(\x) ,
\label{ydhbe}
\ee
where
\bse
&&U_n(\x)~:=~h_n^{-(r+1)}\Enk[\{\nabla K(\bar{\s})\}\trans(\hmbP-\mbP_0)\trans(\x-\X)\{\pis(\X)\}^{-1}TY] ,
\nonumber \\
&&V_{n,N}(\x)~:=~h_n^{-(r+1)}\Enk[\{\nabla K(\bar{\s})\}\trans(\hmbP-\mbP_0)\trans(\x-\X)\hD(\X)TY].
\ese
To control $U_n(\x)$, write
\be
U_n(\x)&~=~& h_n^{-(r+1)}\trace ((\hmbP-\mbP_0)\trans \Enk[(\x-\X)\{\nabla K(\bar{\s})\}\trans\{\pis(\X)\}^{-1}TY]) \nonumber\\
&~=~&h_n^{-(r+1)}\trace[(\hmbP-\mbP_0)\trans\{\bfU_{n,1}(\x)+\bfU_{n,2}(\x)-\bfU_{n,3}(\x)\}],
\label{yun}
\ee
where
\bse
&&\bfU_{n,1}(\x)~:=~\Enk((\x-\X)[\nabla K(\bar{\s}_n)-\nabla K\{h_n^{-1}\mbP_0\trans(\x-\X)\}]\trans\{\pis(\X)\}^{-1}TY),  \\
&&\bfU_{n,2}(\x)~:=~\Enk(\x [\nabla K\{h_n^{-1}\mbP_0\trans(\x-\X)\}]\trans\{\pis(\X)\}^{-1}TY),  \\
&&\bfU_{n,3}(\x)~:=~\Enk(\X [\nabla K\{h_n^{-1}\mbP_0\trans(\x-\X)\}]\trans\{\pis(\X)\}^{-1}TY).
\ese
We know
\be
\ss\E[h_n^{-r}\rho \{h_n^{-1}(\s-\S) \}|Y|]&~=~&\ss\hbox{$\int$}h_n^{-r}\rho\{h_n^{-1}(\s-\bfv) \}\E(|Y|\mid\S=\bfv) f_\S(\bfv)d\bfv \nonumber\\
&~=~&\ss\hbox{$\int$}\rho(\bft )\E(|Y|\mid\S=\s-h_n\bft) f_\S(\s-h_n\bft)d\bft  \nonumber\\
&~=~& O(1).
\label{ygrho}
\ee
where the second step uses change of variables while the last step holds by the boundedness of $\E(|Y|\mid\S=\cdot)f_\S(\cdot)$ from Assumptions \ref{akernel} (ii)--(iii) and the integrability of $\rho(\cdot)$ from Assumption \ref{ahbey} (ii). Moreover, under Assumptions \ref{akernel} (ii)--(iii) and \ref{ahbey} (ii), Theorem 2 of \citet{hansen2008uniform} gives\tcr{:}
\bse
\ss(\Enk[h_n^{-r}\rho \{h_n^{-1}(\s-\S) \}Y]-\E[h_n^{-r}\rho \{h_n^{-1}(\s-\S) \}Y])~=~O_p(\xi_n)~=~o_p(1)\tcr{.}
\ese
This, combined with (\ref{ygrho}), implies\tcr{:}
\be
\ss\Enk[h_n^{-r}\rho \{h_n^{-1}(\s-\S) \}Y]~=~O_p(1).
\label{yexrho}
\ee
Next, we have
\be
&&\phantom{~=~}\sx\Enk (\|[\nabla K(\bar{\s}_n)-\nabla K\{h_n^{-1}\mbP_0\trans(\x-\X)\}]Y\|) \nonumber\\
&&~\leq~ \sx\Enk [\|\bar{\s}_n-h_n^{-1}\mbP_0\trans(\x-\X)\|\rho\{h_n^{-1}\mbP_0\trans(\x-\X)\}|Y|] \nonumber\\
&&~\leq~\sx\Enk [\|(\hmbP-\mbP_0)\trans(\x-\X)\|h_n^{-1}\rho\{h_n^{-1}\mbP_0\trans(\x-\X)\}|Y|] \nonumber\\
&&~\leq~ c\,\|\hmbP-\mbP_0\|_1\sxx\|\x-\X\|_{\infty}\ss\Enk [h_n^{-1}\rho\{h_n^{-1}(\s-\S)\}|Y|]\nonumber \\
&&~=~O_p(h_n^{r-1}\alpha_n),
\label{yalphan}
\ee
where the first step uses the local Lipschitz continuity of $\nabla K(\cdot)$ from Assumption \ref{ahbey} (ii), the second step is due to the definition (\ref{ysbar}) of $\bar{\s}_n$, the third step holds by H\"older's inequality, and the last step is because of Assumptions \ref{al1}, \ref{ahbe} (i) and the equation (\ref{yexrho}). Hence\tcr{,}
\bse
&&\phantom{~=~}\sx\|\bfU_{n,1}(\x)\|_{\infty} \\
&&~\leq~ c\,\sx\Enk (\|\x-\X\|_{\infty}\|[\nabla K(\bar{\s}_n)-\nabla K\{h_n^{-1}\mbP_0\trans(\x-\X)\}]Y\|) \\
&&~\leq~ c\,\sx\Enk (\|[\nabla K(\bar{\s}_n)-\nabla K\{h_n^{-1}\mbP_0\trans(\x-\X)\}]Y\|) =O_p(h_n^{r-1}\alpha_n).
\ese
where the first step holds by the boundedness of $\{\pis(\X)\}^{-1}T$, the second step is due to Assumption \ref{ahbe} (i), and the last step uses (\ref{yalphan}). This, combined with Assumption \ref{al1} and H\"older's inequality, implies\tcr{:}
\be
&&\phantom{~=~}\sx\|(\hmbP-\mbP_0)\trans \bfU_{n,1}(\x)\|_\infty \nonumber\\
&&~\leq~\|\hmbP-\mbP_0\|_1\sx\|\bfU_{n,1}(\x)\|_{\infty}=O_p(h_n^{r-1}\alpha_n^2).
\label{ybdn1}
\ee
Then, under Assumptions \ref{akernel} (ii)--(iii) and \ref{ahbey} (ii), Theorem 2 of \citet{hansen2008uniform} gives
\be
&&\sx\|\bfU_{n,2}(\x)-\E\{\bfU_{n,2}(\x)\}\|_{\infty}~=~O_p(h_n^{r}\xi_n), \label{ydn2}\\
&&\sx\|\bfU_{n,3}(\x)-\E\{\bfU_{n,3}(\x)\}\|_{\infty}~=~O_p(h_n^{r}\xi_n).
\label{ydn3}
\ee
Let $\delta(\s):=f_\S(\s)\kappa_1(\s)$ and $\nabla\delta(\s):=\partial \delta(\s)/\partial \s$. We \tcr{then} have
\be
&&\phantom{~=~}\sx\|\E\{\bfU_{n,2}(\x)\}\|_\infty \nonumber\\
&&~\leq~ \sx\|\x\hbox{$\int$}\delta(\s)[\nabla K\{h_n^{-1}(\mbP_0\trans\x-s)\}]\trans ds\|_\infty \nonumber\\
&&~=~h_n^{r+1}\sx\|\x\hbox{$\int$}\{\nabla\delta(\mbP_0\trans\x-h_n\bft)\}\trans K(\bft)d\bft\|_\infty =O(h_n^{r+1}).
\label{yedn2}
\ee
In the above, the second step uses integration by parts and change of variables, and the last step holds by Assumption \ref{ahbey} (i), the boundedness of $\nabla\delta(\s)$ from Assumptions \ref{akernel} (ii) and (iv), and the integrability of $K(\cdot)$ from Assumption \ref{akernel} (i). Set $\bzeta(\s):=f_\S(\s)\bchi_1(\s)$ and $\nabla\bzeta(\s):=\partial \bzeta(\s)/\partial \s$. Analogous to (\ref{yedn2}), we know
\be
&&\phantom{~=~}\sx\|\E\{\bfU_{n,3}(\x)\}\|_\infty  \nonumber\\
&&~\leq~ \sx\|\hbox{$\int$}\bzeta(\s) [\nabla K\{h_n^{-1}(\mbP_0\trans\x-s)\}]\trans ds\|_\infty \nonumber\\
&&~=~h_n^{r+1}\sx\|\hbox{$\int$}\{\nabla\bzeta(\mbP_0\trans\x-h_n\bft)\}\trans K(\bft)d\bft\|_\infty =O(h_n^{r+1}),
\label{yedn3}
\ee
where the last step holds by the boundedness of $\|\nabla\bzeta(\s)\|_\infty$ from Assumptions \ref{akernel} (ii) and \ref{ahbey} (iii), and the integrability of $K(\cdot)$ from Assumption \ref{akernel} (i). Combining (\ref{ydn2})--(\ref{yedn3}) yields
\bse
\sx\|\bfU_{n,2}(\x)-\bfU_{n,3}(\x)\|_\infty~=~O_p(h_n^{r}\xi_n+h_n^{r+1}),
\ese
which implies that
\bse
&&\phantom{~=~}\sx\|(\mbP_0-\hmbP)\trans\{\bfU_{n,2}(\x)-\bfU_{n,3}(\x)\}\|_\infty \\
&&~\leq~\|\mbP_0-\hmbP\|_1\sx\|\bfU_{n,2}(\x)-\bfU_{n,3}(\x)\|_{\infty} \\
&&~=~O_p(h_n^{r}\xi_n\alpha_n+h_n^{r+1}\alpha_n)\tcr{,}
\ese
using H\"older's inequality and Assumption \ref{al1}. This, combined with (\ref{yun}) and (\ref{ybdn1}), gives
\be
\sx|U_n(\x)|~=~O_p(h_n^{-2}\alpha_n^2+h_n^{-1}\xi_n\alpha_n+\alpha_n).
\label{yunr}
\ee
Then\tcr{,} we consider $V_{n,N}$. Write
\be
V_{n,N}(\x)&~=~& h_n^{-(r+1)}\trace ((\hmbP-\mbP_0)\trans \Enk[(\x-\X)\{\nabla K(\bar{\s})\}\trans\hD(\X)TY]) \nonumber\\
&~=~&h_n^{-(r+1)}\trace[(\hmbP-\mbP_0)\trans\{\bfV^{(1)}_{n,N}(\x)+\bfV^{(2)}_{n,N}(\x)\}],
\label{yvn}
\ee
where
\bse
&&\bfV^{(1)}_{n,N}(\x)~:=~\Enk((\x-\X)[\nabla K(\bar{\s}_n)-\nabla K\{h_n^{-1}\mbP_0\trans(\x-\X)\}]\trans\hD(\X)TY),  \\
&&\bfV^{(2)}_{n,N}(\x)~:=~\Enk((\x-\X) [\nabla K\{h_n^{-1}\mbP_0\trans(\x-\X)\}]\trans\hD(\X)TY).
\ese
We know
\be
&&\phantom{~=~}\ss\E(h_n^{-r}[\rho \{h_n^{-1}(\s-\S) \}Y]^2) \nonumber\\
&&~=~\ss\hbox{$\int$}h_n^{-r}[\rho\{h_n^{-1}(\s-\bfv) \}]^2\E(Y^2\mid\S=\bfv) f_\S(\bfv)d\bfv \nonumber\\
&&~=~\ss\hbox{$\int$}\{\rho(\bft )\}^2\E(Y^2\mid\S=\s-h_n\bft) f_\S(\s-h_n\bft)d\bft = O(1).
\label{ygrhosq}
\ee
where the second step uses change of variables while the last step holds by the boundedness of $\E(Y^2\mid\S=\cdot)f_\S(\cdot)$ from Assumptions \ref{akernel} (ii)--(iii) and the square integrability of $\rho(\cdot)$ from Assumption \ref{ahbey} (ii). Moreover, under Assumptions \ref{akernel} (ii)--(iii) and \ref{ahbey} (ii), Theorem 2 of \citet{hansen2008uniform} gives
\bse
\ss\{\Enk(h_n^{-r}[\rho \{h_n^{-1}(\s-\S) \}Y]^2)-\E(h_n^{-r}[\rho \{h_n^{-1}(\s-\S) \}Y]^2)\}=O_p(\xi_n)~=~o_p(1)\tcr{.}
\ese
This, combined with (\ref{ygrhosq}), implies
\be
\ss\Enk(h_n^{-r}[\rho \{h_n^{-1}(\s-\S) \}Y]^2)~=~O_p(1).
\label{yexrhosq}
\ee
Next, we have
\be
&&\phantom{~=~}\sx\Enk (\|[\nabla K(\bar{\s}_n)-\nabla K\{h_n^{-1}\mbP_0\trans(\x-\X)\}]Y\|^2) \nonumber\\
&&~\leq~\sx\Enk (\|\bar{\s}_n-h_n^{-1}\mbP_0\trans(\x-\X)\|^2[\rho\{h_n^{-1}\mbP_0\trans(\x-\X)\}Y]^2) \nonumber\\
&&~\leq~\sx\Enk (\|(\hmbP-\mbP_0)\trans(\x-\X)\|^2h_n^{-2}[\rho\{h_n^{-1}\mbP_0\trans(\x-\X)\}Y]^2) \nonumber\\
&&~\leq~ c\,\|\hmbP-\mbP_0\|_1^2\sxx\|\x-\X\|_{\infty}^2\ss\Enk (h_n^{-2}[\rho\{h_n^{-1}\mbP_0\trans(\x-\X)\}Y]^2)\nonumber \\
&&~=~O_p(h_n^{r-2}\alpha_n^2),
\label{yalphansq}
\ee
where the first step uses the local Lipschitz continuity of $\nabla K(\cdot)$ from Assumption \ref{ahbey} (ii), the second step is due to the definition (\ref{ysbar}) of $\bar{\s}_n$, the third step holds by H\"older's inequality, and the last step is because of Assumptions \ref{al1}, \ref{ahbe} (i) and the equation (\ref{yexrhosq}). Thus\tcr{,} we have
\be
&&\phantom{~=~}\|\bfV^{(1)}_{n,N}(\x)\|_{\infty} \nonumber\\
&&~\leq~ c\, (\E_{n,k}[\{\hD(\X)\}^2]\sx\Enk (\|[\nabla K(\bar{\s}_n)-\nabla K\{h_n^{-1}\mbP_0\trans(\x-\X)\}]Y\|^2))^{1/2} \nonumber \\
&& ~=~O_p(h_n^{r/2-1}\alpha_n s_N),
\label{yvn1}
\ee
where the first step uses H\"older's inequality and the boundedness of $\sx\|\x-\X\|_\infty $ from Assumption \ref{ahbey} (i), and the last step holds by (\ref{sn}) and (\ref{yalphansq}). Next, we know that
\be
&&\phantom{~=~}|\ss\E_\S([\nabla K_{[j]}\{h_n^{-1}(\s-\S)\}Y]^2)|\nonumber\\
&&~=~|\ss\hbox{$\int$}[\nabla K_{[j]}\{h_n^{-1}(\s-\bfv)\}]^2 E(Y^2\mid\S=\bfv)f_{\S}(\bfv)d\bfv| \nonumber\\
&&~=~h_n^{r}|\ss\hbox{$\int$}\{\nabla K_{[j]}(\bft)\}^2E(Y^2\mid\S=\s-h_n\bft)f_{\S}(\s-h_n\bft)d\bft|=O(h_n^{r}),
\label{yexp1}
\ee
where the second step uses change of variables while the last step is due to the boundedness of $\E(Y^2\mid\S=\cdot)f_\S(\cdot)$ from Assumptions \ref{akernel} (ii)--(iii) and the square integrability of $\nabla K_{[j]}(\cdot)$ from Assumption \ref{akernel} (i). Then, under Assumptions \ref{akernel} (ii)--(iii) and \ref{ahbey} (ii), Theorem 2 of \citet{hansen2008uniform} implies\tcr{:}
\bse
&&\phantom{~=~}\ss|\Enk([\nabla K_{[j]}\{h_n^{-1}(\s-\S)\}Y]^2)-\E_\S([\nabla K_{[j]}\{h_n^{-1}(\s-\S)\}Y]^2)| \\
&&~=~O_p(h_n^{r}\xi_{n})=o_p(h_n^r)\tcr{,}
\ese
where the last step is because we assume $\xi_{n}=o(1)$. This, combined with (\ref{yexp1}), yields
\be
\ss\Enk([\nabla K_{[j]}\{h_n^{-1}(\s-\S)\}Y]^2)~=~O_p(h_n^{r}).
\label{yone1}
\ee
Let $v_{ij}(\x)$ be the $(i,j)$th entry of $\bfV^{(2)}_{n,N}(\x)$ $(i=1,\ldots,p;\,j=1,\ldots,r)$. We know
\bse
&&\phantom{~=~}\sx|v_{ij}(\x)| \\
&&~\equiv~\sx|\Enk[(\x_{[i]}-\X_{[i]}) \nabla K_{[j]}\{h_n^{-1}\mbP_0\trans(\x-\X)\}\hD(\X)TY]| \\
&&~\leq~\ss\Enk[|\nabla K_{[j]}\{h_n^{-1}(\s-\S)\}\hD(\X)Y|] \\
&&~\leq~\{\ss\Enk([\nabla K_{[j]}\{h_n^{-1}(\s-\S)\}Y]^2)\Enk[\{\hD(\X)\}^2]\}^{1/2}=O_p(h_n^{r/2}s_N),
\ese
where the second step uses the boundedness of $\sx\|\x-\X\|_\infty$ from Assumption \ref{ahbe} (i), the third step is due to H\"older's inequality and the last step holds by (\ref{yone1}) and (\ref{sn}). It now follows that
\be
\sx\|\bfV^{(2)}_{n,N}(\x)\|_\infty~=~O_p(h_n^{r/2}s_N).
\label{yvn2}
\ee
Therefore\tcr{,} we have
\bse
&&\phantom{~=~}\sx\|(\mbP_0-\hmbP)\trans\{\bfV^{(1)}_{n,N}(\x)+\bfV^{(2)}_{n,N}(\x)\}\|_\infty \\
&&~\leq~\|\mbP_0-\hmbP\|_1\sx\|\bfV^{(1)}_{n,N}(\x)+\bfV^{(2)}_{n,N}(\x)\|_{\infty} \\
&&~=~O_p(h_n^{r/2-1}\alpha_n^2 s_N+h_n^{r/2}\alpha_n s_N)~=~O_p(h_n^{r/2}\alpha_n s_N),
\ese
where the first step is due to H\"older's inequality, the second step uses (\ref{yvn1}), (\ref{yvn2}) and Assumption \ref{al1}, and the last step is because we assume $h_n^{-1}\alpha_n=o(1)$. Combined with (\ref{yvn}), it gives
\be
\sx |V_{n,N}(\x)|~=~O_p\{h_n^{-(r/2+1)}\alpha_ns_N\}.
\label{yvnr}
\ee
Considering (\ref{ydhbe}), (\ref{yunr}) and (\ref{yvnr}), we know that
\be
&&\phantom{~=~}\sx|\hlo(\x,\hmbP)-\hlo(\x,\mbP_0)| \nonumber\\
&&~=~O_p\{h_n^{-2}\alpha_n^2+h_n^{-1}\xi_n\alpha_n+\alpha_n+h_n^{-(r/2+1)}\alpha_n s_N\}.
\label{yhmbe}
\ee

Further, we control the error from estimating $\pi(\x)$ by $\pihatN(\x)$, i.e., $\hlo(\x,\mbP_0)-\lnk^{(1)}(\x,\mbP_0)$ with
\bse
\lnk^{(1)}(\x,\mbP)~:=~h_n^{-r}\Enk [\{\pis(\X)\}^{-1}TY K_h\{\mbP\trans(\x-\X)\}].
\ese
We have
\be
&&\phantom{~=~}|\ss\E_\S[h_n^{-r}\{K_h(\s-\S)Y\}^2]|\nonumber\\
&&~=~h_n^{-r}|\ss\hbox{$\int$}[K\{h_n^{-1}(\s-\bfv)\}]^2\E(Y^2\mid\S=\bfv)f_{\S}(\bfv)d\bfv| \nonumber\\
&&~=~|\ss\hbox{$\int$}\{K(\bft)\}^2\E(Y^2\mid\S=\s-h_n\bft)f_{\S}(\s-h_n\bft)d\bft|~=~O(1),
\label{yexp}
\ee
where the second step uses change of variables while the last step is due to the boundedness of $\E(Y^2\mid\S=\cdot)f_\S(\cdot)$ from Assumptions \ref{akernel} (ii)--(iii) along with the square integrability of $K(\cdot)$ from Assumption \ref{akernel} (i). Then, under Assumptions \ref{akernel}, Theorem 2 of \citet{hansen2008uniform} gives
\bse
\ss|\Enk[h_n^{-r}\{K_h(\s-\S)Y\}^2]-\E_\S[h_n^{-r}\{K_h(\s-\S)Y\}^2]|~=~O_p(\xi_n)~=~o_p(1),
\ese
where the last step is because we assume $\xi_n=o(1)$. This, combined with (\ref{yexp}), yields
\be
\ss\Enk[h_n^{-r}\{K_h(\s-\S)Y\}^2]~=~O_p(1).
\label{yone}
\ee
Therefore\tcr{,} we know that
\be
&&\phantom{~=~}\sx|\hlo(\x,\mbP_0)-\lnk^{(1)}(\x,\mbP_0)| \nonumber\\
&&~\leq~ c\,\ss \Enk \{|\hD(\X)h_n^{-r}K_h(\s-\S)Y| \}\nonumber\\
&&~\leq~ c\,h^{-r/2}\{\Enk [\{\hD(\X)\}^2] \ss\Enk[h_n^{-r}\{K_h(\s-\S)Y\}^2]\}^{1/2} \nonumber\\
&&~=~O_p(h^{-r/2}s_N), \label{ymnk}
\ee
where the second step is due to H\"older's inequality and the last step holds by (\ref{sn}) and (\ref{yone}).

Combining (\ref{yhmbe}) and (\ref{ymnk}) yields that
\be
&&\phantom{~=~}\sx|\hlo(\x,\hmbP)-\lnk^{(1)}(\x,\mbP_0)| \nonumber\\ &&=O_p\{h_n^{-2}\alpha_n^2+h_n^{-1}\xi_n\alpha_n+\alpha_n+h_n^{-(r/2+1)}\alpha_n s_N+h^{-r/2}s_N\} \nonumber\\
&&~=~O_p\{h_n^{-2}\alpha_n^2+h_n^{-1}\xi_n\alpha_n+\alpha_n+h^{-r/2}s_N\} ~=~O_p\{b_{n,N}^{(2)}\},
\label{yan2}
\ee
where the second step holds by the fact that $h_n^{-(r/2+1)}\alpha_n s_N=o(h^{-r/2}s_N)$ because we assume $h^{-1}\alpha_n=o(1)$.

Now we handle the error $\lnk^{(1)}(\x,\mbP_0)-\ell^{(1)}(\x,\mbP_0)$. Under Assumptions \ref{akernel}, Theorem 2 of \citet{hansen2008uniform} gives
\be
\sx|\lnk^{(1)}(\x,\mbP_0)-\E\{\lnk^{(1)}(\x,\mbP_0)\}|~=~O_p(\xi_n).
\label{ypt2}
\ee
Further, under Assumptions \ref{akernel} (i), (ii) and (iv), standard arguments based on $d$th order Taylor's expansion of $\ell^{(1)}(\x,\mbP_0)$ yield that
\be
\sx|\E\{\lnk^{(1)}(\x,\mbP_0)\}-\ell^{(1)}(\x,\mbP_0)|~=~O(h_n^d).
\label{ypt3}
\ee

Combining (\ref{yan2}), (\ref{ypt2}) and (\ref{ypt3}) yields
\be
\sx|\hlo(\x,\hmbP)-\ell^{(1)}(\x,\mbP_0)|~=~O_p\{b_n^{(1)}+b_{n,N}^{(2)}\}.
\label{ynum}
\ee
Similar arguments imply that
\be
\sx|\hlz(\x,\hmbP)-\ell^{(0)}(\x,\mbP_0)|~=~O_p\{b_n^{(1)}+b_{n,N}^{(2)}\}.
\label{ydeno}
\ee
Therefore\tcr{,} we have
\bse
&&\phantom{~=~}\sx|\mhatnk(\x,\hmbP)-\tmu(\x,\mbP_0)| \nonumber\\
&&~=~\sx|\{\hlz(\x,\hmbP)\}^{-1}\hl^{(0)}(\x,\hmbP)-\{\ell^{(0)}(\x,\mbP_0)\}^{-1}\ell^{(1)}(\x,\mbP_0)| \\
&&~\leq~\sx|\{\hlz(\x,\mbP_0)\}^{-1}\{\hlo(\x,\hmbP)-\ell^{(1)}(\x,\mbP_0)\}|+ \\
&&\phantom{~=~}\sx|[\{\hlz(\x,\mbP_0)\}^{-1}-\{\ell^{(0)}(\x,\mbP_0)\}^{-1}]\ell^{(1)}(\x,\mbP_0)| \\
&&~=~O_p\{b_n^{(1)}+b_{n,N}^{(2)}\},
\ese
where the last step follows from the fact that $b_n^{(1)}+b_{n,N}^{(2)}=o(1)$, and repeated use of (\ref{ynum}) and (\ref{ydeno}) as well as Assumptions \ref{api4} and \ref{akernel} (ii).

\subsection{Proof of Proposition \ref{thphi}}
The function $F(\cdot\mid\S)$ is obviously bounded. For any $\theta_1,\theta_2\in\mbtv$, Taylor's expansion gives
\bse
&&\phantom{~=~}|[\{\pis(\X)\}^{-1}T]^m\{\phis(\X,\theta_1)-\phis(\X,\theta_2)\}| \\
&&~\leq~ c\,|F(\theta_1\mid\S)-F(\theta_2\mid\S)| ~\leq~ c\,\sb f(\theta\mid\S)|\theta_1-\theta_2|\quad (m=0,1),
\ese
where the first step uses the boundedness of $\{\pis(\X)\}^{-1}$ from Assumption \ref{api}. Therefore, the condition \eqref{conditional_density} and Example 19.7 of \citet{van2000asymptotic} give
\be
&&N_{[\,]}\{\eta,\mm,L_2(\P_\X)\}~\leq~ c\,\eta^{-1}, \label{mm} \\
&&N_{[\,]}\{\eta,\mathcal{F}^*,L_2(\P_\X)\}~\leq~ c\,\eta^{-1}\tcr{,} \nonumber
\ee
with $\mathcal{F}^*:=\{\{\pis(\X)\}^{-1}T\phis(\X,\theta):\theta\in\mbtv\}$, which implies that $\mathcal{F}^*$ and $\mm$ are $\P$-Donsker according to Theorem 19.5 of \citet{van2000asymptotic}. Further, we have that, for any sequence $\tvt\to\vt$ in probability,
\bse
&&\phantom{~=~}\E_\X([\{\pis(\X)\}^{-2}T]^m\{\phis(\X,\tvt)-\phis(\X,\vt)\}^2) \\
&&~\leq~ c\,\E_\S[\{F(\tvt\mid\S)-F(\vt\mid\S)\}^2] ~\leq~ c\,(\tvt-\vt)^2\E[\{\sb f(\theta\mid\S)\}^2]\to 0 \;\; (m=0,1)
\ese
in probability, where the first step uses the boundedness of $\{\pis(\X)\}^{-2}$ from Assumption \ref{api}, the second step uses Taylor's expansion as well as the fact that $\tvt\in\mbtv$ with probability approaching one, and the last step holds by the condition \eqref{conditional_density}. Thus applying Lemma 19.24 of \citet{van2000asymptotic} concludes (\ref{unipi1}) and (\ref{unipi2}).

\subsection{Proof of Theorem \ref{thhd}}
Denote $e^{(t)}(\x,\theta,\mbP)=\varphi_t(\mbP\trans\x,\theta)f_\S(\mbP\trans\x)$  $(t=0,1)$. We now derive the convergence rate of $\hateo(\x,\theta,\hmbP)-e^{(1)}(\x,\theta,\mbP)$. The case of $\hatez(\x,\theta,\hmbP) - e^{(0)}(\x,\theta,\mbP)$ is similar.

We first deal with the error from estimating $\mbP_0$ by $\hmbP$, i.e., $\hateo(\x,\theta,\hmbP)-\hateo(\x,\theta,\mbP_0)$. Taylor's expansion gives that, for
\be
\bar{\s}_n~:=~h_n^{-1}\{\mbP_0\trans+\bmu(\hmbP-\mbP_0)\trans\}(\x-\X)
\label{sbar}
\ee
with some $\bmu:=\diag(\mu_1,\ldots,\mu_r)$ and $\mu_j\in(0,1)$ $(j=1,\ldots,r)$,
\be
&&\phantom{~=~}\hateo(\x,\theta,\hmbP)-\hateo(\x,\theta,\mbP_0) \nonumber \\
&&~=~h_n^{-(r+1)}\Enk[\{\nabla K(\bar{\s})\}\trans(\hmbP-\mbP_0)\trans(\x-\X)\{\pihatN(\X)\}^{-1}T\psi(Y,\theta)] \nonumber\\
&&~=~U_n(\x,\theta)+V_{n,N}(\x,\theta) ,
\label{dhbe}
\ee
where
\bse
&&U_n(\x,\theta)~:=~h_n^{-(r+1)}\Enk[\{\nabla K(\bar{\s})\}\trans(\hmbP-\mbP_0)\trans(\x-\X)\{\pis(\X)\}^{-1}T\psi(Y,\theta)] ,
\nonumber \\
&&V_{n,N}(\x,\theta)~:=~h_n^{-(r+1)}\Enk[\{\nabla K(\bar{\s})\}\trans(\hmbP-\mbP_0)\trans(\x-\X)\hD(\X)T\psi(Y,\theta)].
\ese
To control $U_n(\x,\theta)$, write
\be
U_n(\x,\theta)&~=~& h_n^{-(r+1)}\trace ((\hmbP-\mbP_0)\trans \Enk[(\x-\X)\{\nabla K(\bar{\s})\}\trans\{\pis(\X)\}^{-1}T\psi(Y,\theta)]) \nonumber\\
&~=~&h_n^{-(r+1)}\trace[(\hmbP-\mbP_0)\trans\{\bfU_{n,1}(\x,\theta)+\bfU_{n,2}(\x,\theta)-\bfU_{n,3}(\x,\theta)\}],
\label{un}
\ee
where
\bse
&&\bfU_{n,1}(\x,\theta)~:=~\Enk((\x-\X)[\nabla K(\bar{\s}_n)-\nabla K\{h_n^{-1}\mbP_0\trans(\x-\X)\}]\trans\{\pis(\X)\}^{-1}T\psi(Y,\theta)),  \\
&&\bfU_{n,2}(\x,\theta)~:=~\Enk(\x [\nabla K\{h_n^{-1}\mbP_0\trans(\x-\X)\}]\trans\{\pis(\X)\}^{-1}T\psi(Y,\theta)),  \\
&&\bfU_{n,3}(\x,\theta)~:=~\Enk(\X [\nabla K\{h_n^{-1}\mbP_0\trans(\x-\X)\}]\trans\{\pis(\X)\}^{-1}T\psi(Y,\theta)).
\ese
For the function $\rho(\cdot)$ in Assumption \ref{ahbe} (ii), denote $\mathcal{J}_n:=\{h^{-r}_n\rho\{h_n^{-1}(\s-\mbP_0\trans\X)\}:\s\in\ms\}$. Taylor's expansion gives that, for  any $\s_1,\s_2\in\ms$  and some $\bar{\s}:=\s_1+\bmu(\s_2-\s_1)$ with $\bmu:=\diag(\mu_1,\ldots,\mu_r)$ and $\mu_j\in(0,1)$ $(j=1,\ldots,r)$,
\bse
&&\phantom{~=~}h^{-r}_n|\rho\{h_n^{-1}(\s_1-\mbP_0\trans\X)\}-\rho\{h_n^{-1}(\s_2-\mbP_0\trans\X)\}| \\
&&~=~ h_n^{-(r+1)}|[\nabla\rho\{h_n^{-1}(\bar{\s}-\mbP_0\trans\X)\}]\trans(\s_1-\s_2)|\leq c\,h^{-(r+1)}_n\|\s_1-\s_2\|,
\ese
where the second step uses the boundedness of $\nabla\rho(\cdot)$ from Assumption \ref{ahbe} (ii). Therefore Example 19.7 of \citet{van2000asymptotic} implies
\be
N_{[\,]}\{\eta ,\mathcal{J}_n,L_2(\P_\X)\}~\leq~ c\,h_n^{-(r+1)}\eta^{-r}.
\label{bracj}
\ee
Moreover, we have that
\be
\ssx [h^{-r}_n\rho\{h_n^{-1}(\s-\mbP_0\trans\x)\}]~=~O(h_n^{-r}).
\label{supj}
\ee
due to the boundedness of $\rho(\cdot)$ from Assumption \ref{ahbe} (ii). In addition, we know that
\be
\ss\E_\S([h_n^{-r}\rho \{h_n^{-1}(\s-\S) \}]^2)&~=~&h^{-r}\ss\hbox{$\int$}h_n^{-r}[\rho\{h_n^{-1}(\s-\bfv) \}]^2 f_\S(\bfv)d\bfv \nonumber\\
&~=~&h_n^{-r}\ss\hbox{$\int$}\{\rho(\bft )\}^2 f_\S(\s-h_n\bft)d\bft ~=~ O(h_n^{-r}),
\label{varj}
\ee
where the second step uses change of variables while the last step holds by the boundedness of $f_\S(\cdot)$ from Assumption \ref{akernel_qte} (ii) and the square integrability of $\rho(\cdot)$ from Assumption \ref{ahbe} (ii). Based on (\ref{bracj})--(\ref{varj}), applying Lemma \ref{1v2} yields that
\be
&&\phantom{~=~}\ss|\Enk[h^{-r}_n\rho\{h_n^{-1}(\s-\mbP_0\trans\X)\}]-\E_\X[h^{-r}_n\rho\{h_n^{-1}(\s-\mbP_0\trans\X)\}]| \nonumber\\
&&~=~O_p\{n_{\kK^-}^{-1/2}h_n^{-r/2}\log(h_n^{-1})+n_{\kK^-}^{-1}h_n^{-r}(\log\,h_n)^2\}~=~o_p(1),
\label{grho}
\ee
where the second step is because we assume $(nh_n^r)^{-1/2}\log(h_n^{-r})=o(1)$. Then we know
\bse
\ss\E_\S[h_n^{-r}\rho \{h_n^{-1}(\s-\S) \}]&~=~&\ss\hbox{$\int$}h_n^{-r}\rho\{h_n^{-1}(\s-\bfv) \} f_\S(\bfv)d\bfv  \\
&~=~&\ss\hbox{$\int$}\rho(\bft ) f_\S(\s-h_n\bft)d\bft ~=~ O(1).
\ese
where the second step uses change of variables while the last step holds by the boundedness of $f_\S(\cdot)$ from Assumption \ref{akernel_qte} (ii) and the integrability of $\rho(\cdot)$ from Assumption \ref{ahbe} (ii). This, combined with (\ref{grho}), implies\tcr{:}
\be
\ss\Enk[h_n^{-r}\rho \{h_n^{-1}(\s-\S) \}]~=~O_p(1).
\label{exrho}
\ee
Next, we have
\be
&&\phantom{~=~}\sx\Enk [\|\nabla K(\bar{\s}_n)-\nabla K\{h_n^{-1}\mbP_0\trans(\x-\X)\}\|] \nonumber\\
&&~\leq~\sx\Enk [\|\bar{\s}_n-h_n^{-1}\mbP_0\trans(\x-\X)\|\rho\{h_n^{-1}\mbP_0\trans(\x-\X)\}] \nonumber\\
&&~\leq~\sx\Enk [\|(\hmbP-\mbP_0)\trans(\x-\X)\|h_n^{-1}\rho\{h_n^{-1}\mbP_0\trans(\x-\X)\}] \nonumber\\
&&~\leq~ c\,\|\hmbP-\mbP_0\|_1\sxx\|\x-\X\|_{\infty}\ss\Enk [h_n^{-1}\rho\{h_n^{-1}(\s-\S)\}]\nonumber \\
&&~=~O_p(h_n^{r-1}\alpha_n),
\label{alphan}
\ee
where the first step uses the local Lipschitz continuity of $\nabla K(\cdot)$ from Assumption \ref{ahbe} (ii), the second step is due to the definition (\ref{sbar}) of $\bar{\s}_n$, the third step holds by H\"older's inequality, and the last step is because of Assumptions \ref{al1}, \ref{ahbe} (i) and the equation (\ref{exrho}). Hence
\bse
&&\phantom{~=~}\sbx\|\bfU_{n,1}(\x,\theta)\|_{\infty} \\
&&~\leq~ c\,\sx\Enk [\|\x-\X\|_{\infty}\|\nabla K(\bar{\s}_n)-\nabla K\{h_n^{-1}\mbP_0\trans(\x-\X)\}\|] \\
&&~\leq~ c\,\sx\Enk [\|\nabla K(\bar{\s}_n)-\nabla K\{h_n^{-1}\mbP_0\trans(\x-\X)\}\|] ~=~O_p(h_n^{r-1}\alpha_n).
\ese
where the first step holds by the boundedness of $\{\pis(\X)\}^{-1}T\psi(Y,\theta)$, the second step is due to Assumption \ref{ahbe} (i), and the last step uses (\ref{alphan}). This, combined with Assumption \ref{al1} and H\"older's inequality, implies
\be
&&\phantom{~=~}\sbx\|(\hmbP-\mbP_0)\trans \bfU_{n,1}(\x,\theta)\|_\infty \nonumber\\
&&~\leq~\|\hmbP-\mbP_0\|_1\sbx\|\bfU_{n,1}(\x,\theta)\|_{\infty}~=~O_p(h_n^{r-1}\alpha_n^2).
\label{bdn1}
\ee
Then, under Assumptions \ref{akernel_qte} (ii) and \ref{ahbe} (ii), as well as the fact that $\{\{\pis(\X)\}^{-1}T\psi(Y,\theta):\theta\in\mbtv\}$ is a VC class with a bounded envelope function $\sb[\{\pis(\X)\}^{-1}T|\psi(Y,\theta)|]$
from Assumption \ref{api}, Lemma B.4 of \citet{escanciano2014uniform} gives that
\be
&&\sbx\|\bfU_{n,2}(\x,\theta)-\E\{\bfU_{n,2}(\x,\theta)\}\|_{\infty}~=~O_p(h_n^{r}\gamma_n), \label{dn2}\\
&&\sbx\|\bfU_{n,3}(\x,\theta)-\E\{\bfU_{n,3}(\x,\theta)\}\|_{\infty}~=~O_p(h_n^{r}\gamma_n).
\label{dn3}
\ee
Let $\delta(\s,\theta):=f_\S(\s)\varphi_1(\s,\theta)$ and $\nabla\delta(\s,\theta):=\partial \delta(\s,\theta)/\partial \s$. We have
\be
&&\phantom{~=~}\sbx\|\E\{\bfU_{n,2}(\x,\theta)\}\|_\infty \nonumber\\
&&~\leq~ \sbx\|\x\hbox{$\int$}\delta(\s,\theta)[\nabla K\{h_n^{-1}(\mbP_0\trans\x-s)\}]\trans ds\|_\infty \nonumber\\
&&~=~h_n^{r+1}\sbx\|\x\hbox{$\int$}\{\nabla\delta(\mbP_0\trans\x-h_n\bft,\theta)\}\trans K(\bft)d\bft\|_\infty ~=~O(h_n^{r+1}).
\label{edn2}
\ee
In the above, the second step uses integration by parts and change of variables, while the last step holds by Assumption \ref{ahbe} (i), the boundedness of $\nabla\delta(\s,\theta)$ from Assumptions \ref{akernel_qte} (ii)--(iii), as well as the integrability of $K(\cdot)$ from Assumption \ref{akernel_qte} (i). Set $\bzeta(\s,\theta):=f_\S(\s)\bfeta_1(\s,\theta)$ and $\nabla\bzeta(\s,\theta):=\partial \bzeta(\s,\theta)/\partial \s$. Analogous to (\ref{edn2}), we know
\be
&&\phantom{~=~}\sbx\|\E\{\bfU_{n,3}(\x,\theta)\}\|_\infty  \nonumber\\
&&~\leq~ \sbx\|\hbox{$\int$}\bzeta(\s,\theta) [\nabla K\{h_n^{-1}(\mbP_0\trans\x-s)\}]\trans ds\|_\infty \nonumber\\
&&~=~h_n^{r+1}\sbx\|\hbox{$\int$}\{\nabla\bzeta(\mbP_0\trans\x-h_n\bft,\theta)\}\trans K(\bft)d\bft\|_\infty ~=~O(h_n^{r+1}),
\label{edn3}
\ee
where the last step holds by the boundedness of $\|\nabla\bzeta(\s,\theta)\|_\infty$ from Assumptions \ref{akernel_qte} (ii) and \ref{ahbe} (iii), as well as the integrability of $K(\cdot)$ from Assumption \ref{akernel_qte} (i). Combining (\ref{dn2})--(\ref{edn3}) yields
\bse
\sbx\|\bfU_{n,2}(\x,\theta)-\bfU_{n,3}(\x,\theta)\|_\infty~=~O_p(h_n^{r}\gamma_n+h_n^{r+1}),
\ese
which implies that
\bse
&&\phantom{~=~}\sbx\|(\mbP_0-\hmbP)\trans\{\bfU_{n,2}(\x,\theta)-\bfU_{n,3}(\x,\theta)\}\|_\infty \\
&&~\leq~\|\mbP_0-\hmbP\|_1\sbx\|\bfU_{n,2}(\x,\theta)-\bfU_{n,3}(\x,\theta)\|_{\infty} \\
&&~=~O_p(h_n^{r}\gamma_n\alpha_n+h_n^{r+1}\alpha_n)\tcr{,}
\ese
using H\"older's inequality and Assumption \ref{al1}. This, combined with (\ref{un}) and (\ref{bdn1}), gives
\be
\sbx|U_n(\x,\theta)|~=~O_p(h_n^{-2}\alpha_n^2+h_n^{-1}\gamma_n\alpha_n+\alpha_n).
\label{unr}
\ee
Then\tcr{,} we consider $V_{n,N}$. Write
\be
V_{n,N}(\x,\theta)&~=~& h_n^{-(r+1)}\trace ((\hmbP-\mbP_0)\trans \Enk[(\x-\X)\{\nabla K(\bar{\s})\}\trans\hD(\X)T\psi(Y,\theta)]) \nonumber\\
&~=~&h_n^{-(r+1)}\trace[(\hmbP-\mbP_0)\trans\{\bfV^{(1)}_{n,N}(\x,\theta)+\bfV^{(2)}_{n,N}(\x,\theta)\}],
\label{vn}
\ee
where
\bse
&&\bfV^{(1)}_{n,N}(\x,\theta)~:=~\Enk((\x-\X)[\nabla K(\bar{\s}_n)-\nabla K\{h_n^{-1}\mbP_0\trans(\x-\X)\}]\trans\hD(\X)T\psi(Y,\theta)),  \\
&&\bfV^{(2)}_{n,N}(\x,\theta)~:=~\Enk((\x-\X) [\nabla K\{h_n^{-1}\mbP_0\trans(\x-\X)\}]\trans\hD(\X)T\psi(Y,\theta)).
\ese
We have
\be
&&\phantom{~=~}\sbx\|\bfV^{(1)}_{n,N}(\x,\theta)\|_{\infty} \nonumber\\
&&~\leq~ c\,\sx|\hD(\x)|\sx\Enk [\|\nabla K(\bar{\s}_n)-\nabla K\{h_n^{-1}\mbP_0\trans(\x-\X)\}\|] \nonumber \\
&& ~=~O_p(h_n^{r-1}\alpha_n),
\label{vn1}
\ee
where the first step uses the boundedness of $\sx\|\x-\X\|_\infty T\psi(Y,\theta)$ from Assumption \ref{ahbe} (i), and the last step holds by (\ref{alphan}) and (\ref{dsup}) in Assumption \ref{api}. Next, we know that
\be
&&\phantom{~=~}|\ss\E_\S([\nabla K_{[j]}\{h_n^{-1}(\s-\S)\}]^2)|\nonumber\\
&&~=~|\ss\hbox{$\int$}[\nabla K_{[j]}\{h_n^{-1}(\s-\bfv)\}]^2 f_{\S}(\bfv)d\bfv| \nonumber\\
&&~=~h_n^{r}|\ss\hbox{$\int$}\{\nabla K_{[j]}(\bft)\}^2f_{\S}(\s-h_n\bft)d\bft|=O(h_n^{r}),
\label{exp1}
\ee
where the second step uses change of variables while the last step is due to the boundedness of $f_\S(\cdot)$ from Assumption \ref{akernel_qte} (ii) and the square integrability of $\nabla K_{[j]}(\cdot)$ from Assumption \ref{akernel_qte} (i). Then, under Assumptions \ref{akernel_qte} (ii) and \ref{ahbe} (ii), Lemma B.4 of \citet{escanciano2014uniform} implies\tcr{:}
\bse
\ss|\Enk([\nabla K_{[j]}\{h_n^{-1}(\s-\S)\}]^2)-\E_\S([\nabla K_{[j]}\{h_n^{-1}(\s-\S)\}]^2)|~=~O_p(h_n^{r}\gamma_{n})~=~o_p(h_n^r)
\ese
where the last step is because we assume $\gamma_{n}=o(1)$. This, combined with (\ref{exp1}), yields
\be
\ss\Enk([\nabla K_{[j]}\{h_n^{-1}(\s-\S)\}]^2)~=~O_p(h_n^{r}).
\label{one1}
\ee
Let $v_{ij}(\x,\theta)$ be the $(i,j)$th entry of $\bfV^{(2)}_{n,N}(\x,\theta)$ $(i=1,\ldots,p;\,j=1,\ldots,r)$. We know
\bse
&&\phantom{~=~}\sbx|v_{ij}(\x,\theta)| \\
&&~\equiv~\sbx|\Enk[(\x_{[i]}-\X_{[i]}) \nabla K_{[j]}\{h_n^{-1}\mbP_0\trans(\x-\X)\}\hD(\X)T\psi(Y,\theta)]| \\
&&~\leq~\ss\Enk[|\nabla K_{[j]}\{h_n^{-1}(\s-\S)\}\hD(\X)|] \\
&&~\leq~\{\ss\Enk([\nabla K_{[j]}\{h_n^{-1}(\s-\S)\}]^2)\Enk[\{\hD(\X)\}^2]\}^{1/2}~=~O_p(h_n^{r/2}s_N),
\ese
where the second step uses the boundedness of $\sx\|\x-\X\|_\infty T\psi(Y,\theta)$ from Assumption \ref{ahbe} (i), the third step is due to H\"older's inequality and the last step holds by (\ref{one1}) and (\ref{sn}). Therefore it follows that
\be
\sbx\|\bfV^{(2)}_{n,N}(\x,\theta)\|_\infty~=~O_p(h_n^{r/2}s_N).
\label{vn2}
\ee
Therefore, we have
\bse
&&\phantom{~=~}\sbx\|(\mbP_0-\hmbP)\trans\{\bfV^{(1)}_{n,N}(\x,\theta)+\bfV^{(2)}_{n,N}(\x,\theta)\}\|_\infty \\
&&~\leq~\|\mbP_0-\hmbP\|_1\sbx\|\bfV^{(1)}_{n,N}(\x,\theta)+\bfV^{(2)}_{n,N}(\x,\theta)\|_{\infty} \\
&&~=~O_p(h_n^{r-1}\alpha_n^2+h_n^{r/2}\alpha_n s_N),
\ese
where the first step is due to H\"older's inequality and the last step uses (\ref{vn1}), (\ref{vn2}) and Assumption \ref{al1}. Combined with (\ref{vn}), it gives
\be
\sbx |V_{n,N}(\x,\theta)|~=~O_p\{h_n^{-2}\alpha_n^2+h_n^{-(r/2+1)}\alpha_ns_N\}.
\label{vnr}
\ee
Considering (\ref{dhbe}), (\ref{unr}) and (\ref{vnr}), we know that
\be
&&\phantom{~=~}\sbx|\hateo(\x,\theta,\hmbP)-\hateo(\x,\theta,\mbP_0)| \nonumber\\
&&~=~O_p\{h_n^{-2}\alpha_n^2+h_n^{-1}\gamma_n\alpha_n+\alpha_n+h_n^{-(r/2+1)}\alpha_n s_N\}.
\label{hmbe}
\ee

Further, we control the error from estimating $\pi(\x)$ by $\pihatN(\x)$, i.e., $\hateo(\x,\theta,\mbP_0)-\enk^{(1)}(\x,\theta,\mbP_0)$ with
\bse
\enk^{(1)}(\x,\theta,\mbP)~:=~h_n^{-r}\Enk [\{\pis(\X)\}^{-1}T\psi(Y,\theta) K_h\{\mbP\trans(\x-\X)\}].
\ese
We have
\be
&&\phantom{~=~}|\ss\E_\S[h_n^{-r}\{K_h(\s-\S)\}^2]|\nonumber\\
&&~=~h_n^{-r}|\ss\hbox{$\int$}[K\{h_n^{-1}(\s-\bfv)\}]^2f_{\S}(\bfv)d\bfv| \nonumber\\
&&~=~|\ss\hbox{$\int$}\{K(\bft)\}^2f_{\S}(\s-h_n\bft)d\bft|~=~O(1),
\label{exp}
\ee
where the second step uses change of variables while the last step is due to the boundedness of $f_\S(\cdot)$ from Assumption \ref{akernel_qte} (ii) and the square integrability of $K(\cdot)$ from Assumption \ref{akernel_qte} (i). Then, under Assumptions \ref{akernel_qte} (i)--(ii) , Lemma B.4 of \citet{escanciano2014uniform} implies\tcr{:}
\bse
\ss|\Enk[h_n^{-r}\{K_h(\s-\S)\}^2]-\E_\S[h_n^{-r}\{K_h(\s-\S)\}^2]|~=~O_p(\gamma_{n})~=~o_p(1),
\ese
where the last step is because we assume $\gamma_{n}=o(1)$. This, combined with (\ref{exp}), yields
\be
\ss\Enk[h_n^{-r}\{K_h(\s-\S)\}^2]~=~O_p(1).
\label{one}
\ee
Therefore\tcr{,} we know that
\be
&&\phantom{~=~}\sbx|\hateo(\x,\theta,\mbP_0)-\enk^{(1)}(\x,\theta,\mbP_0)| \nonumber\\
&&~\leq~ c\,\ss \Enk \{|\hD(\X)h_n^{-r}K_h(\s-\S)| \}\nonumber\\
&&~\leq~ c\,h^{-r/2}\{\Enk [\{\hD(\X)\}^2] \ss\Enk[h_n^{-r}\{K_h(\s-\S)\}^2]\}^{1/2} \nonumber\\
&&~=~O_p(h^{-r/2}s_N)\tcr{,} \label{mnk}
\ee
where the first step uses the boundedness of $T\psi(Y,\theta)$, the second step is due to H\"older's inequality and the last step holds by (\ref{sn}) and (\ref{one}).

Combining (\ref{hmbe}) and (\ref{mnk}) yields that
\be
&&\phantom{~=~}\sbx|\hateo(\x,\theta,\hmbP)-\enk^{(1)}(\x,\theta,\mbP_0)| \nonumber\\ &&~=~O_p\{h_n^{-2}\alpha_n^2+h_n^{-1}\gamma_n\alpha_n+\alpha_n+h_n^{-(r/2+1)}\alpha_n s_N+h^{-r/2}s_N\} \nonumber\\
&&~=~O_p\{h_n^{-2}\alpha_n^2+h_n^{-1}\gamma_n\alpha_n+\alpha_n+h^{-r/2}s_N\} ~=~O_p\{a_{n,N}^{(2)}\},
\label{an2}
\ee
where the second step holds by the fact that $h_n^{-(r/2+1)}\alpha_n s_N=o(h^{-r/2}s_N)$ because we assume $h^{-1}\alpha_n=o(1)$.

Now\tcr{,} we handle the error $\enk^{(1)}(\x,\theta,\mbP_0)-e^{(1)}(\x,\theta,\mbP_0)$. Under Assumptions \ref{akernel_qte} (i)--(ii) and the fact that $\{\{\pis(\X)\}^{-1}T\psi(Y,\theta):\theta\in\mbtv\}$ is a VC class with a bounded envelope function $\sb[\{\pis(\X)\}^{-1}T\psi(Y,\theta)]$
from Assumption \ref{api}, Lemma B.4 of \citet{escanciano2014uniform} gives that
\be
\sbx|\enk^{(1)}(\x,\theta,\mbP_0)-\E\{\enk^{(1)}(\x,\theta,\mbP_0)\}|~=~O_p(\gamma_n).
\label{pt2}
\ee
Further, under Assumptions \ref{akernel_qte}, standard arguments based on $d$th order Taylor's expansion of $e^{(1)}(\x,\theta,\mbP_0)$ yield that
\be
\sbx|\E\{\enk^{(1)}(\x,\theta,\mbP_0)\}-e^{(1)}(\x,\theta,\mbP_0)|~=~O(h_n^d).
\label{pt3}
\ee

Combining (\ref{an2}), (\ref{pt2}) and (\ref{pt3}) yields
\be
\sbx|\hateo(\x,\theta,\hmbP)-e^{(1)}(\x,\theta,\mbP_0)|~=~O_p\{a_{n}^{(1)}+a_{n,N}^{(2)}\}.
\label{num}
\ee
Similar arguments imply that
\be
\sx|\hatez(\x,\hmbP)-e^{(0)}(\x,\mbP_0)|~=~O_p\{a_{n}^{(1)}+a_{n,N}^{(2)}\},
\label{deno}
\ee
where $\hatez(\x,\mbP)\equiv\hatez(\x,\theta,\mbP)$ and $\ e^{(0)}(\x,\mbP)\equiv e^{(0)}(\x,\theta,\mbP)$.
Therefore\tcr{,} we have
\bse
&&\phantom{~=~}\sbx|\phihatnk(\x,\theta,\hmbP)-\tphi(\x,\theta,\mbP_0)| \nonumber\\
&&~=~\sbx|\{\hatez(\x,\hmbP)\}^{-1}\hatez(\x,\theta,\hmbP)-\{e^{(0)}(\x,\mbP_0)\}^{-1}e^{(1)}(\x,\theta,\mbP_0)| \\
&&~\leq~\sbx|\{\hatez(\x,\mbP_0)\}^{-1}\{\hateo(\x,\theta,\hmbP)-e^{(1)}(\x,\theta,\mbP_0)\}|+ \\
&&\phantom{~=~}\sbx|[\{\hatez(\x,\mbP_0)\}^{-1}-\{e^{(0)}(\x,\mbP_0)\}^{-1}]e^{(1)}(\x,\theta,\mbP_0)| \\
&&~=~O_p\{a_{n}^{(1)}+a_{n,N}^{(2)}\},
\ese
where the last step follows from the fact that $a_{n}^{(1)}+a_{n,N}^{(2)}=o(1)$, and repeated use of (\ref{num}) and (\ref{deno}) as well as Assumptions \ref{api} and \ref{akernel_qte} (ii).

\subsection{Proof of Proposition \ref{thbn}}
Considering
\bse
\phihatnk(\x,\theta,\hmbP)~\equiv~ \{\hatez(\x,\theta,\hmbP)\}^{-1}\hateo(\x,\theta,\hmbP)\equiv\{\hatez(\x,\hmbP)\}^{-1}\hateo(\x,\theta,\hmbP)\tcr{,}
\ese
with
\bse
\hateo(\x,\theta,\mbP)~\equiv~ h_n^{-r}\Enk[\{\pihatN(\X)\}^{-1}T \{I(Y<\theta)-\tau\}K_h\{\mbP\trans(\x-\X)\},
\ese
it is obvious that, given $\cl$,
\bse
\{\phihatnk(\X,\theta,\hmbP):\theta\in\mbtv\}\subset\{\phihatnk(\X,\theta_i,\hmbP):i=1,\ldots,n+1\},
\ese
for any $\theta_1<Y_{(1)}$, $\theta_i\in[Y_{(i-1)},Y_{(i)})$ $(i=2,\ldots,n)$ and $\theta_{n+1}\geq Y_{(n)}$, where $Y_{(i)}$ is the $i$th order statistic of $\{Y_i:i=1,\ldots,n\}$. Therefore the set $\{\phihatnk(\X,\theta,\hmbP):\theta\in\mbtv\}$ contains at most $(n+1)$ different functions given $\cl$. This, combined with (\ref{mm}), implies the set
\bse
\mp_{n,k}~\equiv~\{\phihatnk(\X,\theta,\hmbP)-\phis(\X,\theta):\theta\in\mbtv\}
\ese
satisfies $N_{[\,]}\{\eta,\mp_{n,k}\mid\cl,L_2(\P_\X)\}\leq c\,(n+1)\eta^{-1}$.

\section{Additional simulation results}\label{sm_simulations}
We \tcr{present here} 
in Tables \ref{table_supp_efficiency} (efficiency) and \ref{table_supp_infernce} (inference) the results of \tcr{our} simulation\tcr{s for the} cases with the null and double index outcome models (d)--(e)\tcr{; s}ee 
Section \ref{sec_simulations} for detailed descriptions of the simulation setups. In the null model (d) where $Y$ and $\X$ are independent, it is apparent that the unlabeled data cannot help the estimation in theory, so the supervised and SS methods \tcr{not surprisingly} have close efficiencies. When the outcome model is (e), our SS estimators show significant superiority over the supervised competitors and even outperform the ``oracle'' supervised estimators most of time. As regards inference in the models (d) and (e), our methods still produce satisfactory results analogous \tcr{in pattern} to those in Table \ref{table_inferece} of Section \ref{sec_simulations}. The quantities in Tables \ref{table_supp_efficiency} and \ref{table_supp_infernce} again confirm the advantage of our SS estimators compared to their supervised counterparts in terms of robustness and efficiency, which have already been demonstrated \tcr{in detail} by the simulation results in Section \ref{sec_simulations}.

\begin{table}
\def~{\hphantom{0}}
\caption{
Efficiencies of the ATE and the QTE estimators relative to the corresponding oracle supervised estimators when $p=10$; \tcg{see Remark \ref{remark_interpretation_RE} for interpretations of these relative efficiencies.} Here\tcr{,} $n$ denotes the labeled data size, $p$ the number of covariates, $q$ the model sparsity, $m(\X)\equiv\E(Y\mid\X)$, $\pi(\X)\equiv\E(T\mid\X)$, $\hat{\pi}(\X)$ \tcr{--} the estimated propensity score, Lin \tcr{--} logistic regression of $T$ vs. $\X$\tcr{,} and Quad \tcr{--} logistic regression of $T$ vs. $(\X\trans,\X_{[1]}^2,\ldots,\X_{[p]}^2)\trans$; KS$_1/$KS$_2$ represents kernel smoothing on the one$/$two direction(s) selected by linear regression$/$
\tcr{sliced} inverse regression; PR \tcr{denotes} parametric regression\tcr{,} and ORE \tcr{denotes the} oracle relative efficiency. The \textbf{\tcn{blue}} color \tcr{indicates} 
the best efficiency in each case.}{
\resizebox{\textwidth}{!}{
\begin{tabular}{ccc||ccc|ccc||ccc|ccc||c}
\hline
\multicolumn{3}{c||}{\multirow{2}{*}{ATE}}   & \multicolumn{6}{c||}{$n=200$}                      & \multicolumn{6}{c||}{$n=500$}                      & \multirow{3}{*}{ORE} \\
\cline{4-15}
&  &  & \multicolumn{3}{c|}{Supervised} & \multicolumn{3}{c||}{\textbf{SS}}& \multicolumn{3}{c|}{Supervised} & \multicolumn{3}{c||}{\textbf{SS}}&                      \\
$m(\X)$              & $\pi(\X)$            & $\hat{\pi}(\X)$      & KS$_1$  & KS$_2$ & PR   & KS$_1$ & KS$_2$ & PR   & KS$_1$  & KS$_2$ & PR   & KS$_1$ & KS$_2$ & PR   &                      \\ \hline
\multirow{6}{*}{(d)} & (i)       & Lin             & 0.89    & 0.83   & 0.87 & \tcn{\bf 0.95}   & 0.94   & 0.91 & 0.93    & 0.95   & 0.94 & 0.93   & \tcn{\bf 0.97}   & 0.93 & 1.00                 \\
&           & Quad            & 0.68    & 0.50   & 0.64 & 0.95   & \tcn{\bf 0.96}   & 0.92 & 0.87    & 0.87   & 0.87 & 0.93   & \tcn{\bf 0.96}   & 0.93 & 1.00                 \\
& (ii)      & Lin             & 0.86    & 0.85   & 0.87 & 0.92   & \tcn{\bf 0.93}   & 0.92 & 0.96    & 0.94   & 0.97 & 0.99   & \tcn{\bf 1.00}   & 0.97 & 1.00                 \\
&           & Quad            & 0.75    & 0.77   & 0.67 & 0.92   & \tcn{\bf 0.94}   & 0.92 & 0.93    & 0.91   & 0.92 & 1.00   & \tcn{\bf 1.01}   & 0.98 & 1.00                 \\
& (iii)     & Lin             & 0.85    & 0.84   & 0.85 & 0.88   & \tcn{\bf 0.91}   & 0.86 & 0.93    & 0.95   & 0.94 & 0.94   & \tcn{\bf 0.96}   & 0.94 & 1.00                 \\
&           & Quad            & 0.71    & 0.72   & 0.72 & 0.90   & \tcn{\bf 0.92}   & 0.87 & 0.92    & 0.93   & 0.93 & 0.94   & \tcn{\bf 0.97}   & 0.95 & 1.00                 \\ \hline
\multirow{6}{*}{(e)} & (i)       & Lin             & 0.76    & 0.75   & 0.41 & 1.73   & \tcn{\bf 1.80}   & 0.77 & 0.86    & 0.87   & 0.64 & 2.02   & \tcn{\bf 2.04}   & 0.88 & 5.41                 \\
&           & Quad            & 0.68    & 0.70   & 0.29 & 1.74   & \tcn{\bf 1.78}   & 0.76 & 0.84    & 0.83   & 0.57 & 2.02   & \tcn{\bf 2.03}   & 0.88 & 5.41                 \\
& (ii)      & Lin             & 0.73    & 0.63   & 0.24 & \tcn{\bf 1.18}   & 0.94   & 0.34 & 0.81    & 0.71   & 0.15 & \tcn{\bf 1.35}   & 1.18   & 0.19 & 3.93                 \\
&           & Quad            & 0.69    & 0.59   & 0.27 & \tcn{\bf 1.25}   & 1.00   & 0.38 & 0.85    & 0.76   & 0.18 & \tcn{\bf 1.41}   & 1.23   & 0.21 & 3.93                 \\
& (iii)     & Lin             & 0.75    & 0.71   & 0.41 & \tcn{\bf 1.60}   & 1.57   & 0.72 & 0.74    & 0.77   & 0.53 & 1.32   & \tcn{\bf 1.43}   & 0.65 & 4.78                 \\
&           & Quad            & 0.74    & 0.75   & 0.52 & \tcn{\bf 1.83}   & 1.75   & 0.92 & 0.79    & 0.82   & 0.56 & 1.53   & \tcn{\bf 1.67}   & 0.85 & 4.78                 \\ \hline
\multicolumn{16}{c}{}                                                                                                                                                           \\
\hline
\multicolumn{3}{c||}{\multirow{2}{*}{QTE}}   & \multicolumn{6}{c||}{$n=200$}                      & \multicolumn{6}{c||}{$n=500$}                      & \multirow{3}{*}{ORE} \\
\cline{4-15}
&  &  & \multicolumn{3}{c|}{Supervised} & \multicolumn{3}{c||}{\textbf{SS}}& \multicolumn{3}{c|}{Supervised} & \multicolumn{3}{c||}{\textbf{SS}}&                      \\
$m(\X)$              & $\pi(\X)$            & $\hat{\pi}(\X)$      & KS$_1$  & KS$_2$ & PR   & KS$_1$ & KS$_2$ & PR   & KS$_1$  & KS$_2$ & PR   & KS$_1$ & KS$_2$ & PR   &                      \\ \hline
\multirow{6}{*}{(d)} & (i)       & Lin             & 0.87    & 0.86   & 0.78 & 0.92   & \tcn{\bf 0.95}   & 0.79 & 0.93    & 0.92   & 0.92 & 0.98   & \tcn{\bf 0.98}   & 0.92 & 1.00                 \\
&           & Quad            & 0.72    & 0.73   & 0.55 & 0.92   & \tcn{\bf 0.95}   & 0.79 & 0.89    & 0.88   & 0.89 & \tcn{\bf 0.99}   & 0.99   & 0.92 & 1.00                 \\
& (ii)      & Lin             & 0.87    & 0.86   & 0.89 & 0.93   & \tcn{\bf 0.94}   & 0.89 & 0.92    & 0.90   & \tcn{\bf 0.99} & 0.95   & 0.93   & 0.97 & 1.00                 \\
&           & Quad            & 0.71    & 0.71   & 0.71 & 0.94   & \tcn{\bf 0.96}   & 0.90 & 0.89    & 0.89   & 0.95 & 0.96   & 0.94   & \tcn{\bf 0.98} & 1.00                 \\
& (iii)     & Lin             & 0.83    & 0.82   & 0.85 & \tcn{\bf 0.92}   & 0.92   & 0.83 & 0.94    & 0.93   & 0.95 & 0.96   & \tcn{\bf 0.97}   & 0.96 & 1.00                 \\
&           & Quad            & 0.81    & 0.78   & 0.71 & 0.95   & \tcn{\bf 0.95}   & 0.83 & 0.92    & 0.92   & 0.94 & 0.97   & \tcn{\bf 0.99}   & 0.95 & 1.00                 \\ \hline
\multirow{6}{*}{(e)} & (i)       & Lin             & 0.82    & 0.79   & 0.78 & \tcn{\bf 1.30}   & 1.23   & 1.13 & 0.85    & 0.84   & 0.89 & 1.37   & 1.34   & \tcn{\bf 1.42} & 1.85                 \\
&           & Quad            & 0.65    & 0.68   & 0.61 & \tcn{\bf 1.30}   & 1.24   & 1.11 & 0.87    & 0.86   & 0.85 & 1.39   & 1.35   & \tcn{\bf 1.42} & 1.85                 \\
& (ii)      & Lin             & 0.61    & 0.55   & 0.49 & \tcn{\bf 0.92}   & 0.73   & 0.65 & 0.81    & 0.71   & 0.40 & \tcn{\bf 1.16}   & 0.97   & 0.48 & 1.78                 \\
&           & Quad            & 0.62    & 0.56   & 0.48 & \tcn{\bf 0.99}   & 0.80   & 0.70 & 0.82    & 0.73   & 0.44 & \tcn{\bf 1.23}   & 1.04   & 0.53 & 1.78                 \\
& (iii)     & Lin             & 0.75    & 0.70   & 0.73 & 1.13   & 1.08   & \tcn{\bf 1.22} & 0.82    & 0.82   & 0.85 & \tcn{\bf 1.34}   & 1.33   & 1.18 & 1.93                 \\
&           & Quad            & 0.78    & 0.74   & 0.84 & 1.28   & 1.23   & \tcn{\bf 1.44} & 0.86    & 0.87   & 0.85 & \tcn{\bf 1.45}   & 1.44   & 1.31 & 1.93                \\ \hline
\end{tabular}
}}
\label{table_supp_efficiency}
\end{table}

\begin{table}
\def~{\hphantom{0}}
\caption{
Inference based on the SS estimators \underline{\tcr{using} kernel smoothing on the direction selected by linear regression \tcr{(KS$_1$)}} \tcr{as the choice of the working outcome model, for the ATE and the QTE,} when $n=500$ and $p=10$. Here\tcr{,} ESE is the empirical standard error, \tcr{Bias is the empirical bias,} ASE \tcr{is} the average of the estimated standard errors\tcr{,} and CR \tcr{is} the \tcr{empirical} coverage rate of the 95\% confidence intervals. \tcr{All o}ther notations are the same as in Table \ref{table_supp_efficiency}. The \textbf{{\color{navyblue} blue}} color
\tcr{highlights settings where} the propensity scor\tcr{e} 
and the outcome mode\tcr{l} 
are \tcr{both} correctly specified, while the \textbf{boldfaces} \tcr{denote ones where} 
the propensity scor\tcr{e is} 
correctly specified but the outcome 
mode\tcr{l is} not.}{
\begin{tabular}{ccc|cccc|cccc}
\hline
&  &  & \multicolumn{4}{c|}{ATE}   & \multicolumn{4}{c}{QTE}   \\
$m(\X)$              & $\pi(\X)$            & $\hat{\pi}(\X)$      & ESE  & Bias & ASE  & CR   & ESE  & Bias & ASE  & CR   \\ \hline
& (i)   & {\color{navyblue} \textbf{Lin}}  & {\color{navyblue} \textbf{0.08}} & {\color{navyblue} \textbf{0.00}} & {\color{navyblue} \textbf{0.07}} & {\color{navyblue} \textbf{0.94}} & {\color{navyblue} \textbf{0.09}} & {\color{navyblue} \textbf{0.01}} & {\color{navyblue} \textbf{0.10}} & {\color{navyblue} \textbf{0.96}} \\
&       & {\color{navyblue} \textbf{Quad}} & {\color{navyblue} \textbf{0.08}} & {\color{navyblue} \textbf{0.00}} & {\color{navyblue} \textbf{0.07}} & {\color{navyblue} \textbf{0.94}} & {\color{navyblue} \textbf{0.09}} & {\color{navyblue} \textbf{0.01}} & {\color{navyblue} \textbf{0.10}} & {\color{navyblue} \textbf{0.95}} \\
& (ii)  & Lin                                  & 0.07                                 & 0.00                                 & 0.07                                 & 0.95                                 & 0.08                                 & 0.01                                 & 0.09                                 & 0.94                                 \\
&       & Quad                                 & 0.06                                 & 0.00                                 & 0.07                                 & 0.95                                 & 0.08                                 & 0.01                                 & 0.09                                 & 0.95                                 \\
& (iii) & Lin                                  & 0.07                                 & 0.00                                 & 0.07                                 & 0.94                                 & 0.08                                 & 0.01                                 & 0.09                                 & 0.97                                 \\
\multirow{-6}{*}{(d)} &       & {\color{navyblue} \textbf{Quad}} & {\color{navyblue} \textbf{0.07}} & {\color{navyblue} \textbf{0.00}} & {\color{navyblue} \textbf{0.06}} & {\color{navyblue} \textbf{0.93}} & {\color{navyblue} \textbf{0.08}} & {\color{navyblue} \textbf{0.01}} & {\color{navyblue} \textbf{0.09}} & {\color{navyblue} \textbf{0.96}} \\ \hline
& (i)   & \textbf{Lin}                         & \textbf{0.12}                        & \textbf{0.00}                        & \textbf{0.11}                        & \textbf{0.93}                        & \textbf{0.16}                        & \textbf{0.03}                        & \textbf{0.17}                        & \textbf{0.94}                        \\
&       & \textbf{Quad}                        & \textbf{0.12}                        & \textbf{0.00}                        & \textbf{0.11}                        & \textbf{0.94}                        & \textbf{0.16}                        & \textbf{0.03}                        & \textbf{0.17}                        & \textbf{0.94}                        \\
& (ii)  & Lin                                  & 0.10                                 & 0.04                                 & 0.11                                 & 0.95                                 & 0.15                                 & 0.06                                 & 0.16                                 & 0.96                                 \\
&       & Quad                                 & 0.10                                 & 0.04                                 & 0.11                                 & 0.95                                 & 0.14                                 & 0.05                                 & 0.16                                 & 0.95                                 \\
& (iii) & Lin                                  & 0.12                                 & 0.00                                 & 0.11                                 & 0.91                                 & 0.15                                 & 0.03                                 & 0.16                                 & 0.96                                 \\
\multirow{-6}{*}{(e)} &       & \textbf{Quad}                        & \textbf{0.11}                        & \textbf{0.00}                        & \textbf{0.10}                        & \textbf{0.91}                        & \textbf{0.14}                        & \textbf{0.02}                        & \textbf{0.15}                        & \textbf{0.95}                        \\
\hline
\end{tabular}
}
\label{table_supp_infernce}
\end{table}

\section{Supplement to the data analysis in Section \ref{sec_data_analysis}} \label{sm_data_analysis}
We present in Table \ref{table_data_analysis} the \tcr{detailed} numerical results of the data analysis in Section \ref{sec_data_analysis}, which
\tcr{were} illustrated \tcr{in} Figures \ref{figure_ate} and \ref{figure_qte}, \tcr{in course of our discussion of the analysis and the results.} 
\begin{table}[H]
\def~{\hphantom{0}}
\caption{$95\%$ confidence intervals of the ATE and the QTE in the HIV Drug Resistance data. Here\tcr{,} $m$ is the position of mutatio\tcr{n} 
regarded as the treatment. In the first row of the table, the notation\tcr{s} \tcr{of the form} \tcr{`A-B'} \tcr{refer to} 
estimating the propensity score and the outcome model by the methods \tcr{`A'} and \tcr{`B'}, respectively. Lin stands for logistic regression of $T$ vs. $\X$; KS$_2$ \tcr{--} kernel smoothing on the two directions selected by 
\tcr{sliced} inverse regression, PR \tcr{--} parametric regression\tcr{;} and RF \tcr{--} random forest. The abbreviations Sup and SS refer to supervised and SS estimators, respectively. The \textbf{\tcn{blue}} color \tcr{indicates} 
the shortest SS confidence interval in each case.}{
\resizebox{\textwidth}{!}{
\begin{tabular}{cc|cc|cc|cc}
\hline
& \multirow{2}{*}{$m$} & \multicolumn{2}{c|}{\bf{Lin-KS$_2$}}                              & \multicolumn{2}{c|}{\bf{Lin-PR}}                                  & \multicolumn{2}{c}{\bf{RF-RF}}                                   \\
&                      & Sup                          & \bf{SS}                           & Sup                          & \bf{SS}                        & Sup                          & \bf{SS}                           \\ \hline
\multirow{8}{*}{ATE} & 39                   & $[ 0.13 , 0.43 ]$            & $[ 0.13 , 0.38 ]$            & $[ 0.10 , 0.41 ]$            & $[ 0.11 , 0.36 ]$            & $[ 0.13 , 0.32 ]$            & $\tcn{\bf [ 0.13 , 0.32 ]}$            \\
& 69                   & $[ 0.12 , 0.44 ]$            & $[ 0.19 , 0.44 ]$            & $[ 0.10 , 0.42 ]$            & $[ 0.18 , 0.43 ]$            & $[ 0.19 , 0.40 ]$            & $\tcn{\bf [ 0.24 , 0.43 ]}$            \\
& 75                   & $[ 0.02 , 0.29 ]$            & $[ 0.08 , 0.32 ]$            & $[ 0.04 , 0.33 ]$            & $[ 0.07 , 0.33 ]$            & $[ 0.14 , 0.33 ]$            & $\tcn{\bf [ 0.17 , 0.35 ]}$            \\
& 98                   & $[ \hbox{-}0.02 ,  0.37   ]$ & $[  0.06 ,  0.37 ]$          & $[ 0.01 , 0.40 ]$            & $[ 0.05 , 0.36 ]$            & $[ 0.10 , 0.29 ]$            & $\tcn{\bf [ 0.13 , 0.33 ]}$            \\
& 123                  & $[ \hbox{-}0.16 ,  0.15   ]$ & $[ \hbox{-}0.12 ,  0.13   ]$ & $[ \hbox{-}0.15 ,  0.17   ]$ & $[ \hbox{-}0.10 ,  0.15   ]$ & $[ \hbox{-}0.15 ,  0.04   ]$ & $\tcn{\bf [ \hbox{-}0.15 ,  0.05   ]}$ \\
& 162                  & $[ \hbox{-}0.16 ,  0.19   ]$ & $[ \hbox{-}0.14 ,  0.12   ]$ & $[ \hbox{-}0.16 ,  0.18   ]$ & $[ \hbox{-}0.14 ,  0.13   ]$ & $[ \hbox{-}0.13 ,  0.07   ]$ & $\tcn{\bf [ \hbox{-}0.12 ,  0.09   ]}$ \\
& 184                  & $[ 2.02 , 2.36 ]$            & $[ 2.08 , 2.35 ]$            & $[ 2.03 , 2.37 ]$            & $[ 2.03 , 2.30 ]$            & $[ 2.08 , 2.30 ]$            & $\tcn{\bf [ 2.12 , 2.31 ]}$            \\
& 203                  & $[ 0.08 , 0.50 ]$            & $[ 0.17 , 0.51 ]$            & $[ 0.00 , 0.45 ]$            & $[ 0.08 , 0.45 ]$            & $[ 0.14 , 0.33 ]$            & $\tcn{\bf [ 0.20 , 0.38 ]}$            \\ \hline
\multirow{8}{*}{QTE} & 39  & $[ 0.07 , 0.43 ]$   & $[ 0.12 , 0.38 ]$   & $[ 0.05 , 0.42 ]$   & $[ 0.09 , 0.36 ]$   & $[ \hbox{-}0.01 ,  0.32 ]$ & $\tcn{\bf [  0.05 ,    0.30 ]}$ \\
& 69  & $[ \hbox{-}0.14 ,  0.16 ]$ & $\tcn{\bf [ \hbox{-}0.06 ,  0.18 ]}$ & $[ \hbox{-}0.14 ,  0.17 ]$ & $[ \hbox{-}0.06 ,  0.19 ]$ & $[ \hbox{-}0.13 ,  0.22 ]$ & $[ \hbox{-}0.06 ,  0.20 ]$   \\
& 75  & $[ \hbox{-}0.06 ,  0.29 ]$ & $\tcn{\bf [ \hbox{-}0.01 ,  0.26 ]}$ & $[ \hbox{-}0.09 ,  0.26 ]$ & $[ \hbox{-}0.04 ,  0.23 ]$ & $[ 0.03 , 0.42 ]$   & $[ 0.11 , 0.39 ]$     \\
& 98  & $[ 0.01 , 0.34 ]$   & $[ 0.00 , 0.29 ]$   & $[ 0.03 , 0.38 ]$   & $[ 0.00 , 0.28 ]$   & $[ \hbox{-}0.04 ,  0.37 ]$ & $\tcn{\bf [  0.02 ,  0.30 ]}$   \\
& 123 & $[ \hbox{-}0.16 ,  0.21 ]$ & $\tcn{\bf [ \hbox{-}0.12 ,  0.15 ]}$ & $[ \hbox{-}0.16 ,  0.22 ]$ & $[ \hbox{-}0.13 ,  0.15 ]$ & $[ \hbox{-}0.17 ,  0.29 ]$ & $[ \hbox{-}0.10 ,  0.18 ]$   \\
& 162 & $[ \hbox{-}0.25 ,  0.07 ]$ & $\tcn{\bf [ \hbox{-}0.23 ,  0.02 ]}$ & $[ \hbox{-}0.23 ,  0.09 ]$ & $[ \hbox{-}0.20 ,  0.05 ]$ & $[ \hbox{-}0.22 ,  0.16 ]$ & $[ \hbox{-}0.15 ,  0.11 ]$   \\
& 184 & $[ 2.16 , 2.50 ]$   & $[ 2.22 , 2.49 ]$   & $[ 2.15 , 2.49 ]$   & $\tcn{\bf [ 2.17 , 2.44 ]}$   & $[ 2.14 , 2.50 ]$   & $[ 2.23 , 2.50 ]$     \\
& 203 & $[ \hbox{-}0.15 ,  0.34 ]$ & $[  0.06 ,  0.41 ]$ & $[ \hbox{-}0.14 ,  0.34 ]$ & $[  0.06 ,  0.40 ]$ & $[ 0.01 , 0.40 ]$   & $\tcn{\bf [ 0.09 , 0.36 ]}$       \\ \hline
\end{tabular}}
}
\label{table_data_analysis}
\end{table}

\end{appendix}


\bibliographystyle{imsart-nameyear} 
\bibliography{myreference-te}       

\end{document}